\documentclass[12pt]{article}

\usepackage[T1]{fontenc}
\usepackage[english]{babel}
\usepackage[dvips]{epsfig}
\usepackage{amsmath}
\usepackage{amssymb}
\usepackage{array}
\usepackage{multicol}
\usepackage{xspace}
\usepackage{graphics}
\usepackage{indentfirst}
\usepackage{latexsym}
\usepackage{setspace}
\usepackage{makeidx}
\usepackage{vmargin}
\usepackage{amsmath,amsthm,amssymb,stmaryrd,wasysym}
\usepackage{color}
\usepackage{wrapfig}
\usepackage{ulem}
\usepackage{multirow}
\usepackage{stackrel}
\usepackage{mathrsfs}
\usepackage{fancyref}
\usepackage{graphicx}
\usepackage{enumerate}
\usepackage{mdwlist}
\usepackage{caption}
\usepackage{fancybox}
\usepackage{pifont}
\usepackage[all]{xy}
\usepackage{makecell}
\usepackage{epigraph}
\usepackage{pifont}
\input amssym.def

\usepackage{fancyhdr}
\makeindex

\def\be{\begin{equation}}
\def\ee{\end{equation}}
\def\bqn{\begin{eqnarray}}
\def\eqn{\end{eqnarray}}
\newcommand*\xbar[1]{
  \hbox{
    \vbox{
      \hrule height 0.5pt 
      \kern0.5ex
      \hbox{
        \kern-0.25em
        \ensuremath{#1}
        \kern-0.25em
      }
    }
  }
} 

\newcommand{\bea}{\begin{eqnarray}}
\newcommand{\eea}{\end{eqnarray}}
\newcommand{\p}{\partial}

\newcommand{\mR}{\mathbb R}

\newcommand{\bi}{\begin{itemize}}
\newcommand{\ei}{\end{itemize}}
\newcommand{\nn}{\nonumber}

\newcommand{\pl}{\left(}
\newcommand{\pr}{\right)}

\newcommand{\crl}{\left[}
\newcommand{\crr}{\right]}

\newcommand{\bR}{\textbf{R}}
\newcommand{\bb}{\textbf{b}}
\newcommand{\bc}{\textbf{c}}

\newcommand{\w}{\wedge}

\newcommand{\half}{\frac{1}{2}}

\newcommand{\Lag}{\mathcal{L}}

\newcommand{\C}{\mathscr{C}}
\newcommand\fonc[1]{C^\infty\pl#1\pr}
\newcommand{\nr}{nonrelativistic }
\newcommand{\rel}{relativistic }

\newcommand{\Augssp}{Augustinian structures}

\newcommand{\vf}{\field{T\M}}

\newcommand{\M}{\mathscr{M}}

\newcommand{\Om}{\Omega}
\newcommand{\om}{\omega}

\newcommand{\un}{^{-1}}

\newcommand{\ie}{\textit{i.e.}\, }
\newcommand{\cf}{\textit{cf.}\, }
\newcommand{\eg}{\textit{e.g.}\, }

\newcommand{\etal}{\textit{et al.}\, }

\newcommand{\br}[2]{\crl #1,#2 \crr}

\newcommand{\NbG}{\N{\mathbf{G}}}
\newcommand\pset[1]{\left \lbrace #1\right \rbrace}
\newcommand\bprop[1]{\begin{prop}#1\end{prop}}
\newcommand\bpropp[2]{\begin{prop}[#1]#2\end{prop}}
\newcommand\bdefi[2]{\begin{defi}[#1]#2\end{defi}}
\newcommand\bnota[1]{\begin{notation}#1\end{notation}}
\newcommand\blem[2]{\begin{lem}[#1]#2\end{lem}}
\newcommand\bcor[2]{\begin{cor}[#1]#2\end{cor}}
\newcommand\bthm[2]{\begin{thm}[#1]#2\end{thm}}
\newcommand\bexa[2]{\begin{exa}[#1]#2\end{exa}}

\newcommand{\Ker}{\text{\rm Ker }}
\newcommand{\Rad}{\text{\rm Rad }}
\newcommand{\Ann}{\text{\rm Ann }}
\newcommand{\End}{\text{\rm End}}
\newcommand{\Dim}{\text{\rm Dim}}

\newcommand{\GALZ}{{\text{\rm Gal}_0}}

\newcommand\N[1]{\overset{N}{#1}}

\newcommand\Np[1]{\overset{N'}{#1}}
\newcommand\Z[1]{\overset{Z}{#1}}

\newcommand\bcase[1]{\bea\begin{cases}#1\end{cases}\eea}
\newcommand\Riem[4]{R^{\hspace{0.3mm}#1}_{\hspace{1.5mm}{#2}\hspace{0.2mm}{#3}\hspace{0.2mm}{#4}}}
\newcommand\Riemd[4]{R^{\hspace{0.3mm}#1\hspace{0.2mm}#2}_{\hspace{4mm}#3\hspace{0.2mm}#4}}
\newcommand\Rieme[4]{R^{\hspace{0.3mm}#1\hspace{2.5mm}#2}_{\hspace{2.5mm}#3\hspace{2.5mm}#4}}
\newcommand\Riemf[2]{R^{\hspace{0.3mm}#1}_{\hspace{1.5mm}{#2}}}

\newcommand\Riemh[4]{R^{\hspace{0.3mm}#1\hspace{2.5mm}#3\hspace{1.5mm}#4}_{\hspace{3mm}#2}}
\newcommand\Riemi[4]{R^{\hspace{0.3mm}#1\hspace{0.5mm}#2\hspace{0.5mm}#3}_{\hspace{7mm}#4}}
\newcommand\Riemj[4]{R_{\hspace{0.3mm}#1\hspace{0.2mm}{#2}\hspace{0.2mm}{#3}\hspace{0.2mm}{#4}}}
\newcommand\Jaco[4]{J^{\hspace{0.3mm}#1}_{\hspace{1.5mm}{#2}\hspace{0.2mm}{#3}\hspace{0.2mm}{#4}}}

\newcommand\bN[1]{\overset{\bar N}{#1}}

\newcommand\form[1]{\Omega^1\pl #1\pr}
\newcommand\forma[2]{\Omega^{#1}\pl #2\pr}
\newcommand\bforma[1]{\field{\vee^2\, #1}}

\newcommand\bform{\field{\vee^2\, T\M}}
\newcommand\bforms{\field{\vee^2\, T^*\M}}
\newcommand\field[1]{\Gamma\pl #1\pr}
\newcommand{\ff}{\form{\M}}

\newcommand{\LCc}{Levi-Civita connection }

\newcommand{\fo}{field of observers }
\newcommand{\fop}{field of observers}
\newcommand{\fos}{fields of observers }

\newcommand{\Cffo}{Coriolis-free field of observers }

\newcommand\Prop[1]{Proposition \ref{#1}}
\newcommand\Defi[1]{Definition \ref{#1}}
\newcommand\Lemma[1]{Lemma \ref{#1}}
\newcommand\Span[1]{\text{\rm Span}\hspace{1.2mm} #1}

\newcommand\Spannn[1]{\text{\rm Span}\left\lbrace #1\right\rbrace}

\newcommand{\prd}{\text{Phys. Rev. D}}

\newcommand{\vectddu}{\field{T^*\M\otimes T^*\M\otimes T\M}}
\newcommand{\vectsymdu}{\field{\vee^2T^*\M\otimes T\M}}
\newcommand{\vectasymdu}{\field{\w^2T^*\M\otimes T\M}}
\newcommand{\vectasymduker}{\field{\w^2T^*\M\otimes \Ker\psi}}
\newcommand{\vectsum}{\forma{2}{\M}\oplus\field{\w^2T^*\M\otimes\Ker\psi}}
\newcommand{\vectsumone}{\ff\oplus\field{\w^2T^*\M\otimes\Ker\psi}}
\newcommand{\dPhiNV}{d\hspace{-1.5mm}\PhiNV\hspace{-1mm}}
\newcommand{\Milneg}{\field{\Ker\psi}}
\newcommand{\GammaV}{\mathscr V\pl\M, \psi, \gamma\pr}
\newcommand{\GammaVr}{\mathscr V\pl\M, g\pr}
\newcommand{\PhiNV}{\overset{N,V}{\Phi}}
\newcommand{\lmn}{{}^\lambda_{\mu\nu}}
\newcommand{\lmna}{{}^\lambda_{[\mu\nu]}}
\newcommand{\FO}{FO\pl\M,\psi\pr}
\newcommand{\A}{\mathscr A}
\newcommand{\D}{\mathscr D}
\newcommand{\V}{\mathscr V}

\newcommand{\W}{\mathscr W}
\newcommand{\mP}{\mathscr P}
\newcommand{\PC}{PC\pl\M,\xi\pr}

\newcolumntype{M}[1]{>{\centering}m{#1}}
\newcommand*{\longhookrightarrow}{\ensuremath{\lhook\joinrel\relbar\joinrel\rightarrow}}

\theoremstyle{definition}

\newlength{\blength}
\renewcommand{\proof}[1]{\vspace{-.05cm}
\begin{list}{\bf Proof:}
{\listparindent=\parindent\parsep=0pt \labelwidth=-0.5cm
\labelsep=\parindent \addtolength{\labelsep}{-\blength}
\addtolength{\labelsep}{1.5cm}
\itemindent=-\blength
\addtolength{\itemindent}{\parindent} \leftmargin=1.0cm}
\item
#1~$\qedsymbol$\end{list}
\vspace{.0cm}}

\theoremstyle{plain}
\newtheorem{thm}{Theorem}[section]
\newtheorem{lem}[thm]{Lemma}

\newtheorem{cor}[thm]{Corollary}
\newtheorem{prop}[thm]{Proposition}

\newtheorem{defi}[thm]{Definition}
\theoremstyle{definition}
\newtheorem{notation}[thm]{Notation}

\newtheorem{exa}[thm]{Example}

\begin{document}
\pagenumbering{gobble}
\thispagestyle{empty}

 \begin{centering}

{\large {\bfseries 
Connections and dynamical trajectories \\\vspace{1mm} 
in generalised Newton-Cartan gravity I.\\\vspace{1mm}
An intrinsic view
}
\\\vspace{2mm}
}

 \vspace{5mm}
Xavier Bekaert \& Kevin Morand\\
\vspace{4mm}
{\small Laboratoire de Math\'ematiques et Physique Th\'eorique}\\
{\small Unit\'e Mixte de Recherche $7350$ du CNRS}\\
{\small F\'ed\'eration de Recherche $2964$ Denis Poisson}\\
{\small Universit\'e Fran\c{c}ois Rabelais, Parc de Grandmont}\\
{\small 37200 Tours, France} \\
\vspace{1mm}{\tt \footnotesize Xavier.Bekaert@lmpt.univ-tours.fr}\\
\vspace{1mm}{\tt \footnotesize Kevin.Morand@lmpt.univ-tours.fr}

\vspace{5mm}

\end{centering}

\begin{abstract}
The ``metric'' structure of nonrelativistic spacetimes consists of a one-form (the absolute clock) whose kernel is endowed with a positive-definite metric.                                                                                                                   
Contrarily to the relativistic case, the metric structure and the torsion do not determine a unique Galilean (\ie compatible) connection. This subtlety is intimately related to the fact that the timelike part of the torsion is proportional to the exterior derivative of the absolute clock. When the latter is not closed, torsionfreeness and metric-compatibility are thus mutually exclusive. We will explore generalisations of Galilean connections along the two corresponding alternative roads in a series of papers. In the present one, we focus on compatible connections and investigate the equivalence problem (\ie the search for the necessary data allowing to uniquely determine connections) in the torsionfree and torsional cases. More precisely, we characterise the affine structure of the spaces of such connections and display the associated model vector spaces. In contrast with the relativistic case, the metric structure does not single out a privileged origin for the space of metric-compatible connections. In our construction, the role of the Levi-Civita connection is played by a whole class of privileged origins, the so-called torsional Newton-Cartan (TNC) geometries recently investigated in the literature. Finally, we discuss a generalisation of Newtonian connections to the torsional case. 

\end{abstract}

\vspace{1.5cm}

\vspace{.5cm}

\vspace{4.5cm}

\pagebreak

\tableofcontents

\pagebreak

\pagenumbering{arabic}

\setcounter{page}{1}
\section{Introduction}

As advocated by \'Elie Cartan after the birth of Einstein's theory, the geometrisation of gravity induced by the equivalence principle is by no means restricted to General Relativity \cite{Cartan1923} (\cf also \cite{Friedrichs1928}). 
In this light, Einstein's and Newton's theories of gravity both admit geometrical formulations which are, in particular, diffeomorphism
invariant.
Since the sixties, the corresponding Newton-Cartan geometry has known a revival of interest among relativists and geometers (\cf \eg \cite{Trautman1963,Havas1964,Dombrowski1964,Kunzle1972,Duval1977,Ehlers1981,Trumper1983} for early contributions)
but it is only recently that Newton-Cartan geometry
has been intensively applied to condensed matter problems\footnote{Among the early applications of Newton-Cartan geometry to condensed matter systems is the pioneering work \cite{Carter1994} on superfluid dynamics. More recently, the related concept of ``nonrelativistic general covariance'' was applied to the unitary Fermi gas \cite{Son2006}. } such as the quantum Hall effect \cite{Son2013} for which it proved a very efficient tool to construct effective field theories or for computing Ward identities.

As celebrated in the famous quote\footnote{``Space[time] tells matter how to move. Matter tells space[time] how to curve.''} of Wheeler, there are two facets of the interaction between the geometry of spacetime and the motion of matter. We will focus on the ``kinematical'' facet, \ie the motion of test particles in a fixed gravitational background and will ignore the ``dynamical''
facet, \ie gravitational field equations. In this restricted case, the equivalence and relativity principles strongly prescribe the geometric structures the spacetime is endowed with. 
On the one hand, 
the equivalence principle imply that dynamical trajectories of free falling observers are geodesics of a suitable connection, the latter providing a notion of parallelism on the spacetime manifold. Furthermore, such unparameterised geodesics define a \textit{projective structure} on spacetime. 
On the other hand, 
the relativity principle\footnote{As emphasised by many authors (\eg \cite{L'evy-Leblond1974}), the so-called ``nonrelativistic'' theories also embody the principle of relativity, the only actual (but decisive) difference between Special and Galilean relativity being the expression of (Lorentz \textit{vs} Galilei) boosts. Although the terminology ``nonrelativistic'' is rather unfortunate, we will use it following common practice.}
further dictates the underlying structure group (Lorentzian \textit{vs} Galilean\footnote{An exhaustive enumeration of homogeneous kinematical groups \cite{Bacry1968} must also include the (homogeneous) Carroll group (\cf \cite{L'evy-Leblond1974,Duval2014e}). }) of the reduced frame bundle. The corresponding invariant tensor(s) define a \textit{metric structure} on spacetime.
An important issue is the interplay between these two structures: metric and connection. More precisely, one should answer the following question:
What are the ingredients (\eg the torsion) one must add to the metric structure in order to fix uniquely the connection?
Providing precise answers to this question (sometimes referred to as the ``equivalence problem'' in the mathematics literature) for some generalisations of Newton-Cartan geometry is the main subject of this paper.

In (pseudo)-Riemannian geometry, the answer is well known and provides a clear relation between the various elements constituting the kinematical content of general relativity which can be summarised in the following diagram:

\begin{figure}[ht]
\centering
   \includegraphics[width=0.4\textwidth]{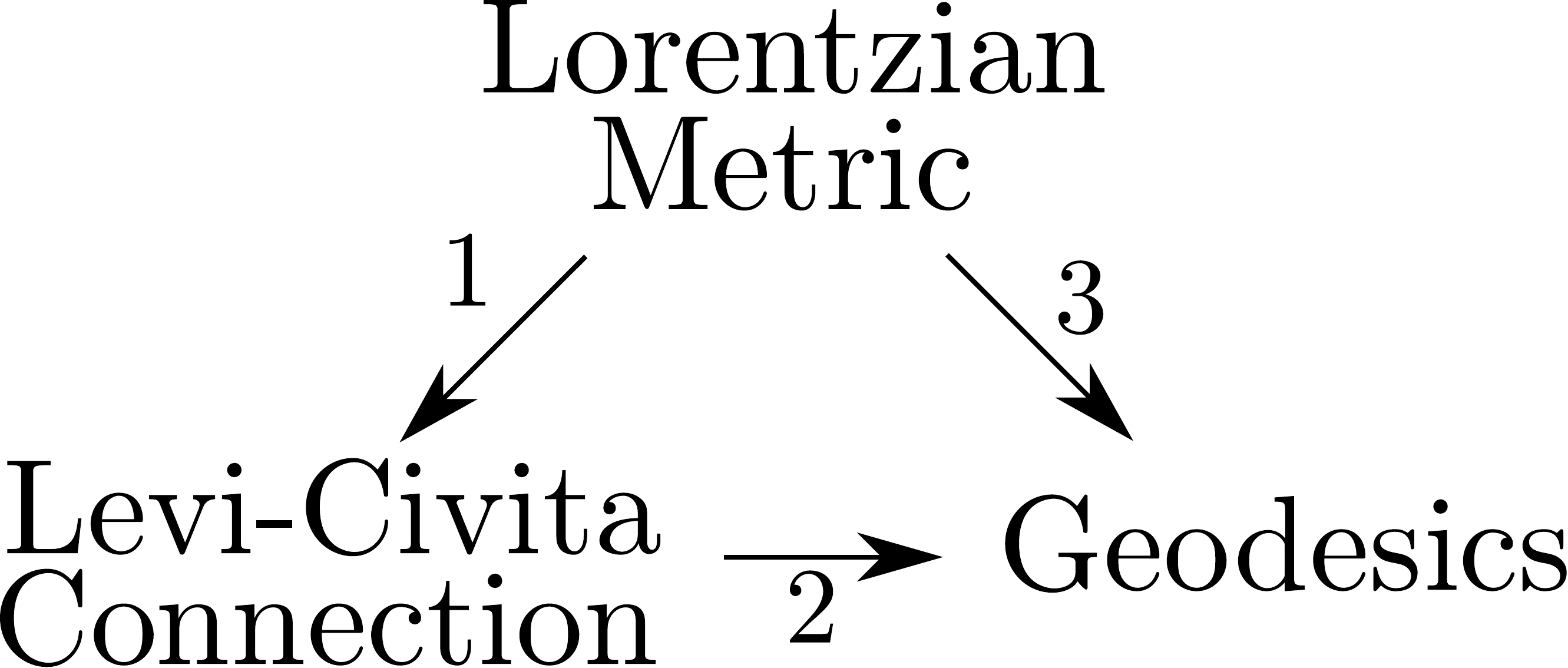}
  \caption{Kinematical content of general relativity\label{diagcinematique}}
\end{figure}

\noindent Let us briefly make some comments in order to present the logic that will be generalised in the less familiar nonrelativistic case.
On top of the triangle sits the metric structure of general relativity: a Lorentzian metric, \ie a field of nondegenerate bilinear forms on the spacetime manifold. This metric structure uniquely determines a compatible torsionfree  connection known as the Levi-Civita connection (Arrow 1). This connection provides the spacetime manifold with a notion of parallelism, thus allowing the definition of a distinguished class of curves: the geodesics (Arrow 2). A geodesic is thus defined as an autoparallel curve with respect to Levi-Civita's parallelism, \ie the tangent vector stays parallel to itself along a geodesic.     
Alternatively, the geodesics can be characterised as curves extremising locally the Lorentzian distance. As a result, the geodesic equation can be obtained as the equation of motion derived from a Lagrangian density built in terms of the metric structure (Arrow 3). 

The relations between these different structures can be abstractly summed up in a commutative diagram which will be our leitmotiv:

\begin{figure}[ht]
\centering
   \includegraphics[width=0.4\textwidth]{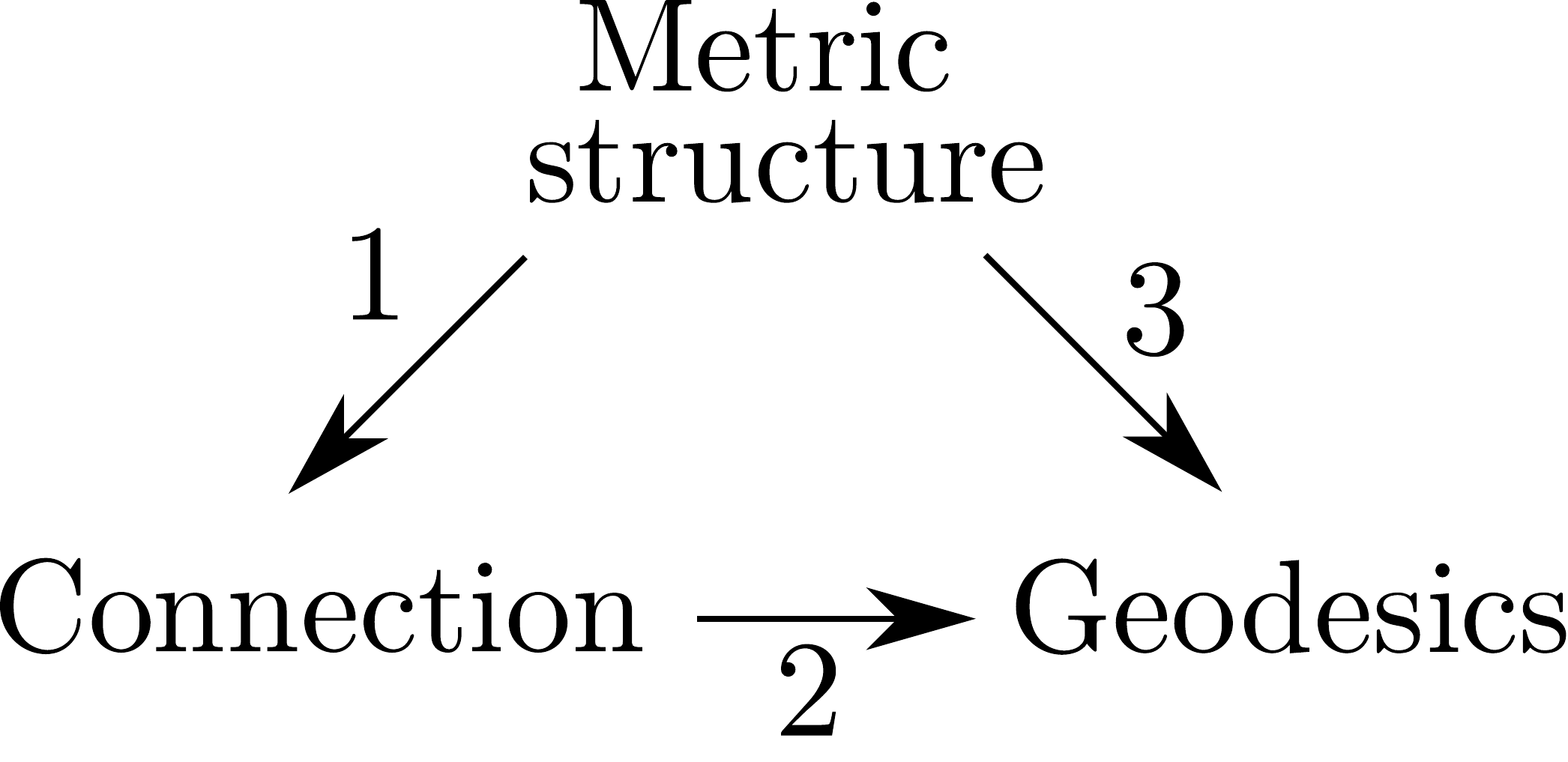}
  \caption{Kinematical content of metrical theories of gravitation\label{diagcinematique2}}
\end{figure}

\noindent Interestingly, the kinematical content of Newton-Cartan gravity (often referred to as Newton-Cartan geometry) can be equally described via a similar diagram, with the important difference that the basic nonrelativistic analogue of the metric structure consists in a degenerate contravariant metric $h^{\mu\nu}$ (a collection of {\it absolute rulers}) 
whose radical is spanned by a nowhere vanishing 1-form $\psi_\mu\neq 0$ (an \textit{absolute clock}): $h^{\mu\nu}\psi_\nu=0$. 
Connections compatible with such a structure are called \textit{Galilean}. 
Two features of nonrelativistic compatible connections $\Gamma^\lambda_{\mu\nu}$ are notably distinct from the relativistic case. 

Firstly, the torsion of a Galilean connection obeys a compatibility condition: its timelike part is proportional to the exterior derivative of the absolute clock ($\nabla_\mu\psi_\nu=0\,\Rightarrow\,\Gamma^\lambda_{[\mu\nu]}\psi_\lambda=\partial_{[\mu}\psi_{\nu]}$). In particular, torsionfree ($\Gamma^\lambda_{[\mu\nu]}=0$) Galilean connections are only defined for closed absolute clocks ($\partial_{[\mu}\psi_{\nu]}=0$). Such absolute clocks are synchronised in the sense that they define a notion of \textit{absolute time} $t$ (locally, $\psi_\mu=\partial_\mu t$). The simultaneity leaves ($t$ = constant) foliate spacetime. 

Secondly, the uniqueness of the torsionfree compatible connection is lost. 
This arbitrariness has a natural physical interpretation: the above ``metric'' structure is too weak to determine the motion of particles. Indeed, motions can be measured via absolute clocks and rulers, but are not constrained by them.
In Newtonian mechanics, the spacetime is a mere container and one should prescribe force fields to determine motion.

Diagram \ref{diagcinematique2} suggests to define a richer metric structure (dubbed here {\it Lagrangian structure}) allowing to restore the uniqueness of the torsionfree compatible connection (Arrow 1). 
This Lagrangian structure defines a unique \textit{Newtonian} connection:

\begin{figure}[ht]
\centering
   \includegraphics[width=0.4\textwidth]{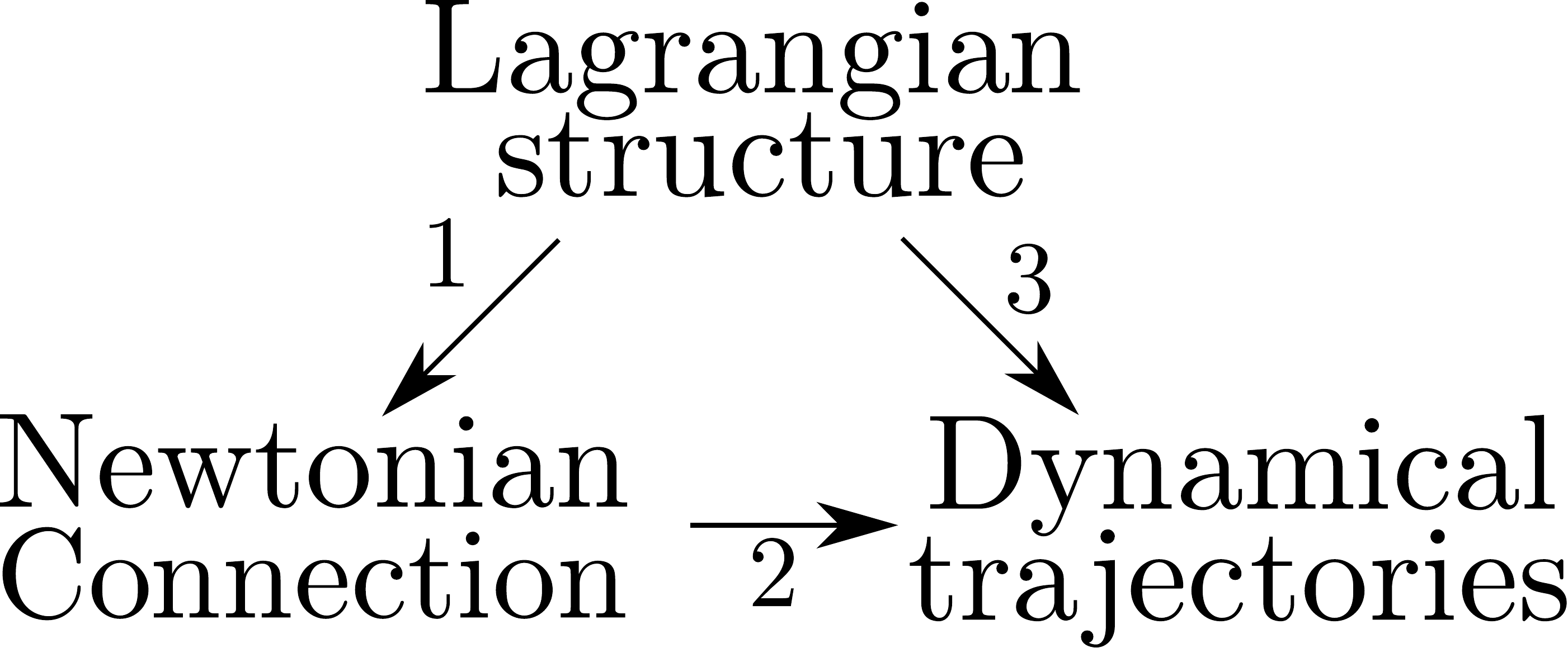}
  \caption{Kinematical content of Newton-Cartan theory\label{diagcinematique3}}
\end{figure}

\noindent  
A {Newtonian} connection endows the spacetime with a notion of parallelism (different from the Levi-Civita one) allowing in turn the definition of self-parallel curves, similarly to the relativistic case. Such curves acquire the interpretation of dynamical trajectories (Arrow 2) for Lagrangians
which are of degree two in the velocities\footnote{This class is natural in Newtonian mechanical systems with holonomic constraints. Recall that a dynamical system with Euclidean coordinates $x^1,\ldots,x^{d+n}$ is said {\it holonomic} if its constraints can be put 
in the form $f^\alpha\pl x^1,\ldots,x^{d+n},t\pr=0$, with $\alpha\in\pset{1,\ldots,n}$ and $n$ the number of independent constraints. The constraints of an holonomic system whose kinetic energy takes the standard form $T=\half \delta_{ab}\dot x^a\dot x^b$ (with $a,b=1,\ldots,d+n$) can always be solved. 
Such a system is therefore equivalent to an unconstrained system with Lagrangian of the form 
$L=\half \gamma_{ij}(q)\dot q^i\dot q^j+A_i(q)\dot q^i-U(q)$ (with $i,j=1,\ldots,d$). The corresponding class of Hamiltonians (of degree two in the momenta) are nowadays called ``natural Hamiltonians'', \cf \cite{Benenti2001}.}:
they can be derived from an action principle built in terms of the Lagrangian structure (Arrow 3).
In a sense, the nonrelativistic analogue of the Lorentzian distance between two events is in fact the value of the action $\int {\cal L}\,dt$, which is
a sort of ``Lagrangian distance''.

When the absolute clock is not closed, metric-compatibility and torsionfreeness are mutually exclusive.
Therefore two alternatives open up: consider either non-Galilean or torsional connections. We will explore these two alternative roads in a series of papers. In a forthcoming work \cite{Bekaert2016}, we will investigate the first option when the absolute clock is twistless, \ie obeys to the Frobenius integrability condition $\psi_{[\mu}\partial_{\nu}\psi_{\rho]}=0$. In such case, the time units vary for each clock and the measured time $\tau$ will depend on the observers.
Nevertheless, spacetime is still foliated by simultaneity slices and a notion of absolute time can be defined. 
{We will show that} one can also define a Lagrangian structure in the case of a twistless absolute clock associated with the action principle $\int {\cal L}\,d\tau$ making use of the measured time $\tau$ instead of the absolute one $t$. Furthermore, the latter Lagrangian structure is conformally related to one for a closed absolute clock. Correspondingly, {we will generalise the} diagram \ref{diagcinematique3} by defining a torsionfree (non Galilean) connection which is uniquely determined by the Lagrangian structure (Arrow 1) and projectively related to a Newtonian connection whose geodesics describe dynamical trajectories (Arrow 2) extremising the corresponding action principle (Arrow 3).

{Before addressing this issue,
we explore the alternative route in the present paper} by considering generalisations of Newton-Cartan gravity characterised by torsional connections which have known a recent surge of interest regarding applications in the geometrisation of condensed matter problems \cite{Geracie2014,TNCcm,Jensen2014} as well as in the context of Lifshitz and Schr\"odinger holography \cite{TNCLif,TNCMilneLif}. In such approaches, the torsion is tuned in order to ensure compatibility with the absolute clock. Of particular mathematical interest for us are the works \cite{Jensen2014,TNCMilneLif} which exhibit a torsional connection compatible with the metric structure, while remaining invariant under local Galilean boosts (called \textit{Milne boosts}) {as it should since a connection is a geometrical object independent of the frame used to represent it}. 
We extend these torsional Newton-Cartan geometries and make use of Lagrangian structures\footnote{Note however that the action principle for the geodesic equation becomes unclear whenever torsion is involved, so that we will not consider the third arrow of diagram \ref{diagcinematique2} in this case. Let us remind here some related subtleties in the presence of torsion.
Two connections defining the same parameterised geodesics differ only by their torsion. However, in the presence of a metric, a torsionful connection defining the same parameterised geodesics as the Levi-Civita connection is not metric compatible. Conversely, a metric compatible torsionful connection does not define the same parameterised geodesics as the Levi-Civita connection. } in order to identify the necessary data which allows to uniquely fix these connections.

\paragraph{Outline}
~\\ 

\noindent The plan of the paper is as follows:
\vspace{2mm}

In Section \ref{NRmetricstructures}, we review various geometric structures of nonrelativistic spacetimes. After a brief reminder of standard definitions and properties regarding relativistic structures, we switch to the investigation of \nr ones by emphasising their points of divergence with their \rel counterparts. 
We focus on a nonrelativistic metric structure (called {\it Leibnizian} structure) defined as a manifold endowed with a degenerate contravariant metric whose radical is spanned by the {absolute clock}. 
The role played by \fos in \nr physics is discussed at length as well as related objects. 
We then discuss two restrictions that can be imposed on the absolute clock, namely closure ({\it Augustinian} structure) or the Frobenius criterion ({\it Aristotelian} structure). 

In Section \ref{Nonrelativistic manifolds}, we discuss the possibility of endowing nonrelativistic metric structures with a notion of parallelism, in the guise of a connection. We first focus on torsionfree connections compatible with the underlying metric structure, thus restricting the scope of the analysis to \Augssp. We thus review the notions of torsionfree {\it Galilean} and {\it Newtonian} connections, with particular attention given to the equivalence problem 
(\ie the search for structures that uniquely determine a given compatible connection). Apart from the standard characterisation of Newtonian connections in terms of equivalence classes of \fo and gauge 1-forms, this motivation will lead us to review the less standard solution of the equivalence problem making use of a {\it Lagrangian} structure. The latter can be thought of as the proper nonrelativistic analogue of the (pseudo)-Riemannian metric structure in that it determines uniquely the compatible torsionfree connection. 

In Section \ref{Torsional connection}, we discuss the extension of ``torsional Newton-Cartan geometry'' \cite{Geracie2014,TNCcm,Jensen2014,TNCMilneLif} to the class of all torsional Galilean connections. Furthermore, we introduce a torsional generalisation of Newtonian connections. We then discuss in details the affine structure of the space of torsional Galilean connections 
and thereby identify the necessary data which allows to uniquely fix torsional Galilean connections. 

The Section \ref{Conclusion} is our conclusion where we briefly summarise our main results and announce some future ones. In a forthcoming paper, we will show how the generalisations of Newton-Cartan geometry we have discussed can be obtained as null dimensional reductions of suitable Lorentzian geometries.

Two appendices close the paper. A detailed discussion of the equivalences between the Trautman and Duval-K\"unzle conditions is provided in Appendix \ref{Curvtfree}. Appendix \ref{proofaffine} consists in a short review on affine spaces while several technical proofs have been relegated to Appendix \ref{proofpropZ}. 
~\\

\paragraph{Notations}
~\\

\noindent Let $V$ be a vector space and $v,w\in V$ two vectors. We will denote by $v\vee w=\half\pl v\otimes w+w\otimes v\pr$ (respectively $v \wedge w=\half\pl v\otimes w-w\otimes v\pr$) the (anti)symmetric product, and similarly for higher products.  
The (anti)symmetrisation of indices is performed with weight one and is denoted by round (respectively, square) brackets,
\eg $\Phi_{(\mu\nu)}\equiv\half\pl\Phi_{\mu\nu}+\Phi_{\nu\mu}\pr$ and
$\Phi_{[\mu\nu]}\equiv\half\pl\Phi_{\mu\nu}-\Phi_{\nu\mu}\pr$.

\noindent The spacetime manifold will be written $\M$ and is of dimension $d+1$.
Let $\cal V$ be a vector bundle over $\M$ with typical fibre the vector space $V$.
By $\Gamma({\cal V})$, we will denote the space of its sections, \ie globally defined $V$-valued fields on $\M$.
{For instance, $\Gamma(\wedge^pT^*\M)=\Omega^p(\M)$ is the space of $p\,$-forms on $\M$.}

\pagebreak
\section{Nonrelativistic metric structures}\label{NRmetricstructures}
\label{NRstructures}

\noindent We start by reviewing some standard material about relativistic structures in order to draw comparison with nonrelativistic ones and fix some terminology. 

\subsection{Relativistic structures}
\label{sectionrelativisticstructures}
\bdefi{Riemannian structure}{A Riemannian structure designates a manifold endowed with a positive-definite metric. }
\noindent Although this definition restricts to the case of signature $\pl+,\ldots,+\pr$, a similar one can be given in the (pseudo)-Riemannian case:
\bdefi{Lorentzian structure}{\label{defilorentzian}A Lorentzian structure consists in a manifold endowed with a nondegenerate metric of signature $\pl-,+,\ldots,+\pr$. }
\noindent These structures are therefore characterised by a metric structure but, as such, are not endowed with a notion of parallel transport. This supplementary notion of parallelism can be implemented under the features of a Koszul 
connection\footnote{We will prefer the denomination ``Koszul connection'' to the more widespread designations of ``affine connection'' or ``covariant derivative'' in order to avoid confusion with the slightly different meanings of these terms in some of the mathematical literature. 
For the sake of completeness, let us remind that a Koszul connection on a vector bundle $E$ over $\M$ is a $C^\infty(\M)$-linear map $\nabla:\Gamma\pl T\M\pr\to \End\big(\field{E}\big)$ such that, for any vector field $X\in\Gamma\pl T\M\pr$, the endomorphism $\nabla_X$ on the space $\field{E}$ of sections obeys to the Leibniz rule:
$\nabla_X\pl f\sigma\pr=X\crl f\crr \sigma+f\nabla_X\sigma$ for any function $f\in C^\infty(\M)$ and section $\sigma\in\field{E}$. If the vector bundle is unspecified, it will be implicitly assumed to be the tangent bundle: $E=T\M$. } compatible with the metric structure. We are thus led to define: 
\bdefi{Riemannian/Lorentzian manifold}{A Riemannian (Lorentzian) manifold consists in a Riemannian (Lorentzian) structure supplemented with a metric-compatible Koszul connection
on the tangent bundle. }
\noindent We will retain this terminology in the sequel and use the word ``structure'' in order to designate a manifold endowed with a metric-like structure while keeping the term ``manifold'' for cases where a Koszul connection is added. However, in the present case the distinction drawn here is only relevant when the Koszul connection has torsion due to the following well-known theorem:
\bthm{Space of metric compatible connections}{\label{fundamentaltheorem}The space of Lorentzian connections compatible with a given Lorentzian structure $\pl\M,g\pr$ forms a vector space which is isomorphic to the vector space $\field{\w^2\, T^*\M\otimes T\M}$ of torsion tensors
and the origin of which is the Levi-Civita connection.
}
\noindent In order to pave the way to the next Section, we now provide a detailed proof of the previous Theorem in a guise suited for its extension to the \nr case. 
\noindent 
We start by recalling that, when no metric structure is involved, the space of Koszul connections on a manifold $\M$ possesses the structure of an affine space modelled on the vector space of 2-covariant, 1-contravariant tensor fields $\vectddu$. This translates the well-known fact that the difference between two Koszul connections on the same manifold is a tensor field $\big($an element of $\vectddu\,\,\big)$ although a Koszul connection is not. Let $\pl\M,g\pr$ be a Lorentzian structure and denote $\mathscr D\pl\M,g\pr$ the space of compatible connections. The compatibility condition $\nabla g=0$ restricts the {difference $S\equiv\Gamma'-\Gamma$ of two compatible connections $\Gamma', \Gamma\in\mathscr D\pl\M,g\pr$ to be such that $S^{(\lambda}_{\mu\nu}\, g^{\rho)\nu}_{}=0$.}
The following Proposition then holds:
\bprop{\label{PropLorentzianaffinespace}The space $\mathscr D\pl\M,g\pr$ of compatible Lorentzian connections possesses the structure of an affine space modelled on the vector space 
\bea\GammaVr\equiv\pset{S\in\vectddu\ /\ S^{(\lambda}_{\mu\nu}\, g^{\rho)\nu}_{}=0}. \nn
\eea
}
\noindent {Indeed, given two Lorentzian connections $\Gamma', \Gamma\in\mathscr D\pl\M,g\pr$, the element $S\equiv\Gamma'-\Gamma$ belongs to $\GammaVr$.} In order to reduce the structure of $\mathscr D\pl\M,g\pr$ from that of an affine space to that of a vector space, one needs to pick an origin $\overset{0}{\Gamma}\in\mathscr D\pl\M,g\pr$ thus allowing to put $\mathscr D\pl\M,g\pr$ and $\GammaVr$ in bijective correspondence by representing each $\Gamma\in\mathscr D\pl\M,g\pr$ as
\bea
\Gamma=\overset{0}{\Gamma}+S\nn
\eea
where $S\in\GammaVr$. 
Obviously, such a choice is arbitrary since any element of $\mathscr D\pl\M,g\pr$ can equivalently be used as origin. However, as is well-known, {the Levi-Civita connection is defined solely in terms of the metric structure and can be taken as a privileged connection. As we will see, the existence of such a naturally privileged connection} will constitute a major point of discrepancy with the \nr case. 

We now provide a line of reasoning {that motivates, retrospectively,} the definition of the Levi-Civita connection, starting with the following Lemma:
\blem{}{\label{LemcanisoLorentzian}The vector space $\GammaVr$ is canonically isomorphic to the space $\vectasymdu$. 
}
\noindent The term canonical is here understood in the sense that the isomorphism only depends on the Lorentzian structure $\pl\M,g\pr$. Explicitly, it is given by 
\bea\varphi:\GammaVr\to\vectasymdu:S^\lambda_{\mu\nu}\mapsto T^\lambda_{[\mu\nu]}=S^\lambda_{[\mu\nu]}\nn\eea
while its inverse takes the form 
\bea\varphi\un:\vectasymdu\to\GammaVr:T^\lambda_{[\mu\nu]}\mapsto S^\lambda_{\mu\nu}=T^\lambda_{[\mu\nu]}+T^\rho_{[\sigma\mu]}g^{\sigma\lambda}g_{\rho\nu}+T^\rho_{[\sigma\nu]}g^{\sigma\lambda}g_{\rho\mu}. \nn\eea 
Proposition \ref{PropLorentzianaffinespace} together with Lemma \ref{LemcanisoLorentzian} then ensure the following:
\bprop{\label{propaffineformLor}The space $\mathscr D\pl\M,g\pr$ of compatible Lorentzian connections possesses the structure of an affine space modelled on the {vector space $\vectasymdu$ of tangent-valued 2-forms.} }
\noindent The next step consists in defining an affine map (\cf \Defi{defiaffinemap}) denoted $\Theta:\mathscr D\pl\M,g\pr\to \vectasymdu$ modelled on the linear map $\varphi:\GammaVr\to\vectasymdu$, \ie such that
\bea
\Theta\pl \Gamma'\pr-\Theta\pl \Gamma\pr=\varphi\pl \Gamma'-\Gamma\pr\nn
\eea
for all $\Gamma', \Gamma\in\mathscr D\pl\M,g\pr$. Note that the fact that $\varphi$ is a bijective map ensures that $\Theta$ is too. In particular,
there exists a (necessarily unique) element $\overset{0}{\Gamma}\in\Ker \Theta$, which is given by $\overset{0}{\Gamma}=\Gamma-\varphi^{-1}\pl\Theta(\Gamma)\pr$ for any $\Gamma\in \mathscr D\pl\M,g\pr$. 
This element $\overset{0}{\Gamma}$ provides an origin for $\mathscr D\pl\M,g\pr$ which thereby acquires a structure of vector space. 

\noindent From the expression of $\varphi$, a natural choice consists in defining:
\bea
\Theta:\mathscr D\pl\M,g\pr\to \vectasymdu:\Gamma^\lambda_{\mu\nu}\mapsto  \Gamma^\lambda_{[\mu\nu]}\nn. 
\eea
\noindent Geometrically, the map $\Theta$ associates to each Lorentzian connection its torsion tensor field. Recall that given a Koszul connection $\nabla$, the associated torsion tensor field

\noindent $T\in\field{\w^2\, T^*\M\otimes T\M}$ is defined by its action on vector fields $X,Y\in\vf$ as \bea T\pl X,Y\pr=\nabla_XY-\nabla_YX-\br{X}{Y}.\nn \eea 
In components, the previous equality reads $T\lmn\equiv 2\, \Gamma\lmna$. 
\noindent Given the previous results, the following Theorem arises as a corollary of \Prop{propaffine}:
\bthm{Fundamental Theorem of (pseudo)-Riemannian geometry}{\label{fundamentaltheorem2}There is a unique torsionfree Koszul connection compatible with a given (pseudo)-Riemannian metric called the Levi-Civita connection. }
\noindent The Levi-Civita connection thus provides the affine space $\mathscr D\pl\M,g\pr$ with an origin, so that the latter acquires a structure of vector space. The map $\Theta$ is thus an isomorphism of vector spaces which puts the elements of $\mathscr D\pl\M,g\pr$ in bijective correspondence with {tangent-valued} 2-forms $T\in\vectasymdu$. 
\noindent We stress that Theorem \ref{fundamentaltheorem2} involves no restriction on the metric structure, so that {\it any} Lorentzian structure induces a unique torsionfree Koszul connection. As we will see, this property is lost when one deals with degenerate metric structures. 

\noindent In local coordinates, if one writes 
$\nabla_\mu Y^\lambda=\p_\mu Y^\lambda+\Gamma^\lambda_{\mu\nu}Y^\nu$, then the components $\overset{0}{\Gamma}{}^\lambda_{\mu\nu}$ defining the Levi-Civita connection are the usual Christoffel symbols:
\bea
\overset{0}{\Gamma}{}^\lambda_{\mu\nu}=\half g^{\lambda\rho}\pl\p_\mu g_{\rho\nu}+\p_\nu g_{\rho\mu}-\p_\rho g_{\mu\nu}\pr. \label{Christoffel}
\eea
Note that the Christoffel symbols are canonical, in the sense defined above {(the Christoffel symbols are defined in terms of the metric only).} Making use of the explicit form of the isomorphism $\varphi\un$ then allows to represent each Lorentzian connection $\Gamma\in\mathscr D\pl\M,g\pr$ using the Levi-Civita connection by its associated torsion tensor field $T\in\vectasymdu$ as
\bea
\Gamma^\lambda_{\mu\nu}=\half g^{\lambda\rho}\pl\p_\mu g_{\rho\nu}+\p_\nu g_{\rho\mu}-\p_\rho g_{\mu\nu}\pr+\half\crl T^\lambda_{[\mu\nu]}+T^\rho_{[\sigma\mu]}g^{\sigma\lambda}g_{\rho\nu}+T^\rho_{[\sigma\nu]}g^{\sigma\lambda}g_{\rho\mu}\crr.\label{eqChristoffeltorsion}
\eea
The previous expression can be reformulated as the Koszul formula:
\bea
2\,g\pl\nabla_XY,Z\pr&=&X\crl g\pl Y,Z\pr\crr+Y\crl g\pl X,Z\pr\crr-Z\crl g\pl X,Y\pr\crr\nn\\
&&+g\pl\br{X}{Y},Z\pr-g\pl\br{Y}{Z},X\pr-g\pl\br{X}{Z},Y\pr\label{Kform}\\
&&+g\pl T\pl X,Y\pr,Z\pr-g\pl T\pl Y,Z\pr,X\pr-g\pl T\pl X,Z\pr,Y\pr  \nn
\eea
with $X,Y,Z\in\vf$. 

\noindent We emphasise that, given a particular metric structure, there is no restriction on the possible torsion tensor field $T$ which can span the whole vector space of vector-field-valued 2-forms.

\noindent We conclude this brief review of \rel structures by mentioning a special class of bases of the tangent space:

\bdefi{Lorentzian basis}{\label{defiLorentzian basis}Let $\pl\M, g\pr$ be a $\pl d+1\pr$-dimensional Lorentzian structure with nondegenerate covariant metric $g$. A Lorentzian basis of the tangent space $T_x\M$ at a point $x\in\M$ is an ordered basis $B_x=\lbrace e_{0}|_x, \ldots, e_{d}|_x\rbrace$ which is orthonormal with respect to $g_x$. }
\noindent The basis vectors $e_{a}|_x\in T_x\M$, with $a\in\pset{0, \ldots, d}$ thus satisfy the condition $g_x\pl e_{a}|_x, e_{b}|_x\pr=\eta_{ab}$, with $\eta_{ab}$ the Minkowski metric. 
The denomination Lorentzian is justified by the fact that at each point $x\in\M$, the group of endomorphisms of $T_x\M$ mapping each Lorentzian basis into another one is isomorphic to the Lorentz group $O(d,1)$.

\subsection{Nonrelativistic structures}

\noindent As mentioned in the introduction, a distinguishing feature of nonrelativistic spacetimes is the existence of a degenerate metric\footnote{Throughout this work, the term ``metric'' will be used in a slightly broader sense than the customary one in the physics literature. Namely, we will employ the term to designate a field of covariant or contravariant symmetric bilinear forms of constant rank being either degenerate or nondegenerate. } structure \cite{Cartan1923,Friedrichs1928}, in the guise of a contravariant degenerate metric (absolute rulers) whose radical is spanned by a given 1-form (absolute clock), which must be separately specified. More precisely, one defines:
\begin{defi}[Absolute clock \cite{Dombrowski1964,Kunzle1972}]\label{absclock}
An absolute clock $\psi$ on a manifold $\M$ is a nowhere vanishing 1-form $\psi\in\form{\M}$.
\end{defi}
\noindent An absolute clock allows to distinguish between {\it timelike} tangent vectors $X_x\in T_x\M$ for which $\psi_x\pl X_x\pr\neq0$ from {\it spacelike} tangent vectors $Y_x\in T_x\M$ satisfying $\psi_x\pl Y_x\pr=0$. The distribution $\Ker\psi$ is the vector subbundle of $T\M$ spanned by spacelike vectors.
\begin{defi}[Absolute rulers \cite{Dombrowski1964,Kunzle1972}]\label{absrulers}
A collection of absolute rulers on a manifold $\M$ endowed with an absolute clock $\psi$ is a positive semi-definite contravariant metric $h\in\bform$ on $\M$ whose radical is spanned by the absolute clock \ie
\be\Rad h=\Span{\psi}\,.\label{radi}\ee  
\noindent Alternatively, a collection of absolute rulers can be defined as a field $\gamma\in\bforma{\pl\Ker\psi\pr^*}$ on $\M$ of positive-definite covariant symmetric bilinear forms acting on spacelike vectors. 
\end{defi}
\noindent These two definitions can be shown to be equivalent. In components, the condition \eqref{radi} reads $h^{\mu \nu}\psi_\nu=0$. Armed with these notions of clocks and rulers, we can now define the nonrelativistic analogue of a Riemannian structure as:
\begin{defi}[Leibnizian structure \cite{Dombrowski1964,Kunzle1972,Bernal2003}]\label{Leibniz2}\label{Leibniz1}
A Leibnizian structure consists of a triplet composed by the following elements: 
\begin{itemize}
\renewcommand{\labelitemi}{$\bullet$} 
\item a manifold $\M$
\item an absolute clock $\psi$
\item a collection of absolute rulers $h$ (or, equivalently, $\gamma$)
\end{itemize}
Such a Leibnizian structure will be interchangeably denoted $\mathscr L\pl\M,\psi,h\pr$ or $\mathscr L\pl\M,\psi,\gamma\pr$. 
\end{defi}
\noindent As mentioned previously, Leibnizian structures are purely ``metric'' structures and as such, do not involve a notion of parallelism. Before addressing \nr connections, we must digress a little on the role played by observers in \nr physics. This discussion will justify the introduction of two refinements of Leibnizian structures, namely {\it Aristotelian} and {\it Augustinian} structures. 

\subsection{Observers}

\noindent A map $\lambda:I\subseteq\mR\to\mathscr{N}$ from a subset $I\subseteq\mR$ of the real line into a manifold $\mathscr{N}$ will be called a \textit{parameterised curve} on $\mathscr{N}$, while a 1-dimensional submanifold $\mathscr{C}$ of $\mathscr{N}$ will be called an \textit{unparameterised curve} on $\mathscr{N}$.
A parameterised curve $\lambda:I\to\mathscr{C}$ on an unparameterised curve $\mathscr{C}$ will be called a \textit{parameterisation} of $\mathscr{C}$ when $\lambda$ is invertible. If the unparameterised curve $\mathscr{C}$ on $\mathscr{N}$ is defined by the embedding\footnote{In this paper, an embedding will be defined in the weak sense: an injective immersion. Therefore, strictly speaking a submanifold is here an immersed submanifold.}
$i:\mathscr{C}\hookrightarrow\mathscr{N}$ then
$n\equiv i\circ \lambda$ is called the corresponding parameterised curve on $\mathscr{N}$. In the following, we let $\mathscr L\pl\M,\psi,\gamma\pr$ be a Leibnizian structure. We start by defining the notion of (nonrelativistic) observer and its vector field generalisation:
\bdefi{Observer \cite{Kunzle1972}}{\label{defiobsparam}A (nonrelativistic) observer is a timelike parameterised curve $n:I\subseteq\mR\to\M:s\mapsto n\pl s\pr$ normalised 
such that the tangent vector $N_{n\pl s\pr}\in T_{n\pl s\pr}\M$ (defined\footnote{The vector $D_s\in T\mR_s$ is defined by its action on functions $f\in\fonc{\mR}$ as $D_s\crl f\crr=\frac{\p f}{\p t}\Big |_{t=s}$. } as $N_{n\pl s\pr}\equiv n_* D_s$) satisfies:
\bea\psi_{n\pl s\pr}\pl N_{n\pl s\pr}\pr=1,\ \forall \,  s\in I. \label{eqdefiobserver}\eea
}
\noindent The parameter $s$ will soon acquire the interpretation of (nonrelativistic) proper time of the observer $n$ (\cf \Prop{propobserverpropertime}). In local coordinates, the observer $n$ is a timelike curve $x^\mu(s)$ with parameterisation chosen such that $\psi_\mu\tfrac{dx^\mu}{ds}=1$.
This notion can be generalised to define vector fields whose integral curves are observers:
\bdefi{Field of observers \cite{Kunzle1972}}{\label{defifieldobservers}A field of (nonrelativistic) observers is a vector field $N\in\field{T\M}$ such that $\psi\pl N\pr=1$. The space of all fields of observers on $\M$ is denoted $\FO$.}

\bdefi{Proper time \cite{Bernal2003}}{\label{defipropertime}Let $\mathscr{C}$ be a timelike unparameterised curve on $\M$ defined by the embedding $i:\mathscr{C}\hookrightarrow\M$. 
We will call (nonrelativistic) proper time any function $\tau\in \fonc{\C}$ satisfying $d\tau=i^*\psi$. 
}
\noindent The fact that the submanifold $\mathscr{C}$ is of dimension 1 ensures that the 1-form $i^*\psi$ is closed, so that locally there always exists a function $\tau$ such that $d\tau=i^*\psi$. Obviously, this condition only defines the proper time up to a constant. 
{The parameter $s$ in Definition \ref{defiobsparam} is closely related to the proper time $\tau$ of the unparameterised curve   associated with an observer:}

\bprop{\label{propobserverpropertime}Let $\mathscr{C}$ be an unparameterised curve on $\M$ defined by the embedding $i:\mathscr{C}\hookrightarrow\M$.
Let $\tau\in \fonc{\C}$ be a proper time on $\mathscr{C}$.

\noindent The parameterised curved $n=i\circ \lambda$ defined by the parameterisation $\lambda:I\subseteq\mR\to\mathscr{C}:s\mapsto \lambda\pl s\pr$ is an observer if and only if. 
 \bea
 \tau\circ \lambda\pl s\pr=s+a,\text{ }\forall s\in I
 \eea
 with $a\in\mR$ a constant. 
}
\noindent {The proof is a straightforward application of the previous definitions.}

\noindent The proper time on an unparameterised curve is defined up to a constant thus, 
without loss of generality one may assume $a=0$. In such case, the parameterisation $\lambda$ is the inverse function of the proper time $\tau$, so that it is natural to identify the parameter $s$ with the value $\tau$ of the proper time at the corresponding point on the curve. 

\bdefi{Spacelike projection of vector fields \cite{Dombrowski1964}}{\label{defispacelikeprojection}Let $N\in \FO$ be a field of observers. The field of endomorphisms $P^N:\field{T\M}\to\field{\Ker \psi}$ defined as \bea P^N\pl X\pr= X-\psi\pl X\pr N\label{eqPN}\eea where $X$ is any vector field, is called a spacelike projector of vector fields. }

\noindent The transpose of a spacelike projector can be defined as the field of linear maps\footnote{At each point $x\in \M$, $\Ann N_x$ stands for the annihilator of $\Span N_x$ in $T_x^*\M$ and $\Ann N$ is thus to be understood as the subbundle of $T^*\M$ spanned by 1-forms annihilating the \fo $N$. } $\bar P^N:\form{\M}\to\field{\Ann N}$ defined as $\bar P^N\pl \alpha\pr=\alpha-\alpha\pl N\pr\psi$, with $\alpha\in\form{\M}$. 
In components, these two spacelike projectors read as: $P^\mu{}_\nu=\delta^\mu_\nu-N^\mu\psi_\nu=\bar P_\nu{}^\mu$.

\subsection{Absolute time and spaces}

\noindent As such, a Leibnizian structure does not allow generically a {\it global} definition of absolute time and space since it only provides a set of {\it local} clocks and rulers. This drawback can be circumvented by restricting the {class of absolute clocks.} The suitable restriction comes in two versions, a weak one and a strong one. Denoting $\mathcal D$ the distribution of spacelike hyperplanes $\mathcal D_x\equiv\Ker\psi_x$ ($\forall x\in\M$), the weak version consists in imposing that the distribution $\mathcal D$ is involutive. One is then led to define what we called an Aristotelian structure\footnote{In the terminology of \cite{Bernal2003}, it would be called a Leibnizian structure with \textit{locally synchronizable} absolute clock.}
as: 
\begin{defi}[Aristotelian structure \cite{Bekaert2013c}]\label{Aristotle}
An Aristotelian structure is a Leibnizian structure whose absolute clock induces an involutive distribution, \ie satisfies the Frobenius integrability condition: $\psi\w d\psi=0$. 
\end{defi}
\noindent This supplementary condition ensures, by Frobenius Theorem, that the kernel of $\psi$ defines a foliation of $\M $ by a family of hypersurfaces of codimension one called \textit{absolute spaces}. These are the maximal integral submanifolds of $\mathcal D$, so that the tangent space $T_x\M$ at each point $x$ of the simultaneity slice is isomorphic to $\Ker \psi_x$. Locally, the 1-form $\psi$ can be written as $\psi=\Omega\,dt$ where $\Omega\in\fonc{\M}$ is a positive function called {\it time unit} and the function $t\in\fonc{\M}$ will be referred to as the \textit{absolute time}. The absolute time has a fixed value on each absolute space. Therefore, absolute spaces can be identified with simultaneity slices $t=$const. 
In contradistinction with $\M$, absolute spaces are Riemannian manifolds since they are endowed with the positive-definite metric $\gamma$. 
\begin{figure}[ht]
\centering
   \includegraphics[width=0.7\textwidth]{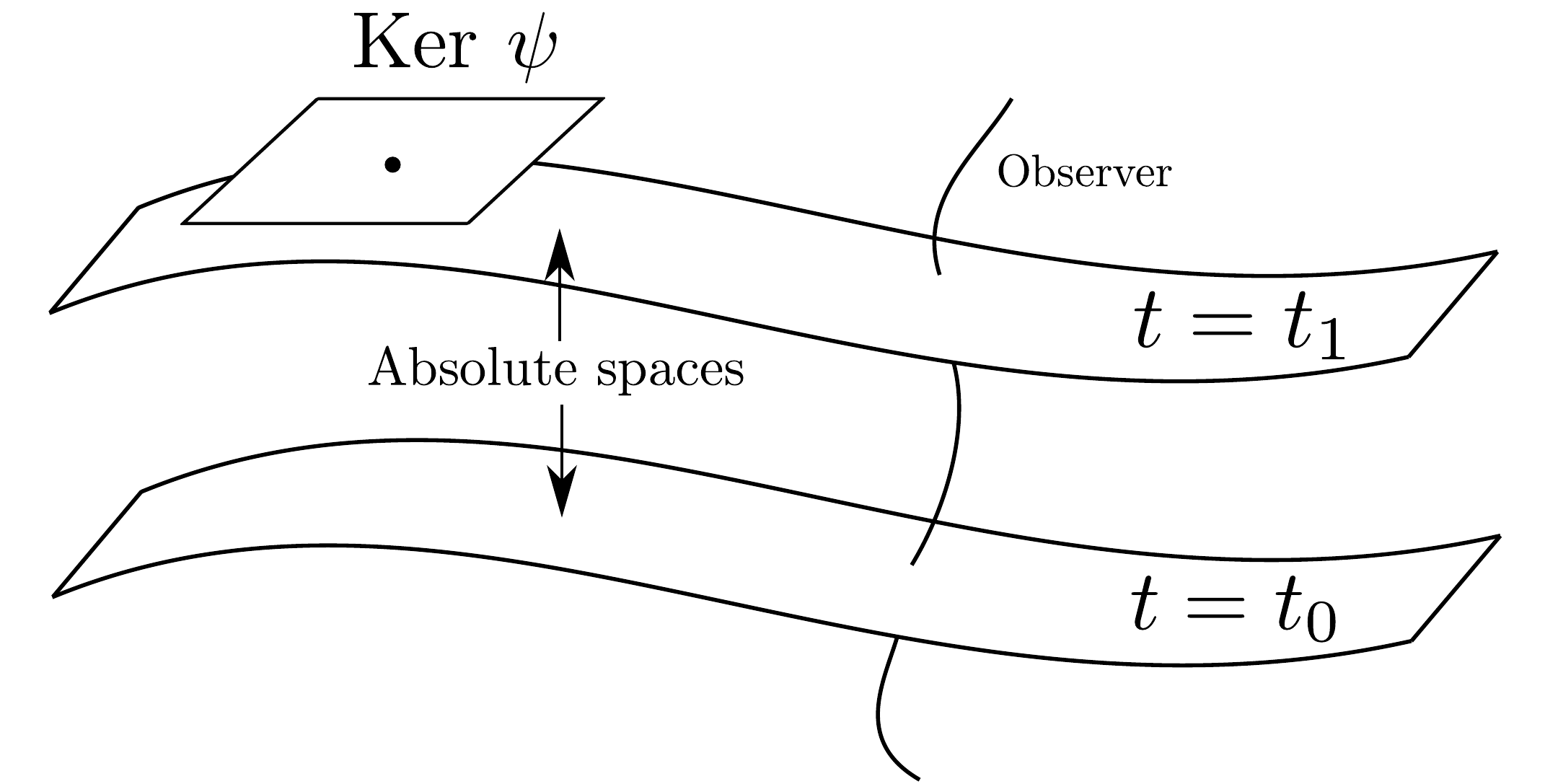}
  \caption{Foliation of an Aristotelian structure by absolute spaces. \label{foliation}}
\end{figure}
\vspace{2mm}
{As pointed out in \cite{Geracie2014}, the causal structure of nonrelativistic spacetimes in the non-Aristotelian case is somewhat pathological: indeed, a Leibnizian structure which is not Aristotelian does not possess a well defined notion of absolute space, as is clear from the definitions, but the situation is even more bizarre since all points in some neighborhood are simultaneous to each other. \footnote{It is very natural to consider two events that can be joined by a spacelike curve to be simultaneous. In fact, for an Aristotelian structure, this provides one way to define the simultaneity slice $\Sigma$ through an event $p\in\M$ which cuts any neighborhood $B$ of $p$ in ``past'' and ``future'' (while $\Sigma$ is ``present''). However, Caratheodory's theorem (\cf \eg \cite{Frankel}) implies that if $\psi\wedge d\psi\neq 0$ at $p$ then there exists a neighborhood $B$ of $p$ such that all points are simultaneous (in the sense of the previous definition) \ie for any point $q$ of $B$, there exists a spacelike curve joining $p$ to $q$.}
}

\noindent Now, let $\mathscr{C}$ be an unparameterised curve on $\M$ defined by the embedding $i:\mathscr{C}\hookrightarrow\M$. The local condition $\psi=\Om\, dt$ allows to write $d\tau\,=\,i^*\psi\,=\,(i^*\Omega)\,d(i^*t)$, where $\tau\in\fonc{\M}$ is a proper time on $\mathscr{C}$ while $i^*\Omega=\Omega\circ i$ and $i^*t=t\circ i$ are the pullbacks on $\mathscr{C}$ of the time unit and absolute time, respectively. {Integrating the pullback of the absolute clock on a curve joining the events $A$ and $B$, one finds the proper time interval $\tau_{A\to B}=\int_A^B i^*\psi$. Any observer on an Aristotelian structure can} make use of the time unit $\Omega$ in order to compare or ``synchronise'' its proper time $\tau$ with the absolute time $t$. 

\noindent The situation regarding synchronisation is even clearer when considering the more restrictive case in which the absolute clock is a closed 1-form. We thus define an {\it Augustinian structure}\footnote{We chose to refer to Augustine of Hippo (also known as ``Saint Augustine'') in order to pay tribute to the role he played regarding the philosophy of time, \cf Book X of his \textit{Confessions}. } as: 
\begin{defi}[Augustinian structure]\label{defiAugustinian}
An Augustinian structure is a Leibnizian structure whose absolute clock $\psi$ is closed. 
\end{defi}

\begin{wrapfigure}{l}{0.15\textwidth}
\vspace{-20pt}
  \begin{center}
   \includegraphics[width=0.12\textwidth]{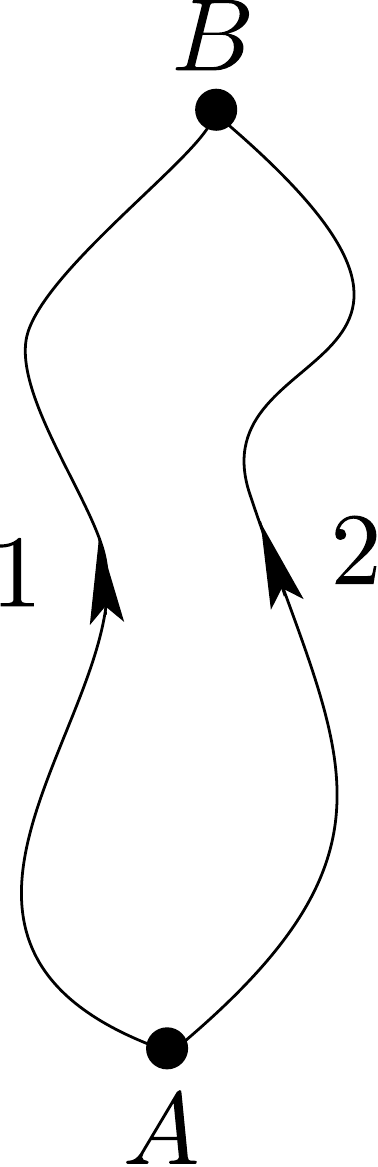}
  \end{center}
\vspace{-10pt}
\end{wrapfigure}
\noindent This stronger condition allows locally to write $\psi=dt$, so that any observer of an Augustinian structure is automatically synchronised\footnote{{Indeed, in the terminology of \cite{Bernal2003} it would be called a Leibnizian structure with \textit{proper time locally synchronizable} absolute clock.}} with the absolute time ($\tau=i^*t=t\circ i$). Consequently, if the spacetime is simply connected then two observers sharing the same endpoints $A,B\in\M$ will agree when comparing the proper time passed when going from $A$ to $B$, since the integral 
\bea
\tau_{A\to B}={\int_A^B} i^*\psi={\int_A^B} d\pl t\circ i\pr=t\pl i\pl B\pr\pr-t\pl i\pl A\pr\pr\nn
\eea
does not depend on the path followed. 

\bexa{Aristotle spacetime}{\label{exaAugustinian}The most simple example of a Leibnizian structure is given by a $\pl d+1\pr$-dimensional Aristotle spacetime characterised by a closed absolute clock and flat absolute spaces:
\bcase{
\psi=dt\\
\gamma=\delta_{ij}\,dx^i\vee dx^j\nn
}
where $i,j\in\pset{1,\dots,d}$ and $\delta_{ij}$ the Kronecker delta. 
{Equivalently, one may consider the following contravariant metric: $h=\delta^{ij}\,\frac{\partial}{\partial x^i}\vee\frac{\partial}{\partial x^j}$ (\cf \cite{Toupin}).} 
}
\noindent {In the Aristotle spacetime, time is absolute and space is Euclidean.} Obviously, this spacetime was the only arena where physical events were conceived to take place before the breakthroughs of non-Euclidean geometry in the 19th century and special relativity in the 20th century.

\noindent The hierarchy 
$$
\mbox{Augustinian}\,\subset\,\mbox{Aristotelian}\,\subset\,\mbox{Leibnizian}
$$
of the three types of nonrelativistic ``metric'' structures introduced so far is summarised in Table \ref{NRstruct}.
\begin{center}
\setlength{\extrarowheight}{2 mm}
\begin{tabular}{|c|c|}
   \hline
\rule[-0.6cm]{0cm}{1.2cm}Nonrelativistic structure & Absolute clock\tabularnewline\hline\hline
 
\rule[-0.6cm]{0cm}{1.2cm}Leibnizian  & Arbitrary $\psi$ \tabularnewline\hline
\rule[-0.6cm]{0cm}{1.2cm}Aristotelian  & Frobenius $\psi\wedge d\psi=0$ \tabularnewline\hline
\rule[-0.6cm]{0cm}{1.2cm}Augustinian  & Closed $d\psi=0$ \tabularnewline\hline
\end{tabular}
\captionof{table}{Absolute clocks of nonrelativistic structures}
\label{NRstruct}
\end{center}

\subsection{Milne boosts}

\noindent Consider an Augustinian structure (locally, $\psi=dt$). One may introduce an \textit{adapted} coordinate system $x^\mu=(t,x^i)$ where the first coordinate is the absolute time and $x^i$ are coordinates on the absolute spaces. In this coordinate system, a field of observers decomposes as $N=\frac{\partial}{\partial t}+v^i\frac{\partial}{\partial x^i}$. The integral curves of $N$ are such that $v^i=\frac{dx^i}{dt}$.
By analogy with the proper velocity spacetime vector,
a field of observers is then sometimes called a ``velocity vector'' (\eg \cite{Son2013}).  
For an Aristotelian structure ($\psi=\Omega\ dt$), the analogous expression reads $N=\frac1{\Omega}\left(\frac{\partial}{\partial t}+v^i\frac{\partial}{\partial x^i}\right)=\frac1{\Omega}\frac{\partial}{\partial t}+\tilde v^i\frac{\partial}{\partial x^i}$
and its integral curves are such that $\tilde v^i=\frac1{\Omega}v^i=\frac{dx^i}{d\tau}$ where $\tau$ is the proper time.

Let us turn back to the general case of a Leibnizian structure.
Given two fields of observers $N'$ and $N$, their difference $V=N'-N$ is a spacelike vector field, \ie it belongs to the kernel of the absolute clock, $\psi\pl V\pr=0$. Therefore, the difference $V\in\field{\Ker\psi }$ is not a field of observer.\footnote{
{Rather, it can be thought as the relative spacelike velocity between two fields of observers, \eg in the case of some adapted coordinates for an Aristotelian structure,
one has $N'-N=v^i\frac{\partial}{\partial x^i}$.} 
} 
This observation prevents the space $\FO$ of all fields of observers $\FO$ from being a vector space. However, $\FO$ possesses a natural structure of {\it affine} space \cite{Bernal2003} with associated vector space $\field{\Ker\psi}$. {Consequently the space of field of observers is a principal homogeneous space for the additive (Abelian) group $\field{\Ker\psi}$, called the {\it Milne group}. In other words, the action of the Milne group $\field{\Ker\psi}$ on the space $\FO$ of field of observers is free and transitive.} The action of $\field{\Ker\psi}$ on $\FO$ as $N\mapsto N+V$ will be referred to as a {\it Milne boost} parameterised by the spacelike vector field $V\in\field{\Ker\psi}$.\footnote{A Milne boost can be alternatively parameterised by a 1-form $\chi\in\ff$ (\cf \eg \cite{Carter1994,Duval1993,Duval2009a}) so that the action reads $N\mapsto N'=N+h\pl\chi\pr$. However, it should be noted that this action of $\ff$ is not free. Given a field of observers $N$, a free action can be recovered by restricting $\chi$ to belong to $\field{\Ann N}$.} Milne boosts are sometimes referred to as 
``local Galilean boosts'', denomination that will be justified in Proposition \ref{propGalileanbasis}.

\noindent Fields of observers are bestowed upon a greater importance in nonrelativistic physics in comparison with the relativistic case, since a great deal of structures can only be defined by making use of a particular choice of field of observers $N$ (thus in a non-canonical way). 
Indeed, since the contravariant metric $h$ of a Leibnizian structure is degenerate, there is no natural covariant metric defined on the whole tangent bundle $T\M$ (remember that the absolute rulers $\gamma$ are only defined on $\Ker \psi$). However, the gift of a field of observers $N$ allows to uniquely define a (degenerate) covariant metric $\overset{N}{\gamma}$ transverse to $N$ as:
\bdefi{Transverse metric}{\label{defitransversemetric}Let $\mathscr L\pl \M, \psi, \gamma\pr$ be a Leibnizian structure and $N\in \FO$ a field of observers on $\M$. The transverse metric $\N{\gamma}\in\bforms$ is defined by its action on vector fields $X,Y\in\field{TM}$ as
\bea
\N{\gamma}\pl X,Y\pr=\gamma\pl P^N\pl X\pr, P^N\pl Y\pr\pr\label{eqtransversemetric}
\eea
where $\gamma\in\bforma{\pl\Ker \psi\pr^*}$ is the collection of absolute rulers and $P^N$ stands for the spacelike projector associated to the field of observers $N$. }
\noindent The right-hand side of eq.\eqref{eqtransversemetric} is well-defined since the image of a spacelike projector lies in $\field{\Ker\psi}$. The epithet ``transverse'' is justified by the fact that $N\in\Rad \N{\gamma}$, \ie $\forall X\in\field{TM}$: $\N{\gamma}(X,N)=0$. Furthermore, it is easy to show that the contraction of $\N{\gamma}$ with the contravariant metric $h$ satisfies the relation: $h\pl\N{\gamma}\pl X\pr\pr=P^N\pl X\pr,  \forall\, X\in\field{T\M}$. 
\noindent In components, we thus have the two relations:
\bea
\begin{cases}
\overset{N}{\gamma}_{\mu\nu}\, N^\nu=0\\ 
\overset{N}{\gamma}_{\nu\lambda}\, h^{\lambda\mu}=\delta^\mu_\nu-N^\mu\psi_\nu\label{eqstransversemetriccond}. 
\end{cases}
\eea
In fact, these two conditions completely determine $\N{\gamma}$, as expressed by: 
\bpropp{\cf \eg \cite{Kunzle1972}}{Let $\mathscr L\pl \M, \psi, \gamma\pr$ be a Leibnizian structure and $N\in \FO$ a field of observers on $\M$. There is a unique covariant metric $\N{\gamma}\in\bforms$ satisfying the conditions \eqref{eqstransversemetriccond}. 
}

\noindent As suggested by the superscript, the covariant metric $\overset{N}{\gamma}$ depends on the choice of field of observers $N$. More precisely, it can be shown that under a change of field of observers $N\mapsto N+V$ via the Milne boost parameterised by the spacelike vector field $V\in\GammaV$, the covariant metric $\overset{N}{\gamma}$ varies as 
\bea\overset{N}{\gamma}_{\mu\nu}\mapsto\overset{N}{\gamma}_{\mu\nu}+\gamma\pl V,V\pr \psi_\mu\psi_\nu-2\, V^\lambda\N{\gamma}_{\lambda(\mu}\psi_{\nu)}. \label{eqtransversemilne}\eea 
A \nr avatar of a Lorentzian basis (\cf \Defi{defiLorentzian basis}) can be formulated:
\bdefi{Galilean basis \cite{Kunzle1972}}{\label{defiGalileanbasis}Let $\mathscr L\pl \M, \psi, \gamma\pr$ be a Leibnizian structure. A Galilean basis of the tangent space $T_x\M$ at a point $x\in\M$ is an ordered basis $B_x=\lbrace N_x, e_{1}|_{x}, \ldots, e_{d}|_{x}\rbrace$ with $N_x$ the tangent vector of an observer and $\lbrace e_{1}|_{x}, \ldots, e_{d}|_{x}\rbrace$ a basis of $\Ker \psi_x$ which is orthonormal with respect to $\gamma_x$. }
\noindent Explicitly, the basis $B_x=\lbrace N_x, e_{1}|_{x}, \ldots, e_{d}|_{x}\rbrace$ must satisfy the conditions:
\vspace{2mm}
\begin{enumerate}
\item $\psi_x \pl N_x\pr=1$
\item$\psi_x\pl e_{i}|_{x}\pr=0\,  ,\forall\, i\in \lbrace1, \ldots, d\rbrace$
\item$\gamma_x\pl e_{i}|_{x}, e_{j}|_{x}\pr=\delta_{ij}\, ,\forall\, i,j\in\pset{ 1, \ldots, d}$. 
\end{enumerate}
\vspace{2mm}
\noindent A basis of $T^*_x\M$ dual to $B_x=\lbrace N_x, e_{i}|_{x}\rbrace$ is given by $B^*_x\equiv\pset{\psi_x, \theta^i_x}$, where the $d$ one-forms $\theta^i_x$ satisfy the requirements $\theta^i_x\pl e_{j}|_{x}\pr=\delta^i_j$ and $\theta^i_x\pl N_x\pr=0$. 

\noindent The reference to Galilei in Definition \ref{defiGalileanbasis} is justified by the following Proposition: 
\bpropp{\cf \eg \cite{Bernal2003}}{\label{propGalileanbasis}At each point $x\in\M$, the set of endomorphisms of $T_x\M$ mapping each Galilean basis into another one forms a group isomorphic to the homogeneous Galilei group.}

\noindent {We detail the proof since it clarifies the interpretation of Milne boosts as local Galilean boosts.}

\proof{ ~\\
Let us denote by $T:T_x\M\to T_x\M$ one of the endomorphisms considered. Since $T$ maps bases into bases, it must be a vector space isomorphism so that it can be represented by an element of $GL\pl T_x\M\pr$ as the invertible matrix 
\bea
T\equiv
\begin{pmatrix}
a & \bc^T \\
\bb & \bR
\end{pmatrix}
\eea
where $a\in \mR$, $\bb,\bc\in\mR^d$ and $\bR\in GL\pl \mR^d\pr$. 
\noindent Let $B_x=\lbrace N_x, e_{ix}\rbrace$ be a Galilean basis of $T_x\M$, the basis $T\pl B_x\pr=\lbrace N'_x, e'_{ix}\rbrace$ reads (dropping the index $x$ for notational simplicity):
\bea
T\pl N,e_i\pr
=\pl N,{e}\pr\begin{pmatrix}
a & \bc^T \\
\bb & \bR
\end{pmatrix}
=
\pl aN+\bb^ie_i,\bc^T_iN+\bR^j_{\, i}e_j\pr. 
\eea
Requiring that $T\pl B_x\pr$ is a Galilean basis (Conditions 1-3 following Definition \ref{defiGalileanbasis}) imposes that $T$ satisfy:
\begin{enumerate}
\item $\psi_x \pl N'_x\pr=1\Rightarrow a=1$
\item$\psi_x\pl e'_{ix}\pr=0\,  ,\forall\, i\in \lbrace1, \ldots, d\rbrace\Rightarrow \bc^T_j=0$
\item$\gamma_x\pl e'_{ix}, e'_{jx}\pr=\delta_{ij}\, ,\forall\, i,j\in\pset{ 1, \ldots, d}\Rightarrow \bR\in O\pl d\pr$. 
\end{enumerate}
The set of matrices representing the set of isomorphisms $T$ is thus of the form
\bea
T=
\begin{pmatrix}
1 & 0 \\
\bb & \bR\label{matrixGalilei}
\end{pmatrix}
\eea
with $\bb\in\mR^d$ and $\bR\in O\pl d\pr$. This set of matrices form a subgroup of $GL\pl T_x\M\pr$ isomorphic to the homogeneous Galilei group $\GALZ\pl d+1\pr$. The homogeneous Galilei group therefore acts regularly on the space of Galilean basis via the group action:
\bea
\pset{N, e_i}\mapsto\pset{N+\bb^ie_i, \bR^{j}_{\, \, i}e_j}. \label{eqgalileiauto}
\eea
}

\noindent Proposition \ref{propGalileanbasis} together with Definition \ref{defiGalileanbasis} can be generalised in a straightforward way from the tangent space at a point of $\M$ to the tangent bundle of $\M$. A Galilean basis of $T\M$ is thus defined as the ordered set of fields $B=\lbrace N, e_{1}, \ldots, e_{n}\rbrace$ with $N$ a field of observers and $\lbrace e_{1}, \ldots, e_{n}\rbrace$ a basis of $\field{\Ker\psi}$, orthonormal with respect to $\gamma$. Two Galilean bases $\lbrace N', e'_i\rbrace$ and $\lbrace N, e_i\rbrace$ are mapped via a local transformation where $\bR:\M\to O\pl d\pr$ parameterises a local rotation and $\bb^i:\M\to \mR^d$ a local Galilean boost. 
Explicitly, one has: 
\bea
\begin{cases}
N'=N+\bb^ie_i\\
e'_i=\bR^j_{\, i}e_j
\end{cases}
\eea
where the first expression is a Milne boost.

\pagebreak
\section{Torsionfree Galilean connections}\label{Nonrelativistic manifolds}

\noindent We now switch to the definition of nonrelativistic manifolds, \ie Leibnizian structures endowed with a compatible Koszul connection and discuss some of the peculiarities arising, which are absent in the relativistic case. 

\subsection{Galilean manifolds}
\label{sectionGalileanmanifolds}
\noindent It should first be noted that the compatibility condition with the metric-like structure must apply to both the absolute rulers {\it and} clock. One then defines:
\bdefi{Galilean manifold \cite{Kunzle1972}}{\label{Galileanmanifold2}Given a Leibnizian structure $\mathscr L\pl \M, \psi, \gamma\pr$, a Galilean manifold is a a quadruple $\mathscr G\pl \M, \psi, \gamma, \nabla\pr$ with $\nabla$ a Koszul connection compatible with the absolute clock $\psi$ and rulers $\gamma$. The Koszul connection $\nabla$ is then referred to as a Galilean connection. 
}
\noindent When the absolute rulers are formulated in terms of a field $h\in\bforma{T\M}$, the compatibility conditions read 
\begin{enumerate}
\item $\nabla\psi=0$
\item $\nabla h=0$. 
\end{enumerate}
These two conditions can be more explicitly stated as:
{\it
\begin{enumerate}
\item $X\crl\psi\pl Y\pr\crr=\psi\pl\nabla_XY\pr$, for all $X,Y\in\field{T\M}$ 
\item $X\crl h\pl \alpha,\beta\pr\crr=h\pl\nabla_X\alpha, \beta\pr+h\pl \alpha,\nabla_X\beta\pr$, for all $X\in\field{T\M}$ and $\alpha,\beta\in\ff$. 
\end{enumerate}}
\noindent When the absolute rulers are formulated in terms of a field $\gamma\in\bforma{\pl\Ker\psi\pr^*}$, the second condition can be restated as $\nabla \gamma=0$ or equivalently: 
\bea
X\crl\gamma\pl V,W\pr\crr=\gamma\pl \nabla_XV,W\pr+\gamma\pl V,\nabla_XW\pr\text{ for all }X\in\field{T\M}\text{ \rm and }V,W\in \field{\Ker\psi}. \nn
\eea

\noindent The right-hand-side of the previous equation is well defined since $V\in\field{\Ker\psi}$ implies $\psi\pl\nabla_XV\pr=0$ (\cf Condition {\it 1.}) which in turn, ensures that $\nabla_XV\in\field{\Ker\psi}$, for all $X\in\field{T\M}$ and $V\in\Milneg$. 

\noindent In components, these two sets of equivalent conditions read:
\bea
\begin{cases}
\nabla_\mu\psi_\nu=0\\
\nabla_\mu h^{\alpha\beta}=0\label{eqconditionsgalileanmanifold}
\end{cases}
\Longleftrightarrow
\begin{cases}
\nabla_\mu\psi_\nu=0\\
\nabla_\mu \gamma_{ij}=0
\end{cases}
\eea
where the last equality is written in adapted coordinates $x^\mu=\pl t,x^i\pr$. 

\vspace{2mm}
\noindent A first peculiarity of a Galilean manifold, in contradistinction with the relativistic case, is the fact that not all the torsion tensors are compatible with a given Leibnizian structure, as the following Proposition shows:
\bpropp{\cf \cite{Kunzle1972,Bernal2003}}{\label{compatibility}Let $\mathscr G\pl \M,\psi,\gamma,\nabla\pr$ be a Galilean manifold and denote $T$ the torsion of the Galilean connection $\nabla$. The following relation holds:
\bea
\psi\big( T\pl X,Y\pr\big)=d\psi\pl X,Y\pr
\label{eqtorsionconstraint}
\eea
for all $X,Y\in \field{T\M}$. 
}
\noindent In components, relation \eqref{eqtorsionconstraint} reads $\psi^{}_\lambda \Gamma^\lambda_{[\mu\nu]}=\p_{[\mu}\psi_{\nu]}$, where the torsion $T$ decomposes as $T\equiv2\, \Gamma^\lambda_{[\mu\nu]}\, dx^\mu\wedge dx^\nu\otimes \p_\lambda$. 

\noindent In particular, the previous Proposition implies that only Augustinian structures (\ie satisfying $d\psi=0$) admit a torsionfree Koszul connection. This is clearly a distinctive feature of nonrelativistic structures as there exists no such restriction in the relativistic case. Furthermore, while in the relativistic case, Theorem \ref{fundamentaltheorem2} ensures that a torsionfree Lorentzian manifold is uniquely determined by the metric structure, in the nonrelativistic case however, the degeneracy of the metric prevents the gift of an Augustinian structure to uniquely fix a compatible torsionfree Koszul connection. As one will see later, this arbitrariness has a natural physical interpretation: the Leibnizian structure merely encodes the properties of a nonrelativistic spacetime which would be a mere container where particles can be placed and measured.
In Newtonian mechanics, their movement will be fixed by prescribed forces, \textit{a priori} independent of the Leibnizian structure (usually taken to be flat, \ie the Aristotle spacetime of Example \ref{exaAugustinian}).
According to the equivalence principle, these dynamical trajectories acquire the interpretation of spacetime geodesics.
In other words, the arbitrariness in the choice of the external forces prescribed on top of the Leibnizian structure corresponds to
the arbitrariness in the choice of a Galilean connection. This freedom is already manifest in the following two paradigmatic examples of Galilean manifolds:
\bexa{Galilei and Newton-Hooke spacetimes}{\label{exGalNH}The Aristotle spacetime (Example \ref{exaAugustinian}) with absolute clock $\psi=dt$ and rulers
$\gamma=\delta_{ij}dx^i\vee dx^j$ 
can be supplemented with a flat connection $\Gamma^{\lambda}_{\mu\nu}=0$ in order to yield the standard \textit{Galilei spacetime} \cite{Toupin}.
Alternatively, one can endow the Aristotle spacetime with the (equally compatible) connection $\Gamma$ whose only nonvanishing components are
$\Gamma^{i}_{00}=-\frac{k}{\tau^2}\,x^i$. This Galilean manifold is referred to as the {\it Newton-Hooke spacetime} (\cf \cite{Aldrovandi1999} for a nice introduction to its physical interpretation as a nonrelativistic cosmological model).
The constant $k$ can take the values $k=+1$ (expanding spacetime), $k=-1$ (oscillating spacetime) or $k=0$ (Galilei spacetime). The corresponding force field is 
simply the one of a harmonic oscillator for $k=\pm1$, i.e. a force linear in the displacement 
(attractive for $k=-1$, repulsive for $k=+1$).  
}

\subsection{Torsionfree Galilean manifolds}
\label{sectionGalileanmanifoldss}

\noindent We now characterise more precisely the space of torsionfree Koszul connections compatible with a given Augustinian structure by mimicking the discussion of Section \ref{sectionrelativisticstructures}. 

\noindent Let $\mathscr S\pl\M,\psi,\gamma\pr$ be an Augustinian structure. The space of torsionfree Galilean connections compatible with $\mathscr S\pl\M,\psi,\gamma\pr$ will be denoted $\mathscr D\pl\M,\psi,\gamma\pr$. Recall that, in the absence of metric structure, the space of torsionfree Koszul connections is an affine space modelled on the vector space $\vectsymdu$. Now, focusing on Galilean connections, the compatibility conditions \eqref{eqconditionsgalileanmanifold} reduce the vector space on which $\mathscr D\pl\M,\psi,\gamma\pr$ is modelled according to the following Proposition:
\bprop{\label{PropGalileanaffinespace}The space $\mathscr D\pl\M,\psi,\gamma\pr$ {of torsionfree Galilean connections possesses the structure of an affine space modelled on the vector space}
\bea\GammaV\equiv\pset{S\in\field{\vee^2T^*\M\otimes T\M}\ /\ \psi_\lambda S^\lambda_{\mu\nu}=0\ \text{ and }\ S^{(\lambda}_{\mu\nu}\, h^{\rho)\nu}_{}=0}. \nn
\eea
}
\noindent Note that $\GammaV$ has dimension $\frac{d\pl d+1\pr}{2}$, for $\M$ a $\pl d+1\pr$-dimensional spacetime. This is in contradistinction with the relativistic case where the compatibility condition with a Lorentzian metric reduces the affine space of torsionfree connections to a single point: the Levi-Civita connection. 

\blem{}{\label{Lemcaniso}The vector space $\GammaV$ is canonically isomorphic to the space $\forma{2}{\M}$ of 2-forms on $\M$. 
}
\noindent Again, the term ``canonical'' designates an object built only in terms of the metric structure $\mathscr S\pl\M,\psi,\gamma\pr$. Explicitly, the canonical isomorphism is given by 
\bea\varphi:\GammaV\to\forma{2}{\M}:S^\lambda_{\mu\nu}\mapsto F_{\mu\nu}\,=\,-2\,\N{\gamma}_{\lambda[\mu}S^\lambda_{\nu]\rho}N^\rho\label{eqdefivarphi}\eea with $N\in \FO$ a field of observers and $\N{\gamma}\in\field{\vee^2\ T^*\M}$ its associated transverse metric while its inverse takes the form 
\bea\varphi\un:\forma{2}{\M}\to\GammaV:F_{\mu\nu}\mapsto S^\lambda_{\mu\nu}=h^{\lambda\rho}\psi_{(\mu}F_{\nu)\rho}. \nn\eea 
It can be checked that the expression $F_{\mu\nu}=-2\,\N{\gamma}_{\lambda[\mu}S^\lambda_{\nu]\rho}N^\rho$ is independent of the choice of field of observers $N$, whenever $S\in\GammaV$, so that $\varphi$ is indeed canonical.  
Proposition \ref{PropGalileanaffinespace} together with Lemma \ref{Lemcaniso} then ensures the following Proposition:
\bprop{\label{propaffineform}The space $\mathscr D\pl\M,\psi,\gamma\pr$ {of torsionfree Galilean connections possesses the structure of an affine space modelled on the vector space $\forma{2}{\M}$ of 2-forms.} }
\noindent Explicitly, given two torsionfree Galilean connections $\Gamma',\Gamma\in\mathscr D\pl\M,\psi,\gamma\pr$, there exists a unique $F\in\forma{2}{\M}$ such that:
\bea
\Gamma'=\Gamma+\varphi\un\pl F\pr\nn \text{ or, equivalently, }
F=\varphi\pl \Gamma'-\Gamma\pr. \nn
\eea
\noindent 
Using the explicit form of $\varphi$, the 2-form $F$ can be expressed in components as:
\bea F_{\alpha \beta}=-2\,\N{\gamma}_{\lambda [\alpha}\nabla'_{\beta]}N^\lambda+2\,\N{\gamma}_{\lambda [\alpha}\nabla^{}_{\beta]}N^\lambda \label{eqformnabla}\eea 
with $N\in \FO$ a field of observers and $\N{\gamma}\in\field{\vee^2\ T^*\M}$ its associated transverse metric. 
The previous expression suggests to introduce the map \bea\N{\Theta}:\mathscr D\pl\M,\psi,\gamma\pr\to\forma{2}{\M}:\nabla\mapsto\N{F}_{\alpha\beta}=-2\N{\gamma}_{\lambda [\alpha}\nabla^{}_{\beta]}N^\lambda. \label{eqdefivarphiN}\eea 
We emphasise that $\N{\Theta}$ is non-canonical, hence the superscript. Eq.\eqref{eqformnabla} can then be rewritten as 
\bea
\N{\Theta}\pl \Gamma'\pr-\N{\Theta}\pl \Gamma\pr=\varphi\pl \Gamma'-\Gamma\pr\label{eqaffinemap}
\eea
so that $\N{\Theta}$ is an affine map modelled on the linear map $\varphi$. 
Following \Prop{propaffine}, the fact that $\N{\Theta}$ is an affine map associated to the linear isomorphism $\varphi$ ensures that $\N{\Theta}$ endows $\mathscr D\pl\M,\psi,\gamma\pr$ with a structure of vector space.

\noindent Before identifying the corresponding origin,
let us provide a more geometric characterisation of the 2-form $\N{F}$ appearing in eq.\eqref{eqdefivarphiN}:
\bdefi{Gravitational fieldstrength measured by a field of observers}{
Let $\mathscr G\pl \M, \psi, \gamma,\nabla\pr$ be a Galilean manifold and $N\in \FO$ a field of observers.
The gravitational fieldstrength measured by the field of observers $N$ with respect to $\nabla$ is defined as the 2-form $\N{F}\in\forma{2}{\M}$ whose action reads
\bea
\N{F}\pl X,Y\pr\equiv  \gamma\pl\nabla_XN, P^N\pl Y\pr\pr-\gamma\pl\nabla_YN, P^N\pl X\pr\pr\label{eqFgeom}
\eea
where $X,Y\in\vf$ are vector fields on $\M$ and $P^N:\field{T\M}\to\field{\Ker \psi}$ designates the spacelike projector (\cf \Defi{defispacelikeprojection}).
\label{defigravfieldstrength}}
\noindent Note that the definition of $\N{F}$ is consistent since $\psi\pl N\pr=1$ and $\nabla\psi=0$ ensure that $\nabla_XN\in\field{\Ker\psi}$, $\forall X\in\vf$. 
In components, eq.\eqref{eqFgeom} reads \cite{Duval1984} \bea\N{F}_{\alpha \beta}\equiv-2\N{\gamma}_{\lambda [\alpha}\nabla_{\beta]}N^\lambda. \nn\eea 

\noindent Expressing the 2-form $\N{F}$ on the Galilean basis $\pset{N,e_i}$ (with $\pset{\psi,\theta^i}$ the associated dual basis) leads to the following decomposition:
\bea
\N{F}=2\,\N{F}\pl N,e_i\pr\psi\wedge\theta^i+\N{F}\pl e_i,e_j\pr \theta^i\wedge\theta^j. \label{eqdecompositionF}
\eea
The first term defines a spacelike vector field $\NbG\in\field{\Ker\psi}$ as $\NbG=\N{F}\pl N,e_i\pr e^i$ (where $e^i\equiv e_j \delta^{ij}$) called the \textit{gravitational force field} measured by the \fo $N$. The second term corresponds to the action of $\N{F}$ on spacelike vector fields and will be referred to as the \textit{Coriolis 2-form} $\N{\boldsymbol{\om}}\in\field{\w^2\pl\Ker\psi\pr^*}$ measured by the \fo $N$. It is defined as $\N{\boldsymbol{\om}}\pl V,W\pr=\N{F}\pl V,W\pr$, with $V,W\in\field{\Ker\psi}$.

\noindent Using eq.\eqref{eqFgeom}, these two definitions can be recast in a more geometric way which justifies further the terminology used: 
\bdefi{Gravitational force field and Coriolis 2-form \cite{Bernal2003}}{\label{defgravvort}Let $\mathscr G\pl \M,\psi,\gamma,\nabla\pr$ be a Galilean manifold and $N\in \FO$ a field of observers. The gravitational force field induced by $\nabla$ on $N$ is the spacelike vector field $\NbG\in\field{\Ker\psi}$:
\bea
\NbG\equiv \nabla_NN. \label{eqgravfield}
\eea
The Coriolis 2-form induced by $\nabla$ on $N$ is the 2-form $\N{\boldsymbol{\om}}\in\field{\w^2\pl\Ker\psi\pr^*}$, acting on $V,W\in\field{\Ker\psi}$ as\footnote{Note that our normalisation for the Coriolis 2-form differs by a factor $\half$ from the one used in \cite{Bernal2003}. }:
\bea
\N{\boldsymbol{\om}}\pl V,W\pr\equiv \gamma\pl \nabla_VN, W\pr-\gamma\pl V,\nabla_WN\pr. \label{eqvorticity}
\eea
 }

\noindent The compatibility condition of the Galilean connection $\nabla$ with the absolute clock $\psi$ (\cf Definition \ref{Galileanmanifold2}) ensures that $\psi\pl\nabla_XN\pr=X\crl\psi\pl N\pr\crr=0$, for all $X\in \field{T\M}$. This expression ensures $\psi\pl\nabla_NN\pr=0$, which in turn guarantees that $\NbG$ is spacelike. 

As one can see from eq.\eqref{eqgravfield}, the gravitational force field represents the obstruction of the \fo $N$ to be geodesic. In turn, for such a \fop, free falling objects (\ie that follow geodesics) appear to experience a gravitational force field. Similarly, the Coriolis 2-form is related to the ``Coriolis force'' associated to rotations of local observers with respect to each other.  
According to the decomposition \eqref{eqdecompositionF}, the gravitational force field $\NbG$ and the Coriolis 2-form $\N{\boldsymbol{\om}}$ associated to the \fo $N$ encode all the information contained in the 2-form $\N{F}$. {Equivalently, this can be seen from the relations $\N{F}(N,V)=\gamma(\NbG,V)$ and $\N{F}(V,W)=\N{\boldsymbol{\om}}\pl V,W\pr$ for any spacelike vector fields $V,W\in\field{\Ker\psi}$.}

\noindent Armed with the previous definitions, we now can articulate the following important Proposition:
\bpropp{Torsionfree special connection \cite{Kunzle1972,Kunzle1976}}{\label{specialobs}Given a field of observers $N\in \FO$, there exists a unique torsionfree Galilean connection $\N{\Gamma}\in\mathscr D\pl\M,\psi,\gamma\pr$ compatible with the Augustinian structure $\mathscr S\pl\M,\psi,\gamma\pr$ such that the gravitational fieldstrength measured by the field of observers $N$ with respect to $\N{\Gamma}$ vanishes. We call $\N{\Gamma}$ the torsionfree special connection associated to $N$. }

\noindent {Torsionfree special connections\footnote{A Galilean manifold $\mathscr G\pl\M,\psi,\gamma,\nabla\pr$ where $\nabla$ is a torsionfree special connection was called a {\it Newton-Cartan-Milne} structure in \cite{Duval2009a}. } play an important role in the condensed matter applications of Newton-Cartan {geometry} (\eg in \cite{Carter1994,Son2013}).}
The space of torsionfree special connections compatible with an Augustinian structure $\mathscr S\pl\M,\psi,\gamma\pr$ will be denoted $\mathscr D_0\pl\M,\psi,\gamma\pr$. An explicit expression of $\N{\Gamma}$ in components is given by (\cf \cite{Kunzle1972,Kunzle1976}):
\bea
\N{\Gamma}{}^\lambda_{\mu\nu}=N^\lambda\p_{(\mu}\psi_{\nu)}+\half h^{\lambda\rho}\pl\p_\mu\overset{N}{\gamma}_{\rho\nu}+\p_\nu\overset{N}{\gamma}_{\rho\mu}-\p_\rho\overset{N}{\gamma}_{\mu\nu}\pr\label{specialChristoffel}. 
\eea
{In physical terms, a Galilean manifold endowed with such a torsionfree special connection decribes a nonrelativistic spacetime where there exists a privileged field (possibly a class) of ``inertial'' observers, \ie measuring no gravitational force field nor Coriolis 2-form. The simplest example is the Galilei spacetime where $N=\frac{\partial}{\partial t}+v^i\frac{\partial}{\partial x^i}$ with $v^i$ constant. In order to account for the general case, the field of forces experienced by $N$ must be separately specified. }

\bthm{\cf \cite{Dombrowski1964,Kunzle1972}}{\label{thmgalileiarbitrariness}Given a field of {observers $N$, the space $\mathscr D\pl\M,\psi,\gamma\pr$ of torsionfree Galilean connections compatible with a given Augustinian structure $\mathscr S\pl\M,\psi,\gamma\pr$ possesses the structure of a vector space, the origin of which is the torsionfree special connection $\N{\Gamma}$, and $\mathscr D\pl\M,\psi,\gamma\pr$ is then isomorphic to the vector space $\forma{2}{\M}$ of 2-forms on $\M$.} }

\noindent Once a field of observers $N\in \FO$ has been picked, any torsionfree Galilean connection $\Gamma\in\mathscr D\pl\M,\psi,\gamma\pr$ can thus be represented as
\bea
{\Gamma}{}^\lambda_{\mu\nu}=\N{\Gamma}{}^\lambda_{\mu\nu}+h^{\lambda\rho}\psi_{(\mu}\N{F}_{\nu)\rho}\label{ChristoffelGalilean}
\eea
where $\N{\Gamma}\in\mathscr D_0\pl\M,\psi,\gamma\pr$ is the torsionfree special connection associated to $N$ and $\N{F}\in\forma{2}{\M}$ the 2-form defined as $\N{F}\equiv\varphi\big( \Gamma-\N{\Gamma}\big)$. 
The superscript acts here as a reminder of the fact that $\N{F}$ varies whenever one chooses a different field of observers to pinpoint an origin to $\mathscr D\pl\M,\psi,\gamma\pr$. 
{\noindent In more physical terms, given a metric structure on spacetime (seen as a ``container'') and a field of observers $N\in \FO$, the arbitrariness of choice in the
torsionfree compatible connection $\nabla$ is encoded into the ``force fields'' (which can be freely specified) measured by these observers. 
The terminology ``gravitational fieldstrength'' (measured by a background field of observers $N$) pursues the analogy between the geodesic equation $v^\nu\nabla_\nu v^\mu=0$ (for a field of observers $v$) and the Lorentz force via its rewriting as $v^\nu\stackrel{N}{\nabla}_\nu v^\mu\,=\,2\,h^{\mu\rho}\N{F}_{\rho\nu}v^\nu$ with the help of \eqref{ChristoffelGalilean}. In the latter equation, the
gravitational fieldstrength $\N{F}$ plays the role of the Faraday tensor while $\stackrel{N}{\nabla}$ stands for the torsionfree special connection associated to the background field of observers $N$.
Accordingly, the gravitational force field $\NbG$ is the analogue of the electric field, while the Coriolis 2-form $\N{\boldsymbol{\om}}$ plays the role of the (Hodge dual to the) magnetic field.}

In order to obtain a precise characterisation of the way the representation of $\Gamma$ gets modified under a change of field of observers, one needs first to understand how torsionfree special connections are related one to another:

\blem{}{\label{lemspecial}Let $N'$ and $N\in \FO$ be two fields of observers related by a Milne boost parameterised by the spacelike vector field $V\in\field{\Ker\psi}$ (\ie $N'=N+V$). 
{
The gravitational fieldstrength, measured by $N$, with respect to the torsionfree special connection $\Np{\Gamma}$ associated to $N'$
is the exact 2-form $\stackrel{N,V}{F}\equiv -d \hspace{-1.5mm}\PhiNV\hspace{-1mm}$, \ie minus the exterior derivative of the 1-form $\hspace{-1.5mm}\PhiNV\hspace{-1mm}\in\Omega^1(\M)$ defined by:} \bea\Phi&:&\FO\times \field{\Ker \psi}\to \ff\nn\\&&\pl N,V\pr\mapsto \hspace{-1.5mm}\PhiNV\hspace{-1mm}\equiv\N{\gamma}\pl V\pr-\half \,{\gamma}\pl V,V\pr\psi.  \label{eqPhi}\eea
}
{
\noindent In components, the respective torsionfree special connections $\Np{\Gamma}$ and $\N{\Gamma}\in\mathscr D_0\pl\M,\psi,\gamma\pr$ are related via
\bea
\Np{\Gamma}{}^\lambda_{\mu\nu}=\N{\Gamma}{}^\lambda_{\mu\nu}+h^{\lambda\rho}\psi_{(\mu} \hspace{-1.5mm}\stackrel{N,V}{F}\hspace{-2.5mm}_{\nu)\rho}\,.\label{gammanprime}
\eea
}
In more abstract terms, the previous Lemma can be restated by saying that the Milne group $\field{\Ker\psi}$ acts transitively on the space $\mathscr D_0\pl\M,\psi,\gamma\pr$ of torsionfree special connections via the group action $\N{\Gamma}\mapsto \N{\Gamma}+\varphi\un\Big( -\dPhiNV\, \Big)$, with $V\in\field{\Ker\psi}$. 
Correspondingly, when one switches the choice of the origin of $\mathscr D\pl\M,\psi,\gamma\pr$ from $\N{\Gamma}$ to $\Np{\Gamma}$, the representation of any torsionfree Galilean connection $\Gamma\in\mathscr D\pl\M,\psi,\gamma\pr$ becomes
\bea
{\Gamma}{}^\lambda_{\mu\nu}=\N{\Gamma}{}^\lambda_{\mu\nu}+h^{\lambda\rho}\psi_{(\mu}\N{F}_{\nu)\rho}=\Np{\Gamma}{}^\lambda_{\mu\nu}+h^{\lambda\rho}\psi_{(\mu}\Np{F}_{\nu)\rho}
\eea
where the 2-forms $\N{F}$ and $\Np{F}$ in $\forma{2}{\M}$ are related according to
\bea
\Np{F}=\N{F}+\dPhiNV. \label{eqtransF}
\eea
\begin{figure}[ht]
\centering
   \includegraphics[width=1\textwidth]{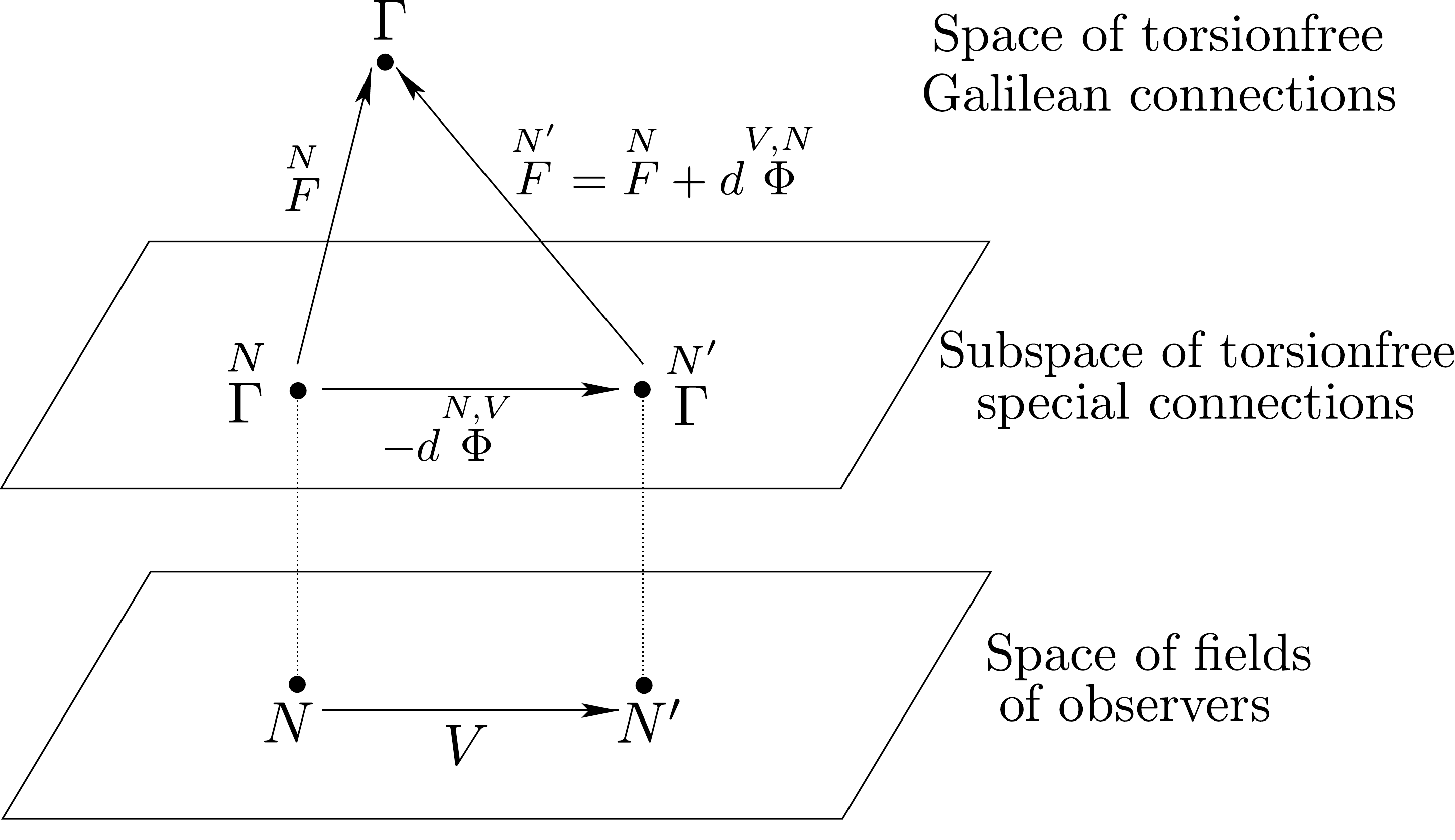}
  \caption{Representation change under a Milne boost}
\end{figure}

\noindent Given a field of observers $N$, this last expression defines an action of the Milne group $\Milneg$ on the vector space of 2-forms. Since the Milne group acts on both $\FO$ and $\forma{2}{\M}$, one can define a group action of the Milne group on the space $FO\pl\M,\psi\pr\times\forma{2}{\M}$ as 
\bea
\pl N,\N{F}\pr{\mapsto}\pl N+V, \N{F}+\dPhiNV\, \pr
\eea
with $N\in FO\pl\M,\psi\pr$, $\N{F}\in\forma{2}{\M}$, $V\in\Milneg$ and the 1-form $\hspace{-1.5mm}\PhiNV\hspace{-1mm}\in\ff$ is defined in eq.\eqref{eqPhi}. 

\bdefi{Gravitational fieldstrength}{A Milne orbit in $FO\pl\M,\psi\pr\times\forma{2}{\M}$ is dubbed a gravitational fieldstrength. The space of gravitational fieldstrengths will be denoted $$\mathscr F\pl\M,\psi,\gamma\pr:=\frac{\FO\times\forma{2}{\M}}{\Milneg}\,.$$}
\noindent We note that, since the Milne group $\Milneg$ acts regularly on $\FO$, for each $f\in\mathscr F\pl\M,\psi,\gamma\pr$ and given any $N\in\FO$, there exists a unique element $\pl N,\N{F}\pr$ in the orbit $f$, where $\N{F}\in\forma{2}{\M}$ is the gravitational fieldstrength measured by the given field of observers $N$\,. The corresponding gravitational fieldstrength is the equivalence class $[N,\N{F}]$.
\bprop{The space $\mathscr F\pl\M,\psi,\gamma\pr$ of gravitational fieldstrengths is an affine space modelled on $\GammaV$. }
\proof{ ~\\
We define the following subtraction map:
\bea
\mathscr F\pl\M,\psi,\gamma\pr\times\mathscr F\pl\M,\psi,\gamma\pr&\to& \GammaV\label{eqsubop}\\
\pl\Big[ N,\N{F}_2\Big],\Big[ N,\N{F_1}\Big]\pr&\mapsto&\varphi\un\pl \N{F}_2-\N{F}_1\pr \text{ for a given }N\in\FO. \nn
\eea
This map is well defined since, choosing a different field of observers $N'=N+V$ with $V\in\Milneg$, the difference $\N{F}_2-\N{F}_1$ becomes $\Np{F}_2-\Np{F}_1=(\N{F}_2+\dPhiNV)-(\N{F}_1+\dPhiNV)=\N{F}_2-\N{F}_1$. Furthermore, the subtraction map defined above satisfies Weyl's axioms defining an affine space, so that $\mathscr F\pl\M,\psi,\gamma\pr$ is indeed an affine space modelled on $\GammaV$. 
}

\noindent Using this terminology, we can further characterise the affine space of torsionfree Galilean connections as follows:
\bprop{\label{corGalilean}The space of torsionfree Galilean connections compatible with a given Augustinian structure possesses the structure of an affine space canonically isomorphic to the affine space of gravitational fieldstrengths. 
}
\proof{ ~\\
Let us define the map:
\bea
\Lambda\,:\,\mathscr D\pl\M,\psi,\gamma\pr&\to&\mathscr F\pl\M,\psi,\gamma\pr
\,:\,\Gamma\mapsto\Big[N,\N{\Theta}\pl \Gamma\pr\Big]. \nn
\eea
Using the subtraction map \eqref{eqsubop}, we compute:
\bea
\Lambda\pl\Gamma'\pr-\Lambda\pl\Gamma\pr&=&\varphi\un\pl \N{\Theta}\pl\Gamma'\pr-\N{\Theta}\pl\Gamma\pr\pr \text{ for some }N\in\FO\nn\\
&=&\Gamma'-\Gamma\nn
\eea
where eq.\eqref{eqaffinemap} has been used. The map $\Lambda$ is thus an affine isomorphism modelled on the identity map in $\GammaV$. 
}

\noindent The former reasonings and the corresponding chain of isomorphisms of vector and affine spaces will be repeated later on for the generalisation to the torsional case (and also in similar constructions of our related work \cite{Bekaert2015}). The logic of the argument is very general and is summarised in Proposition \ref{propaffine} of Appendix \ref{proofaffine}.

\paragraph{Remarks (on Milne invariance and special connections):}
~\\

 \noindent Let us conclude the present Section with some retrospective remarks aiming to draw comparison with the relativistic case. As noted before, the fact that $\N{\Theta}$ is non-canonical prevents to single out a unique origin which would be the analogue of the Levi-Civita connection. Rather, the present construction introduces a \textit{subspace} of privileged origins $\mathscr D_0(\M,\psi,\gamma)$ on which the Milne group $\Milneg$ has been shown to act transitively in Lemma \ref{lemspecial}. We emphasise that the appearance of the Milne group in this context is a mere consequence of our choice to restrict the origin to belong to the space of torsionfree special connections. In fact, it is only the explicit representation of a Galilean connection $\Gamma$ in terms of a given special connection $\N{\Gamma}$ which depends on the choice of $N$. More precisely, each of the two terms in the decomposition \eqref{ChristoffelGalilean} depends on $N$ but their sum does not. Indeed, a Galilean connection is independent from the choice of field of observers $N$ associated with the special connection $\N{\Gamma}$ used as origin in order to represent it. In this sense, any Galilean connection is a Milne invariant object.
This geometric perspective on the Milne invariance of nonrelativistic connections might provide a helpful conceptual tool to readdress the recent discussions on this issue. 

\vspace{4mm}

One way to partially reduce the ambiguity in the definition of the torsionfree Galilean connection is to impose supplementary conditions. The following condition \cite{Kunzle1972,Duval1977} has been proved very useful: 
\bdefi{Duval-K\"unzle condition \cite{Kunzle1972,Duval1977}}{\label{DuvalKunzle}Let $\mathscr G\pl \M,\psi,\gamma,\nabla\pr$ be a Galilean manifold and denote $R$ the curvature of the Galilean connection $\nabla$. The Duval-K\"unzle condition then reads:
\bea
\alpha\Bigl( R\Big( X,h\pl\beta\pr;Y\Big)\Bigr)=\beta\Bigl( R\Big( Y,h\pl\alpha\pr;X\Big)\Bigr)\label{eqDuvalKunzle}
\eea
for all $X,Y\in\field{T\M}$ and $\alpha, \beta\in\form{\M}$.  
}
\noindent This condition on the curvature operator $R$ is written more transparently in components as:
\bea
R\, {}^{\mu\, \, \, \, \nu}_{\, \, \, \alpha\, \, \, \, \beta}=R\, {}^{\nu\, \, \, \, \mu}_{\, \, \, \beta\, \, \, \, \alpha}\nn
\eea
with $R\, {}^{\mu\, \, \, \, \nu}_{\, \, \, \alpha\, \, \, \, \beta}\equiv h^{\nu \rho}\, R\, {}^{\mu}_{\, \, \, \alpha\, \rho\, \beta}$. 
Appendix \ref{Curvtfree} is devoted to the study of the curvature tensor for a Galilean manifold, we discuss in particular some useful identities as well as classic constraints encountered in the literature, focusing on the torsionfree case. 

\subsection{Torsionfree Newtonian manifolds}

\noindent We now turn our attention to the study of torsionfree Galilean manifolds satisfying the Duval-K\"unzle condition (\cf \Defi{DuvalKunzle}). Let $\mathscr S\pl\M,\psi,\gamma\pr$ be an Augustinian structure.  
\bdefi{Torsionfree Newtonian manifold \cite{Kunzle1972,Duval1977}}{\label{defiNewtonianmanifold}A torsionfree Newtonian manifold $\mathscr N\pl\M,\psi,\gamma,\nabla\pr$ is a torsionfree Galilean manifold whose Galilean connection satisfies the Duval-K\"unzle condition. The Koszul connection $\nabla$ is then referred to as a torsionfree Newtonian connection. }
\noindent The space of torsionfree Newtonian connections will be denoted $\hat{\mathscr D}\pl\M,\psi,\gamma\pr$. A non-trivial result \cite{Kunzle1972,Duval1977} consists in the following fact: the map \bea K:\mathscr D\pl\M,\psi,\gamma\pr\to\field{T^*\M\otimes T^*\M\otimes T\M\otimes T\M}:\Gamma\lmn\mapsto R\, {}^{\mu\, \, \, \, \nu}_{\, \, \, \alpha\, \, \, \, \beta}-R\, {}^{\nu\, \, \, \, \mu}_{\, \, \, \beta\, \, \, \, \alpha}\nn\eea despite being non-linear, is an affine map, so that there exists a linear map \bea\theta:\GammaV\to\field{T^*\M\otimes T^*\M\otimes T\M\otimes T\M}\nn\eea such that \bea
K\pl \Gamma'\pr-K\pl \Gamma\pr=\theta\pl \Gamma'-\Gamma\pr,\quad \text {for all }\, \Gamma', \Gamma\in\mathscr D\pl\M,\psi,\gamma\pr. \nn
\eea 
\noindent Since $\theta$ is linear, its kernel, denoted $\hat{\mathscr V}\pl\M,\psi,\gamma\pr$, is a vector subspace of $\GammaV$. Moreover, it can be shown \cite{Kunzle1972,Duval1977} that $\hat{\mathscr V}\pl\M,\psi,\gamma\pr$ is isomorphic to the vector space ${\Omega}^{2}\pl{\M}\pr\cap\Ker d$ of closed 2-forms on $\M$. The explicit form of the isomorphism $\hat\varphi:\hat{\mathscr V}\pl\M,\psi,\gamma\pr\to{\Omega}^{2}\pl{\M}\pr\cap\Ker d$ is obtained by restriction of the isomorphism $\varphi:\GammaV\to\forma{2}{\M}$ (\cf eq.\eqref{eqdefivarphi}). The following Theorem, summing up the previous discussion, can be seen as a specialisation of Proposition \ref{propaffineform}:

\bthm{\cf \cite{Kunzle1972,Duval1977}}{\label{propaffineformNew}The space $\hat{\mathscr D}\pl\M,\psi,\gamma\pr$ of torsionfree Newtonian connections possesses the structure of an affine space modelled on the vector space ${\Omega}^{2}\pl{\M}\pr\cap\Ker d$ of closed 2-forms on $\M$. }

\noindent Furthermore, torsionfree special connections are Newtonian:
\bpropp{\cf \cite{Kunzle1972,Duval1977}}{\label{proptscnewt}The space ${\mathscr D_0}\pl\M,\psi,\gamma\pr$ of torsionfree special connections is a subspace of the space of torsionfree Newtonian connections $\hat{\mathscr D}\pl\M,\psi,\gamma\pr$. }

\noindent A converse statement will be provided in \Prop{Newton=Special}. The previous Proposition guarantees that the restriction $\N{\hat \Theta}:\hat{\mathscr D}\pl\M,\psi,\gamma\pr\to{\Omega}^{2}\pl{\M}\pr\cap\Ker d$ of the isomorphism $\N{\Theta}:\mathscr D\pl\M,\psi,\gamma\pr\to\forma{2}{\M}$ (\cf eq.\eqref{eqdefivarphiN}) is itself an isomorphism. 
In this light, the Duval-K\"unzle condition can be reinterpreted as a geometric characterisation for the closedness of the gravitational fieldstrength $\N{F}$ measured by the field of observers $N$. Torsionfree special connections can then be used in order to represent any torsionfree Newtonian connections as: 
\bea
{\Gamma}{}^\lambda_{\mu\nu}=\N{\Gamma}{}^\lambda_{\mu\nu}+h^{\lambda\rho}\psi_{(\mu}\N{F}_{\nu)\rho}
\eea
with $\Gamma\in\hat{\mathscr D}\pl\M,\psi,\gamma\pr$ a Newtonian connection, $\N{\Gamma}\in\mathscr D_0\pl\M,\psi,\gamma\pr$ the torsionfree special connection associated to the field of observers $N$ and $\N{F}\in{\Omega}^{2}\pl{\M}\pr\cap\Ker d$ a closed 2-form. 

\noindent Applying Poincar\'e Lemma, one can locally write the closed gravitational fieldstrength $\N{F}$ as an exact 2-form so that there exists a class of 1-forms $\overset{N}{A}\in\form{\M}$ satisfying $\N{F}=d\overset{N}{A}$. Two equivalent 1-forms $\N{A'}$ and $\N{A}$ differ by an exact differential: $\N{A'}=\N{A}+df$, with $f\in \fonc{\M}$. Acting on $\ff$ with the action $\N{A}\mapsto\N{A}+df$ of the ``Maxwell group'' $\fonc{\M}$ will be referred to as a \textit{Maxwell gauge transformation}. Locally (\ie on a simply connected neighborhood), the vector space of closed 2-forms is thus isomorphic to the vector space of Maxwell orbits. We will call a Maxwell orbit $[A]$ a {\it principal connection} (\cf Section \ref{paragraphoutofthecave}) and denote $\PC$ the space of principal connections on $\M$.  

\noindent Now, we let $[\N{A}]\in\PC$ be a principal connection on $\M$ such that $\N{F}=d\overset{N}{A}$ for any representative $\N{A}\in[\N A]$. We now investigate how $[\overset{N}{A}]$ transforms under a change of origin. Explicitly, when the origin is switched from the torsionfree special connection associated with the field of observers $N$ to the one associated with $N'=N+V$, the principal connection $[\N{A}]$ gets mapped to $[\overset{N}{A}]\mapsto [\overset{N}{A}+\hspace{-1.5mm}\PhiNV\hspace{-1mm}]$, as follows directly from the transformation of $\N{F}$ \big(\cf eq.\eqref{eqtransF}\big) with $\hspace{-1.5mm}\PhiNV\hspace{-1mm}$ the 1-form defined in eq.\eqref{eqPhi}. The previous relation thus defines an action of the Milne group $\Milneg$ on the space $\FO\times\PC$ as
\bea
\pl N,[\N A]\pr&\mapsto&\pl N+V,[\overset{N}{A}+\hspace{-1.5mm}\PhiNV\hspace{-1mm}]\pr. \nn
\eea

\noindent Similarly to the Galilean case, one is led to define an additional structure supplementing the Augustinian one in order to solve the equivalence problem for Newtonian manifolds. We define the Newtonian analogue of a gravitational fieldstrength as:

\bdefi{Gravitational potential}{\label{defigravpot}Let $\mathscr L\pl \M, \psi, \gamma\pr$ be a Leibnizian structure. An orbit of the Milne group $\Milneg$ in $\FO\times\PC$ is called a gravitational potential.  
}

\noindent In other words, a gravitational potential is an equivalence class where two couples $\Big( N', \overset{N'}{A}\Big)$ and $\Big(N,\N{A}\Big)$ are said to be equivalent if there exists a spacelike vector field $V\in\Milneg$ and a function $f\in C^\infty\pl\M\pr$ such that 
\bea
\begin{cases}
N'=N+V\\
\overset{N'}{A}=\N{A}+\hspace{-1.5mm}\PhiNV\hspace{-1mm}+df. \label{eqMaxwellMilneboost}
\end{cases}
\eea
In a representative $\Big(N,\N{A}\Big)$, the second entry $\N{A}$ is called a 
gravitational gauge 1-form for the field of observers $N$. 

\noindent The space $\mathscr P\pl\M,\psi,\gamma\pr:=\big(\FO\times\PC\big)/\Milneg$ of gravitational potentials possesses a structure of affine space modelled on the space $\PC$ with subtraction map:
\bea
\mathscr P\pl\M,\psi,\gamma\pr\times\mathscr P\pl\M,\psi,\gamma\pr&\to&\PC\nn\\
\pl \big[ N,[A]\big],\big[ N,[\tilde A]\big]\pr&\mapsto&\big[ \tilde A-A\big]. \nn
\eea
\noindent The next Proposition provides a local refinement of Proposition \ref{propaffineformNew}: 
\bprop{\label{corNewtonian}Locally (\ie on a simply connected neighborhood), the space of torsionfree Newtonian connections compatible with a given Augustinian structure is an affine space canonically isomorphic to the affine space of gravitational potentials. 
}

\subsection{Variational approach}\label{paraLagrangianstructures}

\noindent The present section revisits the equivalence problem for Newtonian manifolds (\ie the search for extensions of a given Augustinian structure determining uniquely a Newtonian connection) by displaying an alternative formulation \cite{Trumper1983}, based on Coriolis-free fields of observers (\cf Definition \ref{defgravvort}). We start by stating the following Proposition:

\bprop{\label{propZ}Let $\mathscr N\pl\M,\psi,\gamma,\nabla\pr$ be a Newtonian manifold associated to the gravitational potential $\Big[N,[\N{A}]\Big]$. The field of observers $Z\in \FO$ is Coriolis-free if and only if 
\bea
Z=N-h\pl\N{A}\pr+h\pl df\pr\label{Zdef}
\eea
for a function $f\in C^\infty\pl\M\pr$ and a couple $\Big(N,\N{A}\Big)$ in the equivalence class. 
}
\noindent The proof of Proposition \ref{propZ} can be found in Appendix \ref{proofpropZ}.

\noindent In the following, we let $\mathscr N\pl\M,\psi,\gamma,\nabla\pr$ be a Newtonian manifold associated to the gravitational potential $\Big[N,[\N{A}]\Big]$. The quantity $Z=N-h\Big(\N{A}\Big)$ with $N\in\FO$ and $\N{A}$ an arbitrary representative of $[\N A]$ is invariant under a Milne boost (this fact is shown in the proof of \Prop{propZ}), so that the 1-form $\N{A}$ can be thought of as a compensator field, used in order to construct Milne-invariant objects (\cf Table \ref{tableMilneinvariant} below). Moreover, \Prop{propZ} ensures that any Newtonian manifold admits Coriolis-free fields of observers and even provides an explicit way to construct them: namely, one can go from any field of observers $N\in \FO$ to a Coriolis-free \fo $Z$ via a Milne boost parameterised by the 1-form $\chi=-\overset{N}{A}$. Under such a Milne boost, the gravitational gauge 1-form $\N{A}$ for $N$ gets mapped to a gravitational gauge 1-form $\Z{A}\equiv\half\, \phi\,\psi$ which is along the absolute clock and where the explicit form of the function $\phi\in\fonc{\M}$ is given in:
\bdefi{Gravitational gauge scalar}{Consider a gravitational gauge 1-form $\N{A}$ for the field of observers $N$. 
The function
\be
\phi\equiv 2\overset{N}{A}\pl N\pr-h\Big( \overset{N}{A},\overset{N}{A}\Big)
\ee
is called the gravitational gauge scalar corresponding to $\N{A}$.
}

\noindent This denomination is justified by the form taken by the gravitational force field $\Z{\textbf{G}}\equiv\nabla_ZZ=-\half h\pl d\phi \pr$. {As one can see, the gravitational force field measured by a Coriolis-free \fo derives from a potential (up to a factor, the scalar potential $\phi$).} It can be checked that the gravitational gauge scalar $\phi$ is also a Milne-invariant object. However, it is not gauge invariant, a point which will be adressed in details after the following example. 

\bexa{Galilei and Newton-Hooke spacetimes}{\label{exGalNHZ}The Augustinian structure of these spacetimes is composed of the absolute clock $\psi=dt$ and rulers
$\gamma=\delta_{ij}\,dx^i\vee dx^j$.  
The Galilei and Newton-Hooke spacetimes (Example \ref{exGalNH}) are also endowed with a Newtonian connection, the only nonvanishing components of which are
$\Gamma^{i}_{00}=-\frac{k}{\tau^2}\,x^i$ with $k=0$ for the Galilei spacetime. The field of observers $Z=\frac{\partial}{\partial t}$ is Coriolis-free and measures the gravitational force field $\Z{\textbf{G}}=-\frac{k}{\tau^2}\,\textbf{x}$ which derives from the  gravitational gauge scalar $\phi=\frac{k}{\tau^2}\,x_ix^i$.
The gravitational gauge 1-form for the Coriolis-free field of observers $Z$ is thus $\Z{A}=\frac{k}{2\tau^2}\,x_ix^i\,dt$\,.
Notice that the collection of all Coriolis-free field of observers are obtained from $Z$ by shifting its spatial part by
an irrotational relative velocity field, \ie a gradient $v^i=\partial^i f$.  
}
\vspace{4mm}

\noindent We argued previously that the couple $\pl Z,\half\, \phi\,\psi\pr$ constituted a distinguished representative of the gravitational potential $\Big[N,[\N{A}]\Big]$ characterising the Newtonian manifold $\mathscr N$. Conversely, the whole equivalence class $\Big[N,[\N{A}]\Big]$ can be reconstructed from one of its representatives using relations \eqref{eqMaxwellMilneboost}. Therefore, one can characterise a Newtonian manifold $\mathscr N$ by an Augustinian structure $\mathscr S\pl\M,\psi,\gamma\pr$ together with a couple $\pl Z,\phi\pr$. 

\noindent In order to make a converse statement, one needs first to acknowledge the fact that a given Newtonian manifold does not define a unique \Cffo but rather a class thereof. Indeed, two Coriolis-free fields of observers $Z$ and $Z'\in \FO$ have been seen to be related by a Maxwell transformation $Z'=Z-h\pl df\pr$ with gauge function $f\in\fonc{\M}$ (\cf \Prop{propZ}). This is a direct consequence of the previously mentioned fact that to a given \fo $N$ corresponds a principal connection, \ie a class of 1-forms $[\N{A}]\in\PC$ differing by $\Np{A}=\N{A}+df$, for some function $f$ on $\M$. Consequently, the respective gravitational gauge scalars $\phi$ and $\phi'\in\fonc{\M}$ can be checked to be related according to $\phi'=\phi+2\, df\pl Z\pr-h\pl df,df\pr$. The previous transformation laws induce the following action of the Maxwell group $\fonc{\M}$ on $\FO\times\fonc{\M}$:
\bea
\pl Z,\phi\pr&\mapsto&\pl Z-h\pl df\pr,\phi+2\, df\pl Z\pr-h\pl df,df\pr\pr. \nn
\eea
The previous action allows to reinterpret gravitational potentials as:
\bdefi{Gravitational potential}{Let $\mathscr L\pl \M, \psi, \gamma\pr$ be a Leibnizian structure. A gravitational potential is a Maxwell orbit in $\FO\times\fonc{\M}$. 
}
\noindent We sum up the whole discussion in the following Proposition: 
\bpropp{\cf \cite{Trumper1983}}{\label{propLagNewton}Let $\mathscr S\pl \M, \psi, \gamma\pr$ be an Augustinian structure. The affine space of Newtonian manifolds $\mathscr N\pl\M, \psi, \gamma, \nabla\pr$ is canonically isomorphic to the affine space $\big(\FO\times\fonc{\M}\big)/\fonc{\M}$ of gravitational potentials $\crl Z,\phi\crr$.}
\noindent Having introduced the variables $Z$ and $\phi$, we are now in a position to articulate a converse statement to \Prop{proptscnewt}: 
\bprop{\label{Newton=Special}Locally, any torsionfree Newtonian connection is a torsionfree special connection. More precisely, 
in a neighborhood there always exists a representative of the gravitational potential such that $\phi'=0$ for the corresponding Coriolis-free field of observers $Z'$.}

\noindent The fact that the class of Newtonian and special connections are essentially one and the same in the torsionfree case seems to be known since \cite{Dombrowski1964} but it is rarely emphasised (or even stated at all) in the literature and some confusion surrounds this point. For this reason, we provide a new independent proof (see also the textbook \cite{Malament} for a distinct proof and statement) of this result by showing the statement from the second sentence of Proposition \ref{Newton=Special}.

\proof{ ~\\
The proof of the proposition amounts to show that for any given pair $(Z,\phi)$ one can always find a smooth function $f$ which is solution of 
\be
\phi+2\, df\pl Z\pr-h\pl df,df\pr=0
\label{NSp}
\ee
on a neighborhood. One considers a coordinate system $x^\mu=(t,x^i)$ such that $\psi=dt$. Without loss of generality,
the Coriolis-free field of observers takes the form $Z=\frac{\partial}{\partial t}+v^i\frac{\partial}{\partial x^i}$.
In these coordinates, the left-hand-side of \eqref{NSp} is an affine function of the absolute time derivative
$\frac{\partial f}{\partial t}$. More explicitly, the partial differential equation \eqref{NSp}
can be written in the \textit{normal} form:
\bea
&\frac{\partial f}{\partial t}=F\left(t,x^i,\frac{\partial f}{\partial x^j}\right)\,,&\label{normalform}\\
&\mbox{where}\quad F\left(t,x^i,\frac{\partial f}{\partial x^j}\right)\,\equiv\,\frac12\, h^{ij}(t,x)\frac{\partial f}{\partial x^i}\frac{\partial f}{\partial x^j}-v^i(t,x)\frac{\partial f}{\partial x^i}-\frac12\,\phi(t,x)&\nonumber
\eea
is obviously an analytic function (it is a polynomial of degree 2) in the derivatives $\frac{\partial f}{\partial x^j}$. For technical reasons, we will assume that the functions $h^{ij}(t,x)$, $v^k(t,x)$ and $\phi(t,x)$ are analytic in some neighborhood, so that $F\left(t,x^i,\frac{\partial f}{\partial x^j}\right)$ is analytic in all its arguments. Then, 
according to the existence theorem of Cauchy-Kowalewsky (see \eg \cite{CH}), there exists a unique solution of \eqref{normalform} in some sufficiently small neighborhood for each Cauchy data $f(t_0,x^i)=g(x^i)$ with $g$ analytic.
}

\noindent Another interesting feature of the present formulation is embodied by the following Proposition. We first define the notion of a Lagrangian metric:
\bdefi{Lagrangian metric}{\label{defiLagmetric}Let $\mathscr L\pl \M, \psi, \gamma\pr$ be a Leibnizian structure. 
A covariant metric $g\in\bforms$ on $\M$ satisfying the condition $g\pl X,Y\pr=\gamma\pl X,Y\pr$, for any $X,Y\in\field{\Ker\psi}$ will be called a Lagrangian metric. The space of Lagrangian metrics will be denoted $\Lag\pl \M,\psi,\gamma\pr$. 
}
\bprop{Let $\mathscr L\pl \M, \psi, \gamma\pr$ be a Leibnizian structure. The space $\Lag\pl \M,\psi,\gamma\pr$ of Lagrangian metrics possesses the structure of an affine space modelled on $\ff$. 
}
\proof{ ~\\
We start by showing that $\Lag\pl \M,\psi,\gamma\pr$ is an affine space modelled on the vector space 
\bea
\mathcal V\pl \M,\psi,\gamma\pr\equiv\pset{\tilde g\in\field{\vee^2T^*\M}/\tilde g\pl V,W\pr=0 \text{ for all }V,W\in\Milneg}\nn
\eea
by displaying the following subtraction map
\bea
\Lag\pl \M,\psi,\gamma\pr\times\Lag\pl \M,\psi,\gamma\pr&\to&\mathcal V\pl \M,\psi,\gamma\pr\nn\\
\pl g,g'\pr&\mapsto&g'-g\nn
\eea
which can be shown to satisfy Weyl's axioms. We now conclude the proof by constructing the canonical isomorphism:
\bea
\ensuremath{\boldsymbol \varphi}:\mathcal V\pl \M,\psi,\gamma\pr&\to&\ff\nn\\
\tilde g&\mapsto&\alpha\equiv\tilde g\pl N\pr-\half \tilde g\pl N,N\pr\psi \text{ with }N\in\FO\nn
\eea
together with its inverse
\bea
\ensuremath{\boldsymbol \varphi}\un:\ff&\to&\mathcal V\pl \M,\psi,\gamma\pr\nn\\
\alpha&\mapsto&\tilde g\equiv 2\, \psi\vee\alpha\nn.
\eea
 }
 We now introduce the map
 \bea
 \N{\ensuremath{\boldsymbol \Theta}}:\Lag\pl \M,\psi,\gamma\pr&\to& \ff\nn\\
 g&\mapsto&g\pl N\pr-\half g\pl N,N\pr\psi\nn
 \eea
 which can be checked to be an affine map modelled on $\ensuremath{\boldsymbol \varphi}$ \ie $ \N{\ensuremath{\boldsymbol \Theta}}\pl g'\pr- \N{\ensuremath{\boldsymbol \Theta}}\pl g\pr=\ensuremath{\boldsymbol \varphi}\pl g'-g\pr$ for all $g',g\in\Lag\pl \M,\psi,\gamma\pr$. According to \Prop{propaffine}, for all $N\in\FO$, the map $\N{\ensuremath{\boldsymbol \Theta}}$ endows the space of Lagrangian metrics $\Lag\pl \M,\psi,\gamma\pr$ with a structure of vector space with origin $\Ker \N{\ensuremath{\boldsymbol \Theta}}$ which can be shown to be spanned by the transverse metric $\N{\gamma}\in\Lag\pl \M,\psi,\gamma\pr$ associated to $N$. The map $\N{\ensuremath{\boldsymbol \Theta}}$ is then an isomorphism of vector spaces. Furthemore, for a given $N$, one can represent any element $g\in\Lag\pl \M,\psi,\gamma\pr$ as $g=\N{\gamma}+\ensuremath{\boldsymbol \varphi}\big( \N{A}\big)=\N{\gamma}+2\, \psi\vee\N{A}$, with $\N{A}\equiv \N{\ensuremath{\boldsymbol \Theta}}\pl g\pr\in\ff$. \noindent Under a Milne boost $N\mapsto N+V$, the form $\N{A}$ varies according to $\N A\mapsto\N A+\hspace{-1.5mm}\PhiNV\hspace{-1mm}$. 
\noindent We sum up the discussion in the following Proposition:

\bprop{\label{propbijcorNewtoZphi}Let $\mathscr S\pl \M, \psi, \gamma\pr$ be an Augustinian structure. The affine spaces of
\begin{enumerate}
\item Milne orbits $\Big[N,\N{A}\Big]$, with $N\in \FO$ and $\N{A}\in\ff$
 \item couples $\pl Z,\phi\pr$ with $Z\in \FO$ and $\phi\in \fonc{\M}$
\item Lagrangian metrics $g\in\bforma{T^*\M}$
\end{enumerate}
are canonically isomorphic, \ie
$$\frac{\FO\times\ff}{\Milneg}\,\cong\,\FO\times\fonc{\M}\,\cong\,\Lag\pl \M,\psi,\gamma\pr \,.$$
}
\noindent\Prop{propaffine} ensures the isomorphism between 1 and 3. The somewhat lengthy proof of the isomorphism between 2 and 3 is relegated to Appendix \ref{prooflemfromZphitog}. It rests on the fact that 
the only Lagrangian metric $g\in\Lag\pl \M,\psi,\gamma\pr$ satisfying 
\bea
g\pl Z\pr=\phi\,\psi\nn
\eea
reads as $g\equiv\Z{\gamma}+\phi\, \psi\vee\psi$, with $\Z{\gamma}$ the metric transverse to $Z$. 

\noindent The characterisation of Newtonian manifolds using Coriolis-free fields of observers thus allows to define a covariant metric $g$. Although we are in a nonrelativistic context, the latter metric can be nondegenerate (when the gravitational gauge scalar $\phi$ is nowhere vanishing). Under a Maxwell-gauge transformation $Z\mapsto Z-h\pl df\pr$, the metric $g$ transforms as
\bea
g\mapsto g'=g+2\, \psi\, \vee\,  df \label{eqMaxwellLagrangianmetric}
\eea
thus defining a Maxwell-group action on $\Lag\pl \M,\psi,\gamma\pr$. 
\bdefi{Lagrangian structure}{\label{defiLagstructure}Let $\mathscr L\pl \M, \psi, \gamma\pr$ be a Leibnizian structure. 
A triplet $\mathcal L\pl \M, \psi, \crl g\crr\pr$ where $[g]$ is a Maxwell-orbit in the space $\Lag\pl \M,\psi,\gamma\pr$ of Lagrangian metrics compatible with $\mathscr L\pl \M, \psi, \gamma\pr$ is called a Lagrangian structure. A quadruplet $\mathcal L\pl \M, \psi, \crl g\crr, \nabla\pr$ where $\nabla$ is a Galilean connection compatible with $\mathscr L\pl \M, \psi, \gamma\pr$ will be called a Lagrangian manifold\footnote{Our acception of the term Lagrangian manifolds should not be mistaken with the denomination Lagrangian submanifolds in symplectic geometry.  }. }
\noindent Now, one can combine Propositions \ref{propLagNewton} and \ref{propbijcorNewtoZphi} in order to show:
\bpropp{\cf \cite{Trumper1983}}{\label{propLagNewtong}Let $\mathscr S\pl \M, \psi, \gamma\pr$ be an Augustinian structure. There is a canonical isomorphism between the affine spaces of Newtonian manifolds $\mathscr N\pl\M, \psi, \gamma, \nabla\pr$ and 
Lagrangian structures $\mathcal L\pl \M, \psi, \crl g\crr\pr$.
}
\noindent The following table sums up the Milne-invariant objects introduced in this Section along with their Maxwell-gauge transformation law:

\vspace{2mm}
\hspace{-2.5cm}
 \setlength{\extrarowheight}{2 mm}
\begin{tabular}{|M{2.9cm}|M{3.7cm}|M{4.2cm}|M{4.8cm}|}
   \hline
\rule[-0.5cm]{0cm}{1cm}Type&\rule[-0.5cm]{0cm}{1cm}Name&\rule[-0.5cm]{0cm}{1cm}Definition&\rule[-0.5cm]{0cm}{1cm}Maxwell-gauge transformation law\tabularnewline\hline
\rule[-0.5cm]{0cm}{1cm}$Z\in \FO$&\rule[-0.5cm]{0cm}{1cm}Coriolis-free field of observers&\rule[-0.5cm]{0cm}{1cm}$Z\equiv N-h\pl \overset{N}{A}\pr$&\rule[-0.5cm]{0cm}{1cm}$Z\mapsto Z-h\pl df\pr$\tabularnewline
   \hline 
\rule[-0.5cm]{0cm}{1cm}$\phi\in \fonc{\M}$&\rule[-0.5cm]{0cm}{1cm}Gravitational gauge scalar&\rule[-0.5cm]{0cm}{1cm}$\phi\equiv 2\overset{N}{A}\pl N\pr-h\pl \overset{N}{A},\overset{N}{A}\pr$&\rule[-0.5cm]{0cm}{1cm}$\phi\mapsto\phi+2\, df\pl Z\pr-h\pl df,df\pr$
\tabularnewline\hline
\rule[-0.8cm]{0cm}{2cm}$g\in\bforms$&\rule[-0.8cm]{0cm}{2cm}Lagrangian metric&\rule[-0.8cm]{0cm}{2cm}$g\equiv \overset{N}{\gamma}+2\, \psi\, \vee\, \overset{N}{A}$&\rule[-0.8cm]{0cm}{2cm}$g\mapsto g+2\, \psi\, \vee\,  df$
\tabularnewline\hline
\end{tabular}
\captionof{table}{Milne-invariant objects\label{tableMilneinvariant}}
\vspace{4mm}

The use of the ``Lagrangian'' denomination is justified by the fact that a Lagrangian metric $g$ defines a Lagrangian as $\Lag\equiv \half\, g\pl X,X\pr$ with $X\in \FO$ the tangent vector field associated to an (arbitrary) observer $x:I\subseteq\mR\to\M:\tau\mapsto x\pl \tau\pr$. In components, the Lagrangian then reads 
\bea
\Lag=\half \,g_{\mu \nu}\frac{dx^\mu}{d\tau}\frac{dx^\nu}{d\tau}.\label{LagAugustinian} 
\eea
In order to find the associated equations of motion, it must be kept in mind that the variation of the Lagrangian \eqref{LagAugustinian} is not performed over the whole space of tangent vectors but is constrained to the space of tangent vectors parameterised by the proper time $\tau$, \ie to the space of observers (\cf \Prop{propobserverpropertime}). In the generic case, the constraint $\psi_\mu\frac{dx^\mu}{d\tau}=1$ is linear in the velocities and, in general, non-holonomic (since it is of the form $f\pl x^i, \dot{x}^i,t\pr=0$). However, in the Augustinian case, the absolute clock is closed ($\psi=dt$) so that the constraint can be integrated to give a holonomic constraint (\ie of the form $f\pl x^i,t\pr=0$) which can be resolved by adopting the absolute time $t$ as parameter:
\bea
\Lag=\half\, g_{\mu \nu}\frac{dx^\mu}{dt}\frac{dx^\nu}{dt}.\label{eqLag}
\eea
{In an adapted coordinate system ($t,x^i$), the Lagrangian reads $\Lag=\half \gamma_{ij}\dot x^i\dot x^j+A_i\dot x^i-U$ where we used the relation $g\equiv \overset{N}{\gamma}+2\, \psi\, \vee\, \overset{N}{A}$ with $N\equiv\frac{\partial}{\partial t}$, $\N{A}\equiv-Udt+A_idx^i$ and $\gamma_{ij}$ the components of the collection of absolute rulers $\gamma$. The Lagrangian \eqref{eqLag} is  therefore formally identical to the one describing the motion of a charged particle minimally coupled to an electromagnetic field through the vector potential $A_i$ and the scalar potential $U$ and moving on a Riemannian manifold with metric $\gamma_{ij}$.

\bexa{}{In the particular case of the Aristotle spacetime ($\psi=dt$ and $\gamma=\delta_{ij}\,dx^i\vee dx^j$) endowed with a Coriolis-free field of observers $Z=\frac{\partial}{\partial t}$, the Lagrangian takes the standard form $\Lag=\half \,(\frac{dx^i}{dt}\frac{dx_i}{dt}+\phi)$ since the gravitational gauge 1-form reads $\Z{A}=\frac12\phi\, dt$ for this field of observers. Notice that the usual potential would be $U=-\half\phi$. }}

For a generic Augustinian structure, the Euler-Lagrange equations of motion derived from $\Lag$ take the form \cite{Trumper1983}:
\bea
g_{\alpha\beta}\frac{d^2x^\beta}{dt^2}+\half\crl\p_\mu g_{\nu\alpha }+\p_\nu g_{ \mu \alpha}-\p_\alpha g_{\mu\nu}\crr\frac{dx^\mu}{dt}\frac{dx^\nu}{dt}=0. \nn
\eea
Contracting with $h^{\lambda\alpha}$ and using the relation $g_{\alpha \beta}h^{\lambda\alpha}=\delta^{\lambda}_{\beta}-Z^\lambda\psi_\beta$ (as can be deduced from the expression of the Lagrangian metric $g$) leads to:
\bea
\frac{d^2x^\lambda}{dt^2}-Z^\lambda\psi_\nu\frac{d^2x^\nu}{dt^2} +\half h^{\lambda\alpha}\crl\p_\mu g_{ \nu\alpha}+\p_\nu g_{ \mu \alpha}-\p_\alpha g_{\mu\nu}\crr\frac{dx^\mu}{dt}\frac{dx^\nu}{dt}=0. \label{eqEOMLag}
\eea
Now, differentiating the constraint $\psi_\mu\frac{dx^\mu}{d\tau}=1$, one obtains the relation \bea\psi_\nu\frac{d^2x^\nu}{dt^2}=-\p_{(\alpha}\psi_{\beta)}\frac{dx^\alpha}{dt}\frac{dx^\beta}{dt}\nn\eea which can be substituted in eq.\eqref{eqEOMLag} to give
\bea
\frac{d^2x^\lambda}{dt^2}+\Gamma^\lambda_{\mu \nu}\frac{dx^\mu}{dt}\frac{dx^\nu}{dt}=0\nn
\eea
where the components $\Gamma^\lambda_{\mu \nu}$ read
\bea
\Gamma^\lambda_{\mu\nu}=Z^\lambda\p_{(\mu}\psi_{\nu)}+\half h^{\lambda\rho}\crl\p_\mu g_{\rho\nu}+\p_\nu g_{\rho\mu}-\p_\rho g_{\mu\nu}\crr. \label{GammaZ}
\eea
Using Table \ref{tableMilneinvariant}, one can check that the expression \eqref{GammaZ} is identical to the one of eq.\eqref{ChristoffelGalilean} (with $\N{F}=d\N{A}$), so that the Lagrangian $\Lag$ describes a free particle in geodesic motion with respect to a Newtonian connection, hence providing a concrete implementation of \Prop{propLagNewtong}. 
Note that, although being manifestly Milne-invariant, the Lagrangian $\Lag$ is not invariant under a Maxwell-gauge transformation of $g$ as $g_{\mu\nu}\mapsto g_{\mu\nu}+2\, \psi_{(\mu}\, \p_{\nu)} f$ but transforms by adjonction of a total derivative $\Lag\mapsto\Lag+\frac{df}{dt}$ which only contributes to the boundary term, so that the equations of motion (and thus the expression of $\Gamma^\lambda_{\mu \nu}$) are Maxwell-gauge invariant. Finally, notice that, when the gravitational potential vanishes (according to \Prop{Newton=Special}, this can always be achieved via a Maxwell-gauge transformation), then eq.\eqref{GammaZ} identifies with the expression of the torsionfree special connection associated to $Z$ \big(\cf eq.\eqref{specialChristoffel}\big) since $g=\overset{Z}{\gamma}$ whenever $\phi=0$.

\subsection{Towards the ambient formalism}\label{paragraphoutofthecave}

\noindent Before adressing the issue of torsional Galilean connections, we conclude the present Section by a heuristic discussion regarding the natural emergence of the ambient formalism through the study of Newtonian manifolds, thus paving the way to the more systematic discussion to appear in \cite{Bekaert2015}. 

Let\footnote{In the present section, we anticipate on the notation to be used in \cite{Bekaert2015} where nonrelativistic objects will be topped with a bar. } $\bar{\mathscr N}\pl \bar\M,\bar\psi,\bar\gamma,\bar\nabla\pr$ be a Newtonian manifold where $\bar\M$ is $(d+1)$-dimensional. Pick a \fo $\bar N\in FO\pl\bar \M,\bar\psi\pr$. The characterisation of a Newtonian manifold $\bar{\mathscr N}$ has been seen to require the introduction of a set of 1-forms $\bN{A}\in\form{\bar\M}$ with Maxwell-like transformation law $\bN{A}\mapsto\bN{A}+df$, where $f\in C^\infty\pl\bar\M\pr$. To the bundle-minded physicist, this transformation law suggests to reinterpret the 1-forms $\bN{A}$ as gauge connections for a principal $\pl\mR,+\pr$-bundle of projection $\pi:\M\rightarrow\bar\M$, where $\M$ is a $(d+2)$-dimensional manifold. Recall that, if $\N{A}\in\ff$ is an $\pl\mR,+\pr$-principal connection on $\M$, choosing a section $\sigma:\bar{\mathscr U}\to\M$ (where $\bar{\mathscr U}\subset \bar\M$ is an open subset of $\bar\M$) allows to define a gauge connection $\bN{A}\in\form{\bar{\mathscr U}}$ as $\bN{A}\equiv \sigma^*\N{A}$. Reciprocally, a collection $\pset{\bar{\mathscr U}_\alpha,\bN{A}_\alpha}$ (where the $\bar{\mathscr U}_\alpha$ form an open cover of $\bar\M$ and the set of $\bN{A}_\alpha$ differ by Maxwell-like transformation laws) defines a unique principal connection $\N{A}$. 

The principal $\pl\mR,+\pr$-bundle involves a supplementary ``internal'' direction, the vertical fiber foliation, which is a congruence of integral curves (called \textit{rays}) for the unique fundamental vector field of the principal bundle $\M$, denoted $\xi\in\field{T\M}$ and designated as the {\it wave vector field}. Since $\xi$ is the fundamental vector field, it satisfies $\N{A}\pl\xi\pr=1$ (since 1 is the generator of the Abelian Lie algebra $\mR$).

\noindent Usually (\eg in Yang-Mills theories), the fiber of an Ehresmann bundle is interpreted as an auxiliary geometric object allowing to define an internal symmetry. The key to the ambient approach consists in reinterpreting this additional direction as a new spacetime dimension. 

Now, we investigate how structures on $\bar\M$ can be lifted up to $\M$. First, the absolute clock $\bar\psi\in\form{\bar\M}$ defines a unique closed 1-form $\psi\in\ff$ as $\psi\equiv\pi^*\bar\psi$, called {\it wave covector field}. It can be checked that, since $\pi_*\xi=0$, one has $\psi\pl\xi\pr=0$, 
so that $\xi\in\field{\Ker\psi}$. The \fo $\bar N\in FO\pl\bar\M, \bar\psi\pr$ can be lifted up to $\M$ by defining $N\in \FO$ as the horizontal lift of $\bar N$ with respect to $\N{A}$ (\ie $\pi_*N=\bar N$ and $\N{A}\pl N\pr=0$) while an ambient covariant metric $\N{\gamma}\in\bforma{T^*\M}$ can be defined as the generalised pullback of the transverse metric $\bN{\gamma}\in\bforma{T^*\bar\M}$. It can be checked that $\Rad{\N{\gamma}}\cong\Spannn{\xi,N}$. The kernel of $\psi$ defines an involutive distribution ($\psi$ being closed) whose integral submanifolds are called {\it wavefront worldvolumes}. Each wavefront worldvolume can thus be envisaged as the union of an absolute space with the corresponding fibers. A wavefront wordlvolume $i:\tilde{\mathscr M}\longhookrightarrow\M$ is therefore endowed with a contravariant metric $\tilde\gamma\in\bforma{T^*\tilde\M}$ defined as the generalised pullback $\tilde\gamma\equiv i^*\N\gamma$. Contrarily to its nonrelativistic counterpart, the metric $\tilde\gamma$ is degenerate since $\tilde\gamma\pl\xi\pr=0$ (in the language of \cite{Duval2014e}, the triplet $\pl\tilde{\mathscr M},\xi, \tilde\gamma\pr$ is thus a Carroll metric structure). 

\begin{figure}[ht]
\centering
   \includegraphics[width=0.7\textwidth]{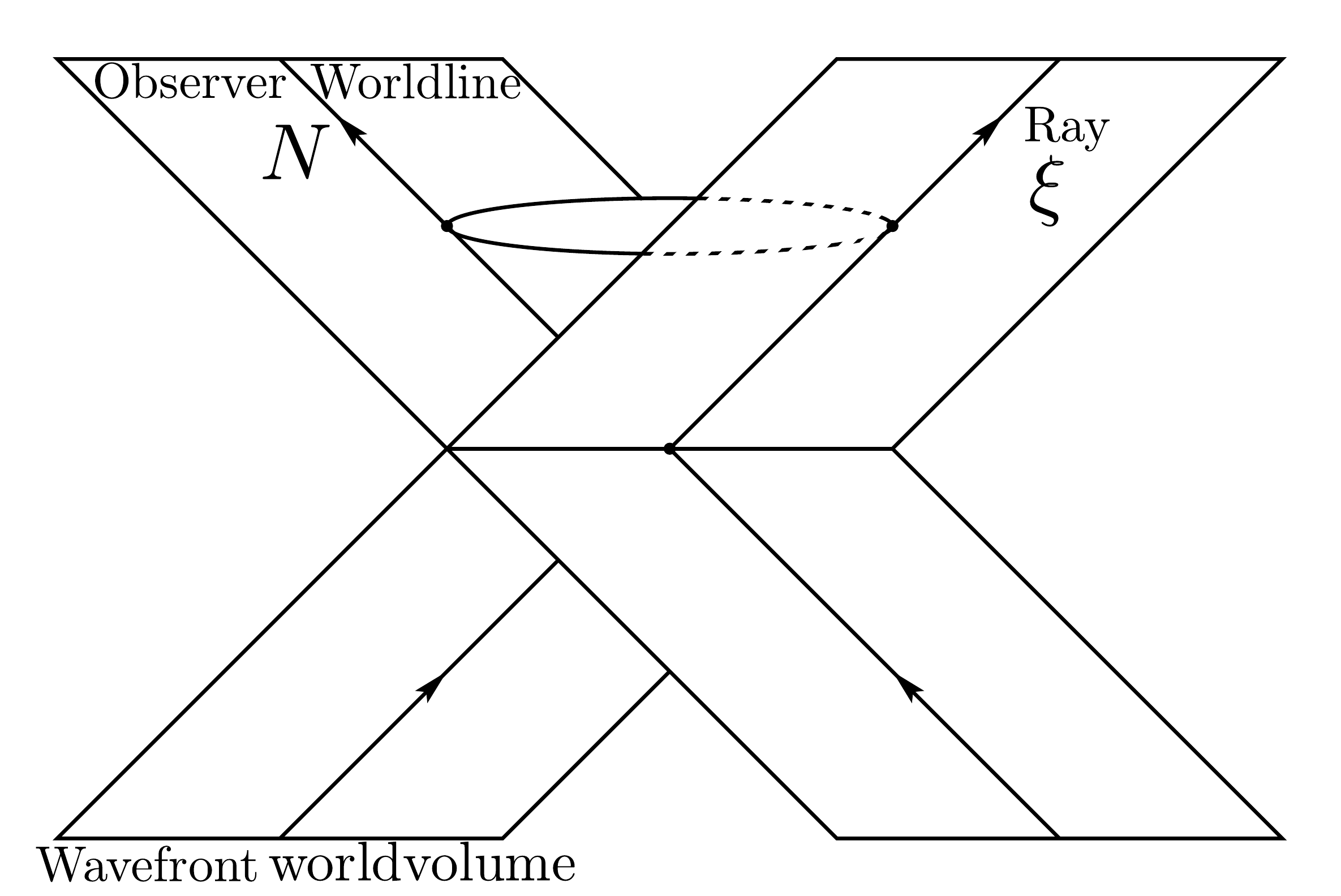}
  \caption{Relativistic light-cone basis\label{observerworldline}}
\end{figure}

According to \Prop{propLagNewtong}, a Newtonian manifold defines a class of Lagrangian metrics $\crl\bar g\crr$ where each metric $\bar g\in\bforma{T^*\bar\M}$ is given by $\bar g\equiv \overset{\bar N}{\gamma}+2\, \bar\psi\, \vee\, \overset{\bar N}{A}$ and transforms under a gauge transformation as $\bar g\mapsto \bar g'=\bar g+2\, \bar\psi\, \vee\,  df$. Similarly to the definition of a principal connection on $\M$, it can be shown that the set $\crl \bar g\crr$ defines a unique covariant metric $g\in\bforma{T^*\M}$ satisfying $\bar g=\sigma^*g$. Explicitly, the metric $g$ can be expressed as $g\equiv \overset{N}{\gamma}+2\, \psi\, \vee\, \overset{N}{A}$. Furthermore, the metric $g$ can be shown to be nondegenerate. The expression for $g$ can be used in order to compute $g\pl\xi,\xi\pr=0$ and $g\pl N,N\pr=0$ (so that $\xi$ and $N$ are null vector fields). Furthermore, $g\pl \xi\pr=\psi$ and $g\pl N\pr=\N{A}$. This implies $g(\xi,N)=1$ so that $\xi$ and $N$ form a lightcone basis (\cf \cite{Bekaert2013c} and Figure \ref{observerworldline}) and $\M$ is thus a Lorentzian manifold. Since $g$ is nondegenerate, it defines a notion of parallelism on $\M$ in the guise of the Levi-Civita connection $\nabla$ and it will be shown in \cite{Bekaert2015} (following \cite{Duval1985}) how the \LCc $\nabla$ projects down to the Newtonian connection $\bar\nabla$ on $\bar\M$. The wavevector field can then be shown to be parallel with respect to $\nabla$, so that $\M$ can be characterised as a Bargmann-Eisenhart wave (\cf the terminology used in \cite{Bekaert2013c}). 

The conclusion emerging from the line of reasoning sketched here is that the usual hierarchy between Bargmann-Eisenhart waves and Newtonian manifolds (where the latter are obtained from the former) can in fact be reversed given that a geometrical understanding of nonrelativistic spacetimes (Newtonian manifolds) leads naturally to the reconstruction of an ambient relativistic spacetime 
(Bargmann-Eisenhart waves). As always in the process of dimensional reduction, a spacetime symmetry of the ambient manifold is interpreted as an internal symmetry on the reduced manifold. A Maxwell gauge symmetry is always found in the reduced theory along one-dimensional orbits independently of the type of curves, whether spacelike (in the usual case \`a la Kaluza-Klein) or lightlike (here).

\pagebreak
\section{Torsional Galilean connections}\label{Torsional connection}

\noindent So far, we chose to restrict the scope of our analysis to \nr structures endowed with torsionfree connections. Such a restriction is quite natural when one is dealing with \nr metric structures whose absolute clock is closed (Augustinian structures) since, in this case, there exist torsionfree connections which are furthermore compatible with the metric structure (\cf Theorem \ref{thmgalileiarbitrariness}), similarly to the \rel case. Nonetheless, the introduction of torsional connections acquires increased relevance when considering \nr metric structures with non-closed absolute clock, since then the torsionfree condition and metric compatibility become mutually exclusive (\cf \Prop{compatibility}). 
In \cite{Bekaert2016}, when considering parallelism for Aristotelian structures, we will be brought to favour the torsionfree condition at the expense of the metric compatibility by considering connections projectively equivalent to Galilean/Newtonian ones. However, the alternative route is equally worth of exploration, as proven by the recent surge of interest in generalisations of Newton-Cartan geometry characterised by torsional connections (\eg \cite{Geracie2014,TNCcm,Jensen2014,TNCLif,TNCMilneLif}). 
Such approaches focus on a Leibnizian structure endowed with a Koszul connection whose torsion is tuned in order to ensure compatibility with the absolute clock and rulers. In particular, the works \cite{Jensen2014,TNCMilneLif} exhibit {a special class of
torsional connections compatible with the metric 
structure and remaining invariant under Milne boosts. As emphasised in Section \ref{sectionGalileanmanifoldss}, the latter property is automatic when the connection is understood as a geometrical object.
Nevertheless, manifest Milne-invariance of its components may for instance be achieved by making use of the manifestly 
Milne-invariant ``Lagrangian'' variables (\ie $Z$, $\phi$, $g$ in Section \ref{paraLagrangianstructures}). }

\subsection{Torsional Galilean manifolds}\label{torsionmanifs}

\noindent We start this study of \nr torsional manifolds by investigating the structure of the space of torsional Galilean connections, thus providing an analysis similar to the ones conducted in Sections \ref{sectionrelativisticstructures} and \ref{Nonrelativistic manifolds}. We then propose a generalisation of the notion of Newtonian connection to the torsional case, allowing the use of the ``Lagrangian'' variables. 
The present discussion is intended as prolegomena in order to pave the way to the description of the embedding of torsional Galilean manifolds inside relativistic spacetimes in \cite{Bekaert2015}.

\noindent Let $\mathscr L\pl\M,\psi,\gamma\pr$ be a Leibnizian structure. The space of Galilean connections compatible with $\mathscr L\pl\M,\psi,\gamma\pr$ will be denoted $\mathscr D\pl\M,\psi,\gamma\pr$.\footnote{Notice that we will make use of the same symbols to denote the various spaces and maps as in Section \ref{Nonrelativistic manifolds} in order to make more transparent the similitude between the arguments.}
\bprop{\label{PropGalileanaffinespacetr}The space $\mathscr D\pl\M,\psi,\gamma\pr$ possesses the structure of an affine space modelled on the vector space $\GammaV$ defined as\footnote{Note that the vector space $\GammaV$ differs from its torsionfree counterpart (\cf Proposition \ref{PropGalileanaffinespace}) in that its elements do not satisfy any symmetry conditions on their lower indices. }:
\bea\GammaV\equiv\pset{S\in\field{T^*\M\otimes T^*\M\otimes T\M}\ /\ \psi_\lambda S^\lambda_{\mu\nu}=0\ \text{ and }\ S^{(\lambda}_{\mu\nu}\, h^{\rho)\nu}=0}. \nn
\eea
}
\noindent The compatibility conditions \eqref{eqconditionsgalileanmanifold} then reduce the vector space on which the affine space of torsional connections is modelled from $\field{T^*\M\otimes T^*\M\otimes T\M}$ to the subspace $\GammaV$. 
\blem{}{\label{Lemcanisotr}The vector space $\GammaV$ is isomorphic to the vector space $\vectsum$. 
}
\noindent A first discrepancy with the torsionfree case is that the linear isomorphism is not canonical in the presence of torsion but rather depends on the gift of a field of observers. Explicitly, for a given $N\in \FO$ the isomorphism is given by 
\bea\N{\varphi}&:&\GammaV\to\vectsum\nn\\
&&S^\lambda_{\mu\nu}\mapsto \pl F_{\mu\nu}=-2\N{\gamma}_{\lambda[\mu}S^\lambda_{\nu]\rho}N^\rho, U^\lambda_{\mu\nu}=S^\lambda_{[\mu\nu]}\pr\nn\eea with $\N{\gamma}\in\field{\vee^2\ T^*\M}$ the transverse metric associated to $N$ while its inverse takes the form 
\bea\N{\varphi}{}\un&:&\vectsum\to\GammaV\nn\\
&&\pl F_{\mu\nu},U^\lambda_{\mu\nu}\pr\mapsto h^{\lambda\rho}\psi_{(\mu}F_{\nu)\rho}+U^\lambda_{\mu\nu}\,+\,2\,h^{\lambda\sigma}\,U^\rho_{\sigma(\mu}\,\N{\gamma}^{\,}_{\nu)\rho}. \nn\eea 
Note that the expression $F_{\mu\nu}=-2\N{\gamma}_{\lambda[\mu}S^\lambda_{\nu]\rho}N^\rho$ is {independent of the choice of field of observers $N$ \textit{only} in the absence of torsion.} Explicitly, under a Milne boost $N\mapsto N'=N+V$, with $V\in\field{\Ker\psi}$, the isomorphism $\N{\varphi}$ transforms as:
\bea
\Np{\varphi}\pl S\lmn\pr=\N{\varphi}\pl S\lmn+h^{\sigma\lambda}\N{\gamma}_{\alpha\rho}V^\alpha(\psi_\mu S^\rho_{[\nu\sigma]}+\psi_\nu S^\rho_{[\mu\sigma]})\pr\\
\Np{\varphi}{}\un\pl F_{\mu\nu},U^\lambda_{\mu\nu}\pr=\N{\varphi}{}\un\pl F_{\mu\nu}+U^\rho_{\mu\nu}\N{\gamma}_{\rho\alpha}V^\alpha\,,\, U^\lambda_{\mu\nu}\pr. \label{eqtransvarphiNun}
\eea
Proposition \ref{PropGalileanaffinespacetr} together with Lemma \ref{Lemcanisotr} then ensure the following Proposition:
\bprop{\label{propaffineformtor}The space $\mathscr D\pl\M,\psi,\gamma\pr$ possesses the structure of an affine space modelled on $\vectsum$. }
Similarly to the Lorentzian and torsionfree Galilean cases, we wish to define an affine map $\N{\Theta}:\mathscr D\pl\M,\psi,\gamma\pr\to\vectsum$ modelled on the linear map $\N{\varphi}$. As for the map $\mathscr D\pl\M,\psi,\gamma\pr{\to}\,\forma{2}{\M}$, we rely on the torsionfree prescription and map Galilean connections to the gravitational fieldstrength measured by the field of observers $N$ (\cf Definition \ref{defigravfieldstrength}). Regarding the map $\mathscr D\pl\M,\psi,\gamma\pr{\to}\,\vectasymduker$, a natural prescription consists in associating each Galilean connection with the spacelike projection transverse to $N$ of its torsion tensor field $P^N\pl \Gamma^\lambda_{[\mu\nu]}\pr\equiv\Gamma^\lambda_{[\mu\nu]}-N^\lambda \psi_\alpha \Gamma^\alpha_{[\mu\nu]}$. Using eq.\eqref{eqtorsionconstraint}, the last expression becomes $P^N\pl \Gamma^\lambda_{[\mu\nu]}\pr=\Gamma^\lambda_{[\mu\nu]}-N^\lambda\p_{[\mu}\psi_{\nu]}$. The map $\N{\Theta}$ thus takes the explicit form:
\bea
\N{\Theta}:&&\mathscr D\pl\M,\psi,\gamma\pr\to\vectsum\nn\\
&&\nabla\mapsto \pl \N{F}_{\alpha\beta}=-2\N{\gamma}_{\lambda [\alpha}\nabla_{\beta]}N^\lambda, \N{U}{}^\lambda_{\mu\nu}=\Gamma^\lambda_{[\mu\nu]}-N^\lambda\p_{[\mu}\psi_{\nu]}\pr\nn
\eea

It can be checked that $\N{\Theta}$ is an affine map associated to the isomorphism $\N{\varphi}$, \ie we have
\bea
\N{\Theta}\pl \Gamma'\pr-\N{\Theta}\pl \Gamma\pr=\N{\varphi}\pl \Gamma'-\Gamma\pr
\eea
for all $\Gamma', \Gamma\in\mathscr D\pl\M,\psi,\gamma\pr$. 

For a given Galilean manifold $\mathscr G\pl \M, \psi, \gamma,\nabla\pr$ and field of observers $N\in \FO$, we designate the couple $\big(\N{F}, \N U\big)\equiv \N{\Theta}\pl \nabla\pr$ as the {\it torsional gravitational fieldstrength} measured by the field of observers $N$ with respect to $\nabla$. 
This piece of terminology allows to formulate the following Proposition:
\bpropp{Torsional special connection \cite{Bernal2003}}{\label{specialobstor}Given a field of observers $N\in \FO$, there is a unique Galilean connection $\N{\Gamma}\in\mathscr D\pl\M,\psi,\gamma\pr$ compatible with the Leibnizian structure $\mathscr L\pl\M,\psi,\gamma\pr$ such that the torsional gravitational fieldstrength measured by the field of observers $N$ with respect to $\N{\Gamma}$ vanishes. We call $\N{\Gamma}$ the torsional special connection associated to $N$. }
The space of torsional special connections compatible with $\mathscr L\pl\M,\psi,\gamma\pr$ will be denoted $\mathscr D_0\pl\M,\psi,\gamma\pr$. An explicit component expression of $\N{\Gamma}$ is given by (\cf \eg \cite{Geracie2014,TNCcm,TNCLif}):
\bea
\N{\Gamma}{}^\lambda_{\mu\nu}=N^\lambda\p_{\mu}\psi_{\nu}+\half h^{\lambda\rho}\pl\p_\mu\overset{N}{\gamma}_{\rho\nu}+\p_\nu\overset{N}{\gamma}_{\rho\mu}-\p_\rho\overset{N}{\gamma}_{\mu\nu}\pr\label{specialChristoffeltor}. 
\eea
Note that $\N{\Gamma}$ is non-symmetric whenever the absolute clock $\psi$ is not closed, since $\N{\Gamma}{}^\lambda_{[\mu\nu]}=N^\lambda\p_{[\mu}\psi_{\nu]}$. Whenever $\psi$ is closed, the metric structure is Augustinian and $\N{\Gamma}$ reduces to the torsionfree special connection \big(\cf eq.\eqref{specialChristoffel}\big).

The following Theorem (generalising Theorem \ref {thmgalileiarbitrariness}) can be seen as a \nr avatar of Theorem \ref{fundamentaltheorem}:

\bthm{\cf \cite{Bernal2003}}{\label{thmspacecompatiblegalileanconnection}
Given a field of observers $N\in \FO$, the space of Koszul connections compatible with a given Leibnizian structure $\mathscr L\pl \M, \psi, \gamma\pr$ possesses the structure of a vector space isomorphic to the vector space $\Om^2\pl\M\pr\oplus\field{\w^2\, T^*\M\otimes \Ker\psi}$. }
The gift of a field of observers $N$ then singles out a privileged Galilean connection (\ie the associated torsional special connection $\N{\Gamma}$) allowing to represent any Galilean connection ${\Gamma}\in\mathscr D\pl\M,\psi,\gamma\pr$ as 
\bea
{\Gamma}{}^\lambda_{\mu\nu}=\N{\Gamma}{}^\lambda_{\mu\nu}+h^{\lambda\rho}\psi_{(\mu}\N{F}_{\nu)\rho}+\N{U}{}^\lambda_{\mu\nu}
+2h^{\lambda\sigma}\N{U}{}^\rho_{\sigma(\mu}\N{\gamma}_{\nu)\rho}\label{ChristoffelGalileantor}
\eea
where the 2-form $\N{F}\in\forma{2}{\M}$ and the spacelike vector field-valued 2-form $\N{U}\in\vectasymduker$ are characteristic of $\Gamma$, given $N$, and are explicitly defined as $\Big( \N{F}, \N{U} \Big)\equiv \N{\varphi}\Big( \Gamma-\N{\Gamma}\Big)$. 
A \nr equivalent of the Koszul formula \eqref{Kform} can be articulated by reformulating the component expression \eqref{ChristoffelGalileantor} as\footnote{A Koszul formula for Galilean connections
was first obtained in \cite{Bernal2003}. Our expression \eqref{Kformula} presents the advantage of being closer to its relativistic avatar \eqref{Kform}.
}
\bea
2\,\N{\gamma}\pl\nabla_XY,V\pr&=&X\crl \N{\gamma}\pl Y,V\pr\crr+Y\crl \N{\gamma}\pl X,V\pr\crr-V\crl \N{\gamma}\pl X,Y\pr\crr\nn\\
&&+\N{\gamma}\pl\br{X}{Y},V\pr-\N{\gamma}\pl\br{Y}{V},X\pr-\N{\gamma}\pl\br{X}{V},Y\pr\label{Kformula}\\
&&+\N{\gamma}\pl \N{U}\pl X,Y\pr,V\pr-\N{\gamma}\pl \N{U}\pl Y,V\pr,X\pr-\N{\gamma}\pl \N{U}\pl X,V\pr,Y\pr\nn\\
&&+\psi\pl X\pr \N{F}\pl Y,V\pr+\psi\pl Y\pr \N{F}\pl X,V\pr\,\nn
\eea
with $X,Y\in\vf$ and $V\in\field{\Ker\psi}$. 

Similarly to the torsionfree case, a natural question that arises consists in determining the transformation relations of $\pl \N{F}, \N{U} \pr$ under a change of field of observers. The following Lemma generalises Lemma \ref{lemspecial}:
\blem{}{Let $N'$ and $N\in \FO$ be two fields of observers related by a Milne boost parameterised by the spacelike vector field $V\in\field{\Ker\psi}$ (\ie $N'=N+V$) and denote $\Np{\Gamma}$ and $\N{\Gamma}\in\mathscr D_0\pl\M,\psi,\gamma\pr$ their respective torsional special connections. These are related via
\bea
\Np{\Gamma}{}^\lambda_{\mu\nu}=\N{\Gamma}{}^\lambda_{\mu\nu}+\N{\varphi}{}\un\pl -\dPhiNV+\half \N{\gamma}\pl V,V\pr d\psi\,,\, \half V\otimes d\psi\pr \label{eqtranstor}
\eea
where $\Phi:\FO\times \field{\Ker \psi}\to \ff$ is defined in eq.\eqref{eqPhi}. 
}
The following Proposition follows straightforwardly from the transformation relations \eqref{eqtransvarphiNun} and \eqref{eqtranstor}:
\bprop{Under a change of field of observers $N\mapsto N'=N+V$, the map $\N{\Theta}$ gets modified as:
\bea
\Np{\Theta}\pl \nabla\pr=\N{\Theta}\pl \nabla\pr+\pl 2\nabla_{[\mu}\hspace{-1.5mm}\PhiNV\hspace{-1mm}_{\nu]}\,,\, -V^\lambda\p_{[\mu}\psi_{\nu]} \pr\label{eqtranstortheta}
\eea
for all $\nabla\in\mathscr D\pl\M,\psi,\gamma\pr$. 
}
Accordingly, in the representation \eqref{ChristoffelGalileantor} of any $\Gamma\in\mathscr D\pl\M,\psi,\gamma\pr$ the torsional gravitational fieldstrengths measured by the fields of observers $N$ or $N'$ are related by a Milne boost reading for the
2-forms $\Np{F},\N{F}\in\forma{2}{\M}$ as
\bea
\Np{F}_{\mu\nu}=\N{F}_{\mu\nu}+2\nabla_{[\mu}\hspace{-1.5mm}\PhiNV\hspace{-1mm}_{\nu]}\label{eqtransFtor} 
\eea
and for the spacelike vector field-valued 2-forms $\N{U}', \N{U}\in\vectasymduker$ as
\bea
\Np{U}{}^\lambda_{\mu\nu}=\N{U}{}^\lambda_{\mu\nu}-V^\lambda\p_{[\mu}\psi_{\nu]}. \label{eqtransUtor} 
\eea

\noindent Given a field of observers $N\in\FO$, 
one can then define a group action of the Milne group on the space $FO\pl\M,\psi\pr\times\big(\vectsum\big)$ as 
\bea
&&\pl N^\lambda\,,\,\N{F}_{\mu\nu}\,,\, \N{U}{}^\lambda_{\mu\nu}\pr\mapsto\nn\\
&&\quad\pl N^\lambda+V^\lambda\,,\,\N{F}_{\mu\nu}+2\,\p_{[\mu}
\hspace{-1.5mm}\PhiNV\hspace{-1mm}_{\nu ]}+\gamma\pl V,V\pr\p_{[ \mu}\psi_{\nu]}
-\N{U}{}^\lambda_{\mu\nu}\N{\gamma}_{\lambda\alpha}V^\alpha\,,\, \N{U}{}^\lambda_{\mu\nu}-V^\lambda\p_{[\mu}\psi_{\nu]}\pr\nn
\eea
with $N\in FO\pl\M,\psi\pr$, $\N{F}\in\forma{2}{\M}$, $\N{U}\in\vectasymduker$, $V\in\Milneg$ and the 1-form $\hspace{-1.5mm}\PhiNV\hspace{-1mm}\in\ff$ is defined in eq.\eqref{eqPhi}. 

\bdefi{Torsional gravitational fieldstrength}{A Milne orbit $[N,\N{F}, \N{U}]$ in $FO\pl\M,\psi\pr\times\big(\vectsum\big)$ is dubbed a torsional gravitational fieldstrength. The space of torsional gravitational fieldstrengths will be denoted $$\mathscr F\pl\M,\psi,\gamma\pr:=\frac{FO\pl\M,\psi\pr\times\big(\vectsum\big)}{\Milneg}\,.$$ }

\bprop{The space $\mathscr F\pl\M,\psi,\gamma\pr$ of torsional gravitational fieldstrengths is an affine space modelled on $\GammaV$. }
\noindent This is a particular case of the third fact in Proposition \ref{propaffine}.
Using this terminology, we can further characterise the affine space of Galilean connections as follows:
\bprop{\label{corGalileantor}The space of torsional Galilean connections compatible with a given Leibnizian structure possesses the structure of an affine space canonically isomorphic to the affine space of torsional gravitational fieldstrengths. 
}

\noindent As noted in \cite{Bernal2003}, the fibers of the vector bundles  ${\w^2\, T^*\M\otimes T\M}$ and $\pl\w^2\, T^*\M\pr\oplus({\w^2\, T^*\M\otimes \Ker\psi})$ both have dimension $\frac{d\pl d+1\pr^2}{2}$ for a $\pl d+1\pr$-dimensional spacetime $\M$, so that the amount of freedom in the choice of a (potentially torsional) compatible Koszul connection is the same in the \rel and \nr cases. 
In a sense, the constraint on the timelike component of the torsion (which is fixed for a Galilean connection contrarily to the relativistic case) is traded for the freedom in the choice of gravitational fieldstrength.
This statement can be made more precise by displaying the following (non-canonical) isomorphism between $\vectsum$ and the vector space of vector field-valued 2-forms $\vectasymdu$ as: 
\bea
\N{\zeta}:&&\vectsum\to\vectasymdu\nn\\
&&\pl F_{\mu\nu},U^\lambda_{\mu\nu}\pr\mapsto {\cal T}^\lambda_{\mu\nu}=U^\lambda_{\mu\nu}+N^\lambda F_{\mu\nu}\nn
\eea
together with its inverse
\bea
\N{\zeta}\un:&&\vectasymdu\to\vectsum\nn\\
&&{\cal T}^\lambda_{\mu\nu}\mapsto \pl F_{\mu\nu}=\psi_\lambda {\cal T}^\lambda_{\mu\nu}\,,\, U^\lambda_{\mu\nu}={\cal T}^\lambda_{\mu\nu}-N^\lambda\psi_\alpha {\cal T}^\alpha_{\mu\nu}\pr. \nn\eea
\noindent
This observation softens the sharp distinction between \rel and \nr cases that hold in the torsionfree case. An ambient perspective of this fact will be provided in \cite{Bekaert2015}.

\subsection{Torsional Newtonian connections}\label{torsionNewt}

\noindent Having reviewed the characterisation of torsional Galilean manifolds, we are now in a position to look for a torsional generalisation of the notion of Newtonian connection. Recall that in the torsionfree case, the Duval-K\"unzle condition could be interpreted as imposing that the torsionfree gravitational fieldstrength measured by any field of observers is a closed 2-form, \ie $\N{F}\in\Om^2\pl\M\pr \cap \Ker d$. The consistency of this condition was insured by the fact that two torsionfree special connections differ by an exact 2-form, \cf Lemma \ref{lemspecial}.  However, in view of expression \eqref{eqtransFtor}, one deduces that, in contradistinction to the torsionfree case, a Milne boost does not preserve the condition that the 2-form $\N{F}$ is closed ($d\N{F}=0$), since $\nabla_{[\alpha}\hspace{-1.5mm}\PhiNV\hspace{-1mm}_{\beta]}$ is not generically exact. Consequently, we are led to discard the condition $d\N{F}=0$ as a potential candidate aiming at generalising the notion of Newtonian connection since it is inconsistent whenever torsion is involved.    
However, relying on the form of eq.\eqref{eqtransFtor}, a natural condition consists in imposing that $\N{F}$ is covariantly exact, in the following sense:

\bdefi{Covariantly exact differential form}{A differential $p$-form $\alpha\in\Om^p\pl\M\pr$ is said to be covariantly exact for the Koszul connection $\nabla$ if there exists a $\pl p-1\pr$-form $\beta\in\Om^{p-1}\pl \M\pr$ such that $\alpha_{\mu_1\ldots\mu_{p}}=\nabla_{[\mu_1}\, \beta_{\mu_2\ldots\mu_{p}]}$. }
\noindent For the case when $\nabla$ is torsionfree, the notion of a covariantly exact form identifies with the one of an exact form. Based on this notion, one can now articulate a generalised definition of Newtonian connection as:

\bdefi{Torsional Newtonian connection}{Let $\mathscr G\pl \M,\psi,\gamma,\nabla\pr$ be a Galilean manifold whose Koszul connection $\nabla$ is characterised by the torsional gravitational fieldstrength $\Big[N,\N{F}, \N{U}\Big]$. The Koszul connection $\nabla$ is said to be a torsional Newtonian connection if the 2-form $\N{F}\in\Om^2\pl\M\pr$ associated to any $N\in \FO$ is covariantly exact. }
\noindent The transformation law given by eq.\eqref{eqtransFtor} ensures the consistency of the previous definition, since the covariant exactness of a 2-form $\N{F}$ is preserved by the action of the Milne group \eqref{eqtransFtor}. 
From the action of the Milne group on $\FO\times\big(\vectsum\big)$ (\cf eqs.\eqref{eqtransFtor} and \eqref{eqtransUtor}\,), one can define the following action on $\FO\times\big(\vectsumone\big)$:
\bea
\pl N,\N{A}, \N{U}\pr\mapsto \pl N+V, \N{A}+\hspace{-1.5mm}\PhiNV\hspace{-1mm}, \N{U}-\half V\otimes d\psi\pr
\eea
where $V\in\Milneg$. 
\bdefi{Torsional gravitational potential}{A Milne orbit $\Big[ N,\N{A}, \N{U}\Big]$ in $\FO\times\big(\vectsum\big)$ is dubbed a torsional gravitational potential.}

\bprop{\label{corNewtoniantorsional}Torsional Newtonian connections compatible with a given Leibnizian structure are in bijective correspondence with torsional gravitational potentials. 
}

It is worth stressing that, contrarily to the case of a torsionfree Newtonian connection, there is no additional Maxwell gauge symmetry at hand. In the case of vanishing torsion, the origin of the supplementary gauge invariance can be traced back to the closed condition $d\N{F}=0$ which allows to locally write the 2-form $\N{F}$ as the gravitational fieldstrength for the gravitational potential $\N{A}\in\ff$ measured by $N$ \ie $\N{F}=d\N{A}$, so that $\N{F}$ is invariant under a gauge transformation of the form $\N{A}\mapsto\N{A}+df$, with $f\in\fonc{\M}$. However, when $\N{F}$ is covariantly exact, \ie $\N{F}_{\alpha\beta}=\nabla_{[\alpha}\N{A}_{\beta]}$, the torsion term if non-vanishing breaks the invariance. This important distinction motivates the following terminology: a \textit{torsionful Newtonian connection} is a torsional Newtonian connection with non-vanishing torsion. 
Similarly to the torsionfree case, an alternative description of torsional Newtonian connections can be given by making use of the ``Lagrangian'' variables $Z$, $\phi$ and $g$ (\cf Table \ref{tableMilneinvariant}). Starting from the component expression \eqref{ChristoffelGalileantor} with $\N{F}_{[\alpha\beta]}=\nabla_{[\alpha}\N{A}_{\beta]}$, this is achieved by performing a Milne boost parameterised by the 1-form $\chi\equiv-\N{A}$, under which the torsional Newtonian connection takes the manifestly Milne-invariant form:
\bea
\Gamma^\lambda_{\mu\nu}&=&Z^\lambda\p_{(\mu}\psi_{\nu)}+\half h^{\lambda\rho}\crl\p_\mu g_{\rho\nu}+\p_\nu g_{\rho\mu}-\p_\rho g_{\mu\nu}\crr\nn\\&&+\Gamma^\lambda_{[\mu\nu]}+\Gamma^\rho_{[\sigma\mu]}h^{\sigma\lambda}g_{\rho\nu}+\Gamma^\rho_{[\sigma\nu]}h^{\sigma\lambda}g_{\rho\mu}\label{eqNewtonianconnectionTrumper}. 
\eea
By construction, the field of observers $Z$ is the unique vector field which is Coriolis-free with respect to $\nabla$
(as follows by repeating the steps in the proof of Proposition \ref{propZ}, \cf Appendix \ref{Curvtfree}). 
This reformulation, along with Proposition \ref{propbijcorNewtoZphi}, allows to articulate two additional characterisations of torsionful Newtonian manifolds, as formulated in the following Proposition:

\bprop{\label{propgenNewtonconnectionZ}Let $\mathscr L\pl \M, \psi, \gamma\pr$ be a Leibnizian structure. There is a bijective correspondence between torsionful Newtonian manifolds and 
\begin{enumerate}
 \item triplets $\pl Z, \phi, \Z U\pr$, where $Z\in \FO$, $\phi\in\fonc{\M}$ and $\Z U\in\field{\w^2\, T^*\M\otimes \Ker\psi}$. 
 \item  couples $\pl g, T\pr$ where $g\in\bforms$ is a Lagrangian metric and $T\in\field{\w^2\, T^*\M\otimes T\M}$ is a non-vanishing torsion tensor whose timelike part is constrained to satisfy $\psi\big( T\pl X,Y\pr\big)=d\psi\pl X,Y\pr$
for all $X,Y\in \field{T\M}$. 
\end{enumerate}
}
\noindent The two items in the previous Proposition complement the content of Propositions \ref{propLagNewton} and \ref{propLagNewtong}, respectively, to account for the torsionful case. As commented earlier, the torsionfree case is special in that it involves an additional gauge invariance. In the torsionful case where no such symmetry is present\footnote{In other words, the coefficients \eqref{eqNewtonianconnectionTrumper} are not invariant under the transformations recorded in Table \ref{tableMilneinvariant}. }, torsionful Newtonian connections are characterised by individual objects rather than classes thereof. This justifies the formal separation between the torsionfree and torsionful cases in two sets of Propositions. 

According to the first item of \Prop{propgenNewtonconnectionZ} if one is given a Leibnizian structure and a field of observers $Z$, the space of torsional Newtonian connections is in bijective correspondence with $\fonc{\M}\times\field{\w^2\, T^*\M\otimes \Ker\psi}$. Putting $\phi=0$ and $\Z U=0$ (so that $g=\Z\gamma$ and $T=Z\otimes d\psi$) allows to recover the  expression of the torsional special connection associated to $Z$ \big(\cf \eqref{specialChristoffeltor}\big):
\bea
\Z\Gamma{}^\lambda_{\mu\nu}&=&Z^\lambda\p_{\mu}\psi_{\nu}+\half h^{\lambda\rho}\crl\p_\mu \Z{\gamma}_{\rho\nu}+\p_\nu \Z{\gamma}_{\rho\mu}-\p_\rho \Z{\gamma}_{\mu\nu}\crr. 
\eea
Let $\big[ N,\N A\big]$ be the Milne orbit containing $\pl Z,0\pr$ (\ie $\big[ N,\N A\big]\equiv\big[Z,0\big]$). The last expression can be written in a manifestly invariant way (\ie independently of the choice of representative in $\big[ N,\N A\big]$) by substituting the expressions of Table \ref{tableMilneinvariant} as:
\bea
\Z\Gamma{}^\lambda_{\mu\nu}&=&\big( N^\lambda-h^{\lambda\rho} \N{A}_\rho\big)\p_{\mu}\psi_{\nu}\nn+\half h^{\lambda\rho}\Big[\p_\mu \big(\N{\gamma}_{\rho\nu}+\N{A}_{\nu}\psi_\rho+\N{A}_\rho\psi_\nu\big)\\&&+\p_\nu \big(\N{\gamma}_{\rho\mu}+\N{A}_{\mu}\psi_\rho+\N{A}_\rho\psi_\mu\big)-\p_\rho \big(\N{\gamma}_{\mu\nu}+\N{A}_{\nu}\psi_\mu+\N{A}_\mu\psi_\nu\big)\Big] \nn
\eea
thus allowing to recover the components of the torsional connection introduced in the works \cite{Jensen2014,TNCMilneLif}. 
\vspace{2mm}

\noindent The different structures necessary to uniquely determine a given manifold are summarised in the following table, both in the relativistic and nonrelativistic cases: 
\vspace{2mm}

\hspace{-2cm}
\setlength{\extrarowheight}{2 mm}
\begin{tabular}{|c|c|c|}
   \hline
Metric structure&\rule[-0.5cm]{0cm}{1cm}Supplementary structure&Manifold\tabularnewline\hline
Lorentzian&\rule[-0.5cm]{0cm}{1cm}$\times$&Lorentzian\tabularnewline
   \hline \hline
\multirow{2}{*}{Augustinian}&\rule[-0.5cm]{0cm}{1cm}Gravitational fieldstrength $\Big[ N,\N{F}\Big]$&Galilean
\tabularnewline
\cline{2-3}
&\rule[-0.5cm]{0cm}{1cm}Gravitational potential $\Big[N,[\N{A}]\Big]$, $\Big[Z,\phi\Big], \Big[ g\Big]$&Newtonian
\tabularnewline
\hline
\multirow{2}{*}{\rule[-3cm]{0cm}{6cm}Leibnizian}\rule[-2cm]{0cm}{4cm}&
\begin{tabular}{c}\rule[-0.5cm]{0cm}{1cm}Torsional gravitational fieldstrength\\
\rule[-0.5cm]{0cm}{1cm}$\Big[N,\N{F}, \N{U}\Big]$
\end{tabular}
&Torsional Galilean
\tabularnewline
\cline{2-3}
&\rule[-0.5cm]{0cm}{1cm}\rule[-2cm]{0cm}{4cm}
\begin{tabular}{c}\rule[-0.5cm]{0cm}{1cm}Torsional gravitational potential\\
\rule[-0.5cm]{0cm}{1cm}$\Big[N,\N{A}, \N{U}\Big]$, $\pl Z,\phi, \Z{U}\pr$, $\pl g, T\pr$
\end{tabular}
&Torsional Newtonian
\tabularnewline
\hline
\end{tabular}
\captionof{table}{Solutions to the equivalence problem}

\pagebreak
\section{Conclusion}\label{Conclusion}

In this series of papers, we adress two novel generalisations of Newtonian connections for metric structures with non-closed absolute clock by going down the two following crossing roads: {in the present paper} we reviewed {the class of torsional Galilean connections by emphasising its affine space structure.
We also} isolated a subclass (dubbed {\it torsional Newtonian} connections) characterised by a {\it covariantly exact} 2-form. Similarly to their torsionfree counterparts, torsional Newtonian connections can be expressed in terms of {\it Lagrangian} variables which we used in order to make contact with the recently introduced torsional Newton-Cartan geometry \cite{Jensen2014,TNCMilneLif}. We further discussed the geometric origin behind the lack of Maxwell gauge-invariance of the latter. 
In a forthcoming paper \cite{Bekaert2016}, we will present a connection living on the most general \nr metric structure allowing a notion of absolute time ({\it Aristotelian} structure). This torsionfree connection has the nice feature of being Maxwell gauge-invariant and such that its geodesic equation follows from a variational principle, similarly to its Newtonian counterpart to which it can be said projectively {related}, since they define the same unparameterised geodesics. 
\vspace{2mm}

The present analysis restricted to an {\it intrinsic} point of view on \nr connections, with particular emphasis placed on the equivalence problem. In a forthcoming companion paper \cite{Bekaert2015}, we will discuss an {\it ambient} approach to these classes of connections by generalising the Bargmann framework of Duval \etal (\cf \cite{Duval1985,Duval1991}) where \nr Newtonian manifolds were obtained as null dimensional reduction of a specific class of \rel ones. As hinted in Section \ref{paragraphoutofthecave}, this approach is indeed quite natural for Newtonian manifolds but can be extended to embed more general \nr structures.  
In \cite{Bekaert2015} an ambient account of (torsional) Galilean connections will be provided  
by considering a class of relativistic spacetimes endowed with a (possibly torsional) connection parallelising a null Killing vector field. In particular, this setup will allow us to embed torsionfree Galilean manifolds into torsional relativistic spacetimes, thus shedding new light on the torsional origin of the gravitational fieldstrength. 
Finally, in \cite{Bekaert2016} the projectively Newtonian connection will be shown to arise as a projection of the Levi-Civita connection for a class of relativistic spacetimes admitting a null and hypersurface-orthogonal Killing vector field. These spacetimes were studied in \cite{Julia1995} and dubbed {\it Platonic waves} in \cite{Bekaert2013c} where they were shown to be conformally related to the class of \cite{Duval1985,Duval1991}. This ambient construction will notably allow us to formulate at the level of connections the Eisenhart-Lichnerowicz lift \cite{Eisenhart1928} of dynamical trajectories to relativistic geodesics.

\section*{Acknowledgements}

We are very grateful to C.~Duval, J.~Hartong and P.~Horv\'athy for useful exchanges and discussions on nonrelativistic structures.
\pagebreak
\appendix

\section{Curvature of a torsionfree Galilean manifold}\label{Curvtfree}

\noindent Let us recall the definition of the curvature of a Koszul connection for the vector bundle $E$ on $\M$: 
\bea
R\pl X,Y;f\pr=\nabla_X\nabla_Yf-\nabla_Y\nabla_Xf-\nabla_{\br{X}{Y}}f\nn
\eea
with $X,Y\in\field{T\M}$ and $f\in\field{E}$. The components of the Koszul curvature for the tangent bundle $E=T\M$ read: $dx^\lambda\crl R\pl \p_\mu,\p_\nu;\p_\rho\pr\crr\equiv \Riem{\lambda}{\rho}{\mu}{\nu}$. 
\vspace{2mm}

\noindent Compatibility conditions \eqref{eqconditionsgalileanmanifold} for the Galilei connection $\nabla$ impose the following constraints on the Koszul curvature:
\bcase{
\nabla_\mu\psi_\nu=0\Rightarrow\psi_\lambda \Riem{\lambda}{\rho}{\mu}{\nu}=0\nn\\
\nabla_\mu h^{\alpha\beta}=0\Rightarrow h^{\rho \beta}\Riem{\alpha}{\rho}{\mu}{\nu}+h^{\alpha\rho}\Riem{\beta}{\rho}{\mu}{\nu}=0\nn
}
\bnota{In the following we will use a Galilean basis $B\equiv\pset{N, e_i}$ together with its dual $B^*\equiv{\pset{\psi, \theta^i}}$. Now, let $T^\mu_{\nu}$ be the holonomic components of a tensor $T\in\field{T\M\times T^*\M}$. The following notation will prove to be handy:
\bea
\begin{cases}
T^0_\nu\equiv \psi_\mu T^\mu_\nu  \hspace{6mm} T^i_\nu\equiv \theta^i_\mu T^\mu_\nu\nn\\
T^\mu_0\equiv N^\nu T^\mu_\nu  \hspace{5mm} T^\mu_i\equiv e_i^\nu T^\mu_\nu. \nn
\end{cases}
\eea
}
\noindent The previously stated constraints on the Koszul curvature can thus be reexpressed as:
\bea
\Riem{0}{\rho}{\mu}{\nu}=0=\Riemd{(i}{j)}{\mu}{\nu}\,. \label{eqcompatibilityconditions}
\eea
\noindent Taking these constraints into account, the components of the curvature 2-form $\Riemf{\lambda}{\rho}\in\forma{2}{\M}$ can be expanded as:
\bea
\Riemf{\lambda}{\rho}=\Riemf{i}{j}\theta^j_\rho e^\lambda_i+\Riemf{i}{0}\psi_\rho e^\lambda_i. 
\eea
\bpropp{Symmetries of the Galilean curvature}{The curvature tensor of a torsionfree Galilean connection satisfies the following identities: 
\bcase{
\Riem{i}{\rho}{(\mu}{\nu)}=0\\
\Riem{i}{[\rho}{\mu}{\nu]}=0\\
\Rieme{i}{j}{k}{l}=\Rieme{j}{i}{l}{k}. 
}
}
\proof{ ~\\
These equalities follow respectively from the well-known identities
\bea
&&R\pl X,Y;Z,W\pr=-R\pl Y,X;Z,W\pr\,,\nn\\
&&R\pl X,Y;Z\pr+R\pl Y,Z;X\pr+R\pl Z,X;Y\pr=0\,,\nn\\
&&R\pl X,Y;Z,W\pr=R\pl Z,W;X,Y\pr\,,\nn
\eea
where $R\pl X,Y;Z,W\pr\equiv \gamma\big( R\pl X,Y;Z\pr,W\big)$.
}
The second identity of the previous Proposition, known as the first Bianchi identity, can be decomposed further:
\bprop{The first Bianchi identity for the Galilei curvature leads to the following set of equations:
\bcase{
\Riem{l}{[i}{j]}{0}+\half \Riem{l}{0}{i}{j}=0\label{eqBianchiset}\\
\Riem{l}{[i}{j}{k]}=0. 
}
}
\bcor{\cf \cite{Julia1995}}{\label{lemKT}The curvature tensor of a torsionfree Galilean manifold satisfies the relation:
\bea
\Rieme{(i}{j)}{[\mu}{\nu]}=0\nn
\eea
}
\proof{ ~\\
The relation is equivalent to the set:
\bcase{
\Rieme{(i}{j)}{[k}{l]}=0\nn\\
\Rieme{(i}{j)}{[0}{k]}=0.\nn
}
The first relation follows straightforwardly from the all-spacelike first Bianchi identity and compatibility relations \eqref{eqcompatibilityconditions}. 
The second identity is obtained by taking the symmetric part in $(k\leftrightarrow i)$ of the temporal/spacelike Bianchi identity:
\bea
\Riemj{(k}{i)}{j}{0}-\Riemj{(k|}{j}{|i)}{0}+\Riemj{(k|}{0}{|i)}{j}=0. 
\eea
The first term vanishes, leaving $\Rieme{(k}{i)}{[0}{j]}=0$. 
}
\noindent We now focus on the Duval-K\"unzle condition (\cf \Defi{DuvalKunzle}) which, in components reads: 

\bea
R\, {}^{\mu\, \, \, \, \nu}_{\, \, \, \alpha\, \, \, \, \beta}=R\, {}^{\nu\, \, \, \, \mu}_{\, \, \, \beta\, \, \, \, \alpha}. \nn
\eea
\noindent Decomposing on a Galilean basis leads to the set of equations:
\bcase{
\Rieme{i}{j}{k}{l}=\Rieme{j}{i}{l}{k}\\
\Rieme{[i}{j]}{0}{0}=0\\
\Rieme{j}{i}{0}{k}=\Rieme{i}{j}{k}{0}\label{eqKeq}
}
The first equation is already implied by the first Bianchi identity. However the two remaining are non-trivial constraints that reduce the number of independent components from $\frac{1}{12}d^2\pl d+1\pr\pl d+5\pr$ to $\frac{1}{12}\pl d+1\pr^2\crl\pl d+1\pr^2-1\crr$, \ie to the same number as in a $\pl d+1\pr$-dimensional (pseudo)-Riemannian manifold (\cf \eg \cite{Kunzle1972}). 
\bprop{\label{propreformulationK}The Duval-K\"unzle condition can be alternatively written as \bea\Riemf{i}{0}\w\theta_i=0. \eea}
\proof{ ~\\

The alternative formulation $\Riemf{i}{0}\w\theta_i=0$ of the Duval-K\"unzle condition imposes the following constraints
on the components of $\Riemf{i}{0}=\Rieme{i}{j}{0}{0}\,\theta_j\wedge \psi\,+\,\frac12\Riemh{i}{0}{j}{k}\,\theta_j\wedge\theta_k$:
\bcase{
\Rieme{[i}{j]}{0}{0}=0\nn\\
\Riemh{[i}{0}{k}{l]}=0. 
}
The first equality matches the second one from \eqref{eqKeq} so what remains to be proved is the following equivalence:
\bea
\Rieme{j}{i}{0}{k}=\Rieme{i}{j}{k}{0}\Leftrightarrow\Riemh{[i}{0}{j}{k]}=0. 
\eea 
\noindent We start by totally antisymmetrising the first of the identities of the Bianchi set \eqref{eqBianchiset}:
\bea
\Riemi{[i}{j}{k]}{0}+\half\Riemh{[i}{0}{j}{k]}=0. 
\eea 
Expanding the first term leads to:
\bea
\frac{1}{3}\pl 2\Riemi{j}{[k}{i]}{0}-\Riemi{i}{k}{j}{0}\pr+\half\Riemh{[i}{0}{j}{k]}=0. 
\eea
Using again the first Bianchi identity allows to transform the first term on the left-hand side:
\bea
\frac{1}{3}\pl \Riemh{j}{0}{i}{k}-\Riemi{i}{k}{j}{0}\pr+\half\Riemh{[i}{0}{j}{k]}=0. 
\eea
so that $\Rieme{j}{i}{0}{k}=\Rieme{i}{j}{k}{0}\Leftrightarrow\Riemh{[i}{0}{j}{k]}=0$.  
}
\noindent Along the Duval-K\"unzle condition, another constraint on the curvature, dubbed the Trautman condition\footnote{Although the denominations Duval-K\"unzle and Trautman conditions seem customary in the literature, it is amusing to note that in the respective works commonly cited when these conditions are discussed, Trautman wrote what is usually referred to as the Duval-K\"unzle condition (\cf eq.(IV) of \cite{Trautman1963}) while K\"unzle wrote the Trautman condition (\cf eq.(4.14) of \cite{Kunzle1972}). } is frequently encountered in the literature:
\bdefi{Trautman condition, \cf \eg \cite{MTW}}{Let $J$ be the Jacobi curvature operator defined as 
\bea
J\pl X,Y;Z\pr\equiv \half\pl R\pl Z,X;Y\pr+R\pl Z,Y;X\pr\pr
\eea
where $X$, $Y$ and $Z$ are three vector fields. 
The Trautman condition states that the Jacobi operator must be self-adjoint when acting on spacelike vectors, \ie
\bea
\gamma\pl J\pl X,Y;V\pr,W\pr=\gamma\pl J\pl X,Y;W\pr,V\pr
\eea
for all $X,Y\in\field{T\M}$ and for all $V,W\in\field{\Ker\psi}$. 
}
\noindent In components, the Jacobi operator reads $\Jaco{\lambda}{\rho}{\mu}{\nu}=\Riem{\lambda}{(\mu|}{\rho}{|\nu)}=-\Riem{\lambda}{(\mu}{\nu)}{\rho}$ while the Trautman condition imposes:
\bea
\Rieme{[i}{j]}{(\mu}{\nu)}=0. 
\eea
\bpropp{\cf \cite{Julia1995}}{The Duval-K\"unzle and Trautman conditions are equivalent for a torsionfree Galilean manifold.\label{Juliaproof}}
\proof{ ~\\

Decomposing the Duval-K\"unzle operator as:
\bea
\underset{\text{(K)}}{\Rieme{i}{j}{\mu}{\nu}-\Rieme{j}{i}{\nu}{\mu}}&=&\half\pl\Rieme{(i}{j)}{\mu}{\nu}+\Rieme{[i}{j]}{\mu}{\nu}-\Rieme{(j}{i)}{\nu}{\mu}-\Rieme{[j}{i]}{\nu}{\mu}\pr\nn\\
&=&\half\pl\Rieme{(i}{j)}{\mu}{\nu}+\Rieme{[i}{j]}{\mu}{\nu}-\Rieme{(i}{j)}{\nu}{\mu}+\Rieme{[i}{j]}{\nu}{\mu}\pr\nn\\
&=&\underset{\text{(C)}}{\Rieme{(i}{j)}{[\mu}{\nu]}}+\underset{\text{(T)}}{\Rieme{[i}{j]}{(\mu}{\nu)}}
\eea
one recognises the operator (C) obtained in Corollary \ref{lemKT} as well as the Trautman operator (T). Provided Corollary \ref{lemKT}, the Duval-K\"unzle and Trautman conditions are therefore equivalent. 
}
\pagebreak
\section{Affine, vector \& principal homogeneous spaces}\label{proofaffine}

We will briefly review the definition of affine spaces and related concepts. The relation between affine and vector spaces is much more clear when the former are seen as principal homogeneous spaces for the latter.

\bdefi{Principal homogeneous space}{When the action of a Lie group $G$ on a manifold $P$ is regular (\ie free and transitive), one says that $P$ is a principal homogeneous space for the group $G$.
}

\bdefi{Affine space}{An affine space $\A$ modelled on a vector space $\V$ is a principal homogeneous space $\A$ for $\V$ (seen as an Abelian group). More explicitly, the regular action of $\V$ on $\A$ is a map
$$t\,:\,\V\times\A\to\A\,:\,(\mathbf{v},a)\mapsto t_\mathbf{v}a
$$
where the bijections $t_\mathbf{v}:\A\to\A:a\mapsto a+\mathbf{v}$ are called translations of $\A$ by $\bf v$.

Equivalently, an affine space $\A$ modelled on a vector space $\V$ can be defined as a set $\A$ together with a subtraction map
$$-\,:\,\A\times\A\to\V\,:\,(a,b)\mapsto b-a\equiv \overrightarrow{ab}$$
with the following properties
\begin{enumerate}
 \item $\forall a\in\A,\,\forall \mathbf{v}\in\V,\,\exists!\,b\in\A$ such that $b-a=\mathbf{v}$
 \item $\forall a,b,c\in\A$\,:\,$(c-b)+(b-a)=c-a$ ($\Leftrightarrow \overrightarrow{ab}+\overrightarrow{bc}=\overrightarrow{ac}$)  
\end{enumerate}
called Weyl's axioms.
The relation between the regular action and the subtraction map is: $t_{b-a}a=b$ ($\forall a,b\in\A$).
}
\bdefi{Affine map}{\label{defiaffinemap}Let $\A'$ and $\A$ be two affine spaces modelled on the vector spaces $\V'$ and $\V$, respectively. Let $\varphi:\V'\to \V$ be a linear map. The map $\Theta:\A'\to \A$ will be said to be an affine map modelled on $\varphi$ if it satisfies $\Theta\pl a'\pr-\Theta\pl a\pr=\varphi\pl a'-a\pr$ for all $a'$. 

}

By choosing a specific element $O\in\A$, called origin, the space $\A$ is endowed with a structure of vector space, denoted $\A_O$, isomorphic to $\V$.
The addition map on $\A$ takes the form
\bea +\,:\,\A\times\A\to \A\,:\,(a',a)\mapsto t_{\pl a'-O\pr+\pl a-O\pr}\,O\nn \eea
while the multiplication by a scalar reads:
\bea
\mR\times\D\to\D\,:\,(\lambda,a)\mapsto t_{\lambda\pl a-O\pr}\,O. \nn
\eea
The bijection $\V\to\A:\mathbf{v}\mapsto t_\mathbf{v}O$ or its inverse $\A\to\V:a\mapsto a-O\equiv \overrightarrow{Oa}$ provides the isomorphism $\A_O\cong\V$.

\bprop{\label{propaffine}Let:
\begin{itemize}
 \item $\D$ be an affine space modelled on the vector space $\V$ isomorphic to $\W$.
 \item $F$ be a principal homogeneous space modelled on the Lie group $G$ and denote $G\times F \to F:\pl g,f\pr\mapsto g\cdot f$ the group action of $G$ on $F$.
 \item $\varphi:F\times\V\to\W$ be a collection of isomorphisms between $\V$ and $\W$ indexed by elements of $F$, \ie the map $\varphi_f\equiv \varphi\pl f,\cdot\pr:\V\to\W$ is an isomorphism for all $f\in F$.
 \item $\Theta:F\times\D\to\W$ be a map such that for all $f\in F$, the map $\Theta_f\equiv\Theta\pl f,\cdot\pr:\D\to\W$ is an affine map modelled on $\varphi_f$, \ie  $\Theta_f\pl d'\pr-\Theta_f\pl d\pr=\varphi_f\pl d'-d\pr$ for all $d,d'\in\D$. 
\end{itemize}
Then:
\begin{enumerate}
 \item For all $f\in F$, the map $\Theta_f:\D\to\W$ endows $\D$ with a structure of vector space, the origin of which is the unique element of $\Ker \Theta_f$, so that it becomes an isomorphism of vector spaces. 
 \item There is a canonical group action of the group $G$ on the space $F\times\W$ defined as
 \bea
 \rho:G\times\pl F\times\W\pr&\to&\pl F\times\W\pr\nn\\
 \big(g,\pl f,w\pr\big)&\mapsto&\pl g\cdot f,\Theta_{g\cdot f}\pl \Theta_f\un\pl w\pr\pr\pr. \nn
 \eea 
 \item The affine space $\D$ is canonically isomorphic to the space $\mP:= \pl F\times\W\pr/G$ of orbits of $G$ acting on $\pl F\times\W\pr$ via the group action $\rho$. 
\end{enumerate}
}{}

\paragraph{Proof}
\begin{enumerate}
\item We start by showing that, for all $f\in F$, there exists a unique element $d_f\in\D$ such that $\Theta_f\pl d_f\pr=0$. 
\begin{itemize}
 \item Existence: 
 
 \noindent Let $d\in\D$ and define $d_f\in\D$ as $d-d_f=\varphi_f\un\pl \Theta_f\pl d\pr\pr$. Using that $\Theta_f$ is an affine map, we write 
 \bea
 \Theta_f\pl d\pr-\Theta_f\pl d_f\pr=\varphi_f\pl d-d_f\pr =\Theta_f\pl d\pr\nn
 \eea
 so that $\Theta_f\pl d_f\pr=0$. 
 \item Uniqueness:
 
\noindent Let $d_f'$ and $d_f$ be two elements of $\Ker \Theta_f$. Then $ \Theta_f\pl d'_f\pr-\Theta_f\pl d_f\pr=\varphi_f\pl d'_f-d_f\pr=0$ so that, using that $\varphi_f$ is an isomorphism, $d_f'=d_f$. 
 \end{itemize}
\noindent Given $f\in F$, the affine space $\D$ thus acquires the structure of a vector space $\D_{d_f}$ with origin $d_f$. 
Since the map $\Theta_f$ can be written as $\Theta_f\pl d\pr=\varphi_f\pl d-d_f\pr$, the fact that $\varphi_f$ is a linear isomorphism ensures that $\Theta_f$ is too.
\item We first check the Identity condition, and then the Compatibility:
\begin{itemize}
 \item Identity: $\rho\big(e, \pl f,w\pr\big)=\Big( f,\Theta_{f}\pl \Theta_f\un\pl w\pr\pr\Big)=\pl f,w\pr$.
 \item Compatibility: 
 \bea
\rho\Big(g',\rho\big( g,\pl f,w\pr\big)\Big)&=&\rho\Bigg(g',\Big( g\cdot f,\,\Theta_{g\cdot f}\big( \Theta_f\un\pl w\pr\big)\Big)\Bigg)\nn\\
&=&\Bigg( g'\cdot(g\cdot f),\,\Theta_{g'\cdot(g\cdot f)}\Big(\Theta_{g\cdot f}\un\big(\Theta_{g\cdot f}\pl\Theta_f\un\pl w\pr\pr\big)\Big)\Bigg)\nn\\
&=&\Big( (g'g)\cdot f,\,\Theta_{(g'g)\cdot f}\pl\Theta_f\un\pl w\pr\pr\Big)\nn\\
&=&\rho\Big(g'g, \pl f,w\pr\Big). \nn
\eea
\end{itemize}
\item We start by showing that the space $\mP$ of $G$-orbits is an affine space modelled on $\V$ by displaying the following subtraction map:
\bea
\mP\times\mP\to\V\,:\,\pl p',p\pr\mapsto\varphi_f\un\pl\overset{f}{w}{}-\overset{f}{w}{}'\pr\label{eqsubmap}
\eea
where $f$ is an arbitrary element of $F$ and $\overset{f}{w}{}\in\W$ denotes the unique element of $\W$ such that $\pl f,\overset{f}{w}{}\pr$ belongs to the $G$-orbit $p\in\mP$. The existence and uniqueness of $\overset{f}{w}{}\in\W$ are guaranteed by the fact that $G$ acts regularly on $F$. The map \eqref{eqsubmap} is independent of the choice of $f$ since, picking a different representative $f'\in F$ defined as $f'\equiv g\cdot f$, with $g\in G$, the term $\varphi_{f'}\un\pl\overset{f'}{w}{}-\overset{f'}{w}{}'\pr$ reads 
\bea
\varphi_{g\cdot f}\un\pl\overset{g\cdot f}{w}{}-\overset{g\cdot f}{w}{}'\pr&=&\varphi_{g\cdot f}\un\pl\Theta_{g\cdot f}\pl \Theta_f\un\pl \overset{f}{w}{}\pr\pr-\Theta_{g\cdot f}\pl \Theta_f\un\pl \overset{f}{w}{}'\pr\pr\pr\nn\\
&=&\varphi_{g\cdot f}\un\pl\varphi_{g\cdot f}\pl\Theta_f\un\pl \overset{f}{w}{}\pr-\Theta_f\un\pl \overset{f}{w}{}'\pr\pr\pr\nn\\
&=&\Theta_f\un\pl \overset{f}{w}{}\pr-\Theta_f\un\pl \overset{f}{w}{}'\pr\nn\\
&=&\varphi_f\un\pl\overset{f}{w}{}-\overset{f}{w}{}'\pr\nn. 
\eea
Furthermore, the subtraction map \eqref{eqsubmap} can be checked to satisfy Weyl's axioms, so that $\mP$ is an affine space modelled on $\V$. 

We now introduce the map
\bea
\Xi:\D\to\mP:d\mapsto\crl f,\Theta_f\pl d\pr\crr\nn
\eea
where $f\in F$ and $\crl f,\Theta_f\pl d\pr\crr\in\mP$ is the unique $G$-orbit of $F\times\W$ containing the element $\pl f,\Theta_f\pl d\pr\pr$. The map $\Xi$ can be shown to satisfy:
\bea
\Xi\pl d'\pr-\Xi\pl d\pr&=&\crl f,\Theta_f\pl d'\pr\crr-\crl f,\Theta_f\pl d\pr\crr\nn\\
&=&\varphi_f\un\pl \Theta_f\pl d'\pr-\Theta_f\pl d\pr\pr\nn\\
&=&d'-d\nn
\eea
so that $\Xi:\D\to\mP$ is an affine isomorphism modelled on the identity map in $\V$. The inverse map is given by 
\bea
\Xi\un:\mP\to\D:p\mapsto\Theta_f\un\pl\overset{f}{w}{}\pr\nn
\eea
where $f$ is an arbitrary element of $F$ and $\overset{f}{w}{}\in\W$ the unique element of $\W$ such that $\pl f,\overset{f}{w}{}\pr\in p$. 
\end{enumerate}

\pagebreak
\section{Technical proofs}\label{proofpropZ}\label{prooflemfromZphitog}

\paragraph{Proof of Proposition \ref{propZ}}
~\\
Let us first check that the previous definition for $Z$ is well-defined under a change of representative (\cf eq.\eqref{eqMaxwellMilneboost}). This is easily seen as: 
\bea
Z'&=&N'-h\pl\N{A}'\pr+h\pl df'\pr\nn\\
&=&N+h\pl\chi\pr-h\pl\N{A}+\hspace{-1.5mm}\PhiNV\hspace{-1mm}+df\pr+h\pl df'\pr\nn\\
&=&N-h\pl\N{A}\pr+h\bigg( d\pl f'-f\pr\bigg)\nn
\eea
since $\hspace{-1.5mm}\PhiNV\hspace{-1mm}$ is given by \eqref{eqPhi}. 

\noindent Now, let us compute the Coriolis 2-form of a field of observers $Z=N+h\pl\chi\pr$, with $\chi\in\form{\M}$: 
\bea
\Z{\boldsymbol{\om}}\pl V,W\pr&=& \gamma\pl \nabla_VZ, W\pr-\gamma\pl V,\nabla_WZ\pr\nn\\
&=&\N{\boldsymbol{\om}}\pl V,W\pr+\gamma\pl\nabla_V h\pl\chi\pr,W\pr-\gamma\pl\nabla_W h\pl\chi\pr,V\pr\nn
\eea
with $V,W\in \field{\Ker\psi}$. Note that the second and third terms make sense, since $\psi\big(\nabla_V h\pl\chi\pr\big)=V\crl\psi\big( h\pl\chi\pr\big)\crr=0$. Using $\nabla\gamma=0$ allows to reformulate the first term in brackets as $\gamma\pl\nabla_V h\pl\chi\pr,W\pr=V\crl\gamma\pl h\pl\chi\pr,W\pr\crr-\gamma\pl h\pl\chi\pr,\nabla_VW\pr$. Proceeding similarly with the second term in brackets leads to:
\bea
\Z{\boldsymbol{\om}}\pl V,W\pr&=&\N{\boldsymbol{\om}}\pl V,W\pr+\bigg( V\crl\gamma\pl h\pl\chi\pr,W\pr\crr-\gamma\pl h\pl\chi\pr,\nabla_VW\pr-\pl V\leftrightarrow W\pr\bigg)\nn\\
&=&\N{\boldsymbol{\om}}\pl V,W\pr+ V\crl \chi\pl W\pr\crr-W\crl\chi\pl V\pr\crr-\chi\pl\nabla_VW-\nabla_WV\pr\nn\\
&=&\N{\boldsymbol{\om}}\pl V,W\pr+ V\crl \chi\pl W\pr\crr-W\crl\chi\pl V\pr\crr-\chi\pl\br{V}{W}\pr \nn\\
&=&\N{\boldsymbol{\om}}\pl V,W\pr+d\chi\pl V,W\pr\nn
\eea
where one used respectively: in the first step, the equality $\gamma\pl h\pl\alpha\pr,X\pr=\alpha\pl X\pr$, with $\alpha\in\form{\M}$ and $X\in\field{\Ker\psi}$; in the second step, the fact that the Newtonian connection is torsionfree; in the third step, the definition of the exterior derivative of a 1-form. 

Imposing that $\Z{\boldsymbol{\om}}$ vanishes and using the local expression of $\N{\boldsymbol{\om}}$ as $\N{\boldsymbol{\om}}\pl V,W\pr=d\N{A}\pl V,W\pr$ leads to the condition:
\bea
d\pl \N{A}+\chi\pr\pl V,W\pr=0 \hspace{4mm},\hspace{4mm} \forall\, V,W\in\field{\Ker\psi}\label{eqcondZ}. 
\eea
Using the involutivity of the distribution induced by $\Ker\psi$, one can show that the condition \eqref{eqcondZ} implies, locally, that
\bea
\exists f\in C^\infty\pl\M\pr \hspace{4mm}/\hspace{4mm} \chi\pl V\pr=-\N{A}\pl V\pr+df\pl V\pr\nn \hspace{4mm},\hspace{4mm} \forall V\in\field{\Ker\psi}. 
\eea
Therefore, there exists a function $f$ on $\M$ such that $Z=N-h\pl\N{A}\pr+h\pl df\pr$. 

\paragraph{Proof of Proposition \ref{propbijcorNewtoZphi}}
~\\
As noted earlier, the isomorphism between items 1 and 3 is ensured by \Prop{propaffine}. We now prove the isomorphism between items 2 and 3, starting with the implication $\pl Z,\phi\pr\Rightarrow g$:
\blem{}{\label{lemfromZphitog}Let $\mathscr S\pl \M, \psi, \gamma\pr$ be an Augustinian structure, $Z\in \FO$ a field of observers and $\phi\in C^\infty\pl\M\pr$ a function on $\M$. The metric $g\in\bforms$ defined as:
\bea
g\equiv\Z{\gamma}+\phi\, \psi\vee\psi\nn
\eea
with $\Z{\gamma}$ the metric transverse to $Z$, is the only Lagrangian metric satisfying 
\bea
g\pl Z\pr=\phi\,\psi\,. \label{eqpropconverselag}
\eea
}
\proof{ ~\\
Let $g\in\bforms$ be an arbitrary covariant metric on $\M$. The decomposition of $g$ on the Galilean basis $\pset{Z,e_i}$ (with dual basis $\pset{\psi, \theta^i}$) reads:
\bea
g=g\pl Z,Z\pr \psi\vee\psi+2g\pl Z,e_i\pr\psi\vee\theta^i+g\pl e_i,e_j\pr\theta^i\vee\theta^j. \nn
\eea
Requiring that the Lagrangian metric $g$ satisfies the condition \ref{eqpropconverselag} reduces its expression to:
\bea
g=\phi\,\psi\vee\psi+\gamma\pl e_i,e_j\pr\theta^i\vee\theta^j\nn
\eea
where the second term is nothing but $\Z{\gamma}$. 
}

A statement converse to \Lemma{lemfromZphitog}, \ie the implication $g \Rightarrow\pl Z,\phi\pr $, can be formulated as follows:
\blem{}{\label{lemLagZm}Let $\mathscr S\pl \M, \psi, \gamma\pr$ be an Augustinian structure and $g\in\bforms$ be a Lagrangian metric on $\M$. There is a unique couple $\pl Z,\phi\pr$, with $Z\in \FO$ a field of observers and $\phi\in \fonc{\M}$ a function such that:
\bea
g\pl Z\pr=\phi\,\psi.
\eea
}
\proof{ ~\\
We start by proving that the condition $g\pl X,Y\pr=\gamma\pl X,Y\pr$, $\forall\, X,Y\in\field{\Ker\psi}$ implies that $\Rad g\cap\Ker\psi=\lbrace0\rbrace$. Suppose there exists a vector field $v\in\field{T\M}$ such that $g\pl v\pr=\psi\pl v\pr=0$. Since $\psi\pl v\pr=0$, $g\pl v,w\pr=\gamma\pl v,w\pr=0$, $\forall w\in\field{\Ker\psi}$, which leads to a contradiction since $\gamma$ is positive definite. In conclusion, such a vector field $v$ does not exist and $\Rad g\, \cap\Ker\psi=\lbrace0\rbrace$. 

The positive definiteness of $\gamma$ implies also that the dimension of $\Rad g$ is either 0 or 1, so that we will distinguish these two cases:
\vspace{4mm}

\noindent\fbox{$\Dim\pl \Rad g\pr=1$}
\vspace{4mm}

\noindent Let $v\in\field{T\M}$ such that $\Rad g=\Span{v}$. The defining relation for $Z$ and $\phi$ then implies $g\pl Z,v\pr=\phi\, \psi\pl v\pr=0$, which in turn ensures $\phi=0$, since $\psi\pl v\pr\neq 0$ in virtue of the precedent discussion. Then, one obtains $g\pl Z\pr=0$ so that $Z\in\Rad g$, \ie $Z\sim v$. The normalization condition $\psi\pl Z\pr=1$ fixes $Z$ uniquely.  
\vspace{4mm}

\noindent\fbox{$\Dim\pl \Rad g\pr=0$}
\vspace{4mm}

\noindent Since the metric $g$ is now assumed to be nondegenerate, one can introduce its inverse  $g\un\in\bform$. Acting on each side of the defining equation for $Z$ and $\phi$ with $g\un$, one gets $Z=\phi\, g\un\pl \psi\pr$. Acting now with $\psi$ on each side leads to $\phi\, g\un\pl\psi,\psi\pr=1$, so that $\phi=${\Large$\frac{1}{g\un\pl \psi,\psi\pr}$}. Plugging back into the expression for $Z$ leads to $Z=$ {\Large$\frac{g\un\pl \psi\pr}{g\un\pl \psi,\psi\pr}$}. We summarise our results in the following table: 
\vspace{2mm}
\hspace{1cm}
\setlength{\extrarowheight}{2 mm}
\begin{tabular}{|c|c|c|}
   \hline
\rule[-0.6cm]{0cm}{1.2cm}$\Dim\pl\Rad g\pr$&Definition of $\phi$&Definition of $Z$\tabularnewline\hline
\rule[-0.6cm]{0cm}{1.2cm}1&0&$\lbrace Z\in\Rad g, \psi\pl Z\pr=1\rbrace$
\tabularnewline
   \hline 
\rule[-0.6cm]{0cm}{1.2cm}0&{$\phi=$\Large$\frac{1}{g\un\pl \psi,\psi\pr}$}&{$Z=$\Large $\frac{g\un\pl \psi\pr}{g\un\pl \psi,\psi\pr}$}\tabularnewline
   \hline 

\end{tabular}
\captionof{table}{Lagrangian variables}
\vspace{2mm}

\noindent Note that $\phi=0$ if and only if $\Dim\pl\Rad g\pr=1$. 
}

\pagebreak

\bibliographystyle{/Users/Morand/Desktop/bibliographie/hunsrt.bst}
\bibliographystyle{hep.bst}

\providecommand{\href}[2]{#2}\begingroup\raggedright\endgroup
\end{document}